\documentclass[a4paper,fleqn]{cas-sc}

\usepackage{textcomp} % 引入宏包用于输出特殊符号
\usepackage{amsmath,amsfonts}
\usepackage{algorithm}
\usepackage{amssymb}
\usepackage{bbding}
\usepackage{array}
\usepackage[caption=false,font=footnotesize,labelfont=rm,textfont=rm]{subfig}
\usepackage{stfloats}
\usepackage{url}
\usepackage{verbatim}
\usepackage{graphicx}
\usepackage{amssymb}
\usepackage{amsmath}
\usepackage{supertabular}
\usepackage{multirow}
\usepackage{threeparttable}
\usepackage{booktabs}
\usepackage{mathrsfs}
\usepackage[noend]{algpseudocode}
\usepackage{longtable}
\usepackage[numbers,sort&compress]{natbib}
\usepackage{adjustbox}
\usepackage{tabularx}
\usepackage{tcolorbox}
\usepackage{caption}
\usepackage{setspace}
\def\tsc#1{\csdef{#1}{\textsc{\lowercase{#1}}\xspace}}
\tsc{WGM}
\tsc{QE}
\begin{document}
\let\WriteBookmarks\relax
\def\floatpagepagefraction{1}
\def\textpagefraction{.001}
\let\printorcid\relax

%\shorttitle{<short title of the paper for running head>}    

%\shortauthors{<short author list for running head>}  

\title [mode = title]{Deep reinforcement learning-based joint real-time energy scheduling for green buildings with heterogeneous battery energy storage devices}  

% Author

% eg: \author[1,3]{Author

\author[First,Second]{Chi Liu}
% \ead{220110009@fzu.edu.cn}

\author[First,Second]{Zhezhuang Xu}
% \ead{zzxu@fzu.edu.cn}

% \ead{meng.yuan@ieee.org}

\author[First,Second]{Jiawei Zhou}
% \ead{230120084@fzu.edu.cn}

\author[Third]{Yazhou Yuan}
% \ead{yzyuan@ysu.edu.cn}

\author[Third]{Kai Ma}
% \ead{kma@ysu.edu.cn}

\author[First,Fourth]{Meng Yuan}
\cormark[1]

%\credit{<Credit authorship details>}

\cortext[cor]{Corresponding author (meng.yuan@ieee.org).}

\address[First]{College of
Electrical Engineering and Automation, Fuzhou University, Fuzhou, 350108, Fujian, China}

\address[Second]{Key Laboratory of Industrial Automation Control Technology and Information Processing,
Education Department of Fujian Province, Fuzhou, 350000, Fujian, China}

\address[Third]{School of Electrical Engineering, Yanshan University, Qinhuangdao, 066004, Hebei, China}

\address[Fourth]{Department of Electrical Engineering, Chalmers University of Technology, Gothenburg, 41296, Sweden}

% \cortext[1]{Corresponding author} 

\begin{abstract}
Green buildings (GBs) equipped with renewable energy resources and building energy management systems (BEMS), enabling efficient energy utilization and playing a vital role in sustainable societal development. Electric vehicles (EVs), as flexible energy storage resources, enhance system flexibility and arbitrage potential when integrated with stationary energy storage system (ESS) for real-time energy scheduling within BEMS. However, the distinct degradation and operational characteristics of ESS and EVs hinder the development of effective energy scheduling strategies. To address this challenge, this paper presents a model-free deep reinforcement learning (DRL) joint real-time energy scheduling method based on a combined battery system (CBS). Firstly, we construct the CBS, which incorporates both ESS and EVs, to improve energy utilization efficiency. We analyze the characteristics of diverse battery types within the CBS and establish precise degradation models for ESS and EVs, providing reliable cost estimates for BEMS operations. For energy management, we prioritize the travel demands of EV users to encourage their participation in the scheduling process. Meanwhile, the collaborative operation of ESS and EVs under varying scheduling windows and degradation costs is considered. To optimize local energy consumption and operational costs, we introduce a prediction model that facilitates effective energy interaction between the CBS and BEMS. Finally, to tackle the challenges of heterogeneous state features, action coupling, and declining learning efficiency in a multi-storage environment, an energy scheduling algorithm based on DRL is proposed. This algorithm integrates double networks and a dueling mechanism to optimize Q-value estimation and action selection, while prioritized experience replay enhances efficient learning. Experimental results demonstrate the effectiveness of proposed approach, achieving a 37.94\%-40.01\% reduction in system operating costs compared to mixed-integer linear programming (MILP) approach. 
\end{abstract}



% Keywords
\maketitle

% Main text
\section{Introduction}
% Improtance of green building
Against the backdrop of the global climate crisis and accelerated urbanization, buildings are the primary sources of urban energy consumption and carbon emissions \cite{hwang2025dt}. Efficient utilization and optimized management of building energy are crucial for achieving sustainable urban and societal development \cite{liu2025building} \cite{yadollahi2024optimal}. Traditional management practices in building energy systems have resulted in significant energy waste \cite{onile2023energy}. Green buildings (GB) equipped with building energy management system (BEMS) present a promising solution to these environmental and energy challenges \cite{liu2025building}. BEMS integrates distributed energy resources (DERs) like photovoltaic (PV) systems and energy storage system (ESS), to achieve operational coordination and supply-demand complementarity, significantly enhancing the energy utilization \cite{zhai2025risk}. This approach also supports efforts to mitigate climate change, enhance social benefits, and meet resident demands \cite{tavana2021integrated}. Therefore, the development of effective energy management strategies to promote energy-saving and low-carbon operations of BEMS is of paramount importance \cite{yoon2018multiple}.

% Improtance of ESD, two catagories: 1. is stationary energy storage battery, 2. EV battery. -> Benefit of EV, and related E2G technology. Current research on energy management based on stationary and EV batteries.
Building upon the capabilities of BEMS, energy storage devices (ESDs) are critical to achieving real-time energy optimization, offering both flexibility and reliability in managing supply-demand dynamics \cite{yoon2018multiple}. In the context of BEMS, ESDs can be broadly classified into stationary ESS and mobile storage resources such as electric vehicles (EVs), which can provide flexibility through vehicle-to-grid (V2G) integration. By utilizing information on energy prices and supply-demand dynamics, stationary ESS can store excess energy during off-peak periods, and release it during peak demand times, alleviating pressure on the power grid (PG) \cite{8931747}. This not only reduces the operating costs of the system but also supports the balance of electricity supply and demand. ESS collaborate effectively with distributed energy components in BEMS, mitigating system disturbances caused by uncertainties in PV generation and user demands. Additionally, ESS serve as a backup power source during grid failures, enhancing system resilience and reliability. For the mobile ESDs, the EVs utilize V2G technology, allowing energy to be injected into the grid and widely adopted in home energy management systems \cite{ortega2014optimal} \cite{wan2018model}. V2G technology transforms EVs from passive energy consumers into active participants in energy management, creating a bidirectional flow of energy and value that significantly enhances the system's operational flexibility and resilience \cite{8048016TII}. Commercial building EV charging stations inherently aggregate larger-scale EV fleets compared to decentralized home energy systems, presenting unique opportunities to leverage battery mobility and scalability for enhanced energy management efficiency.

% Compared to home energy management, EV charging stations in commercial buildings aggregate more EV resources. It is important to fully utilize the mobility and scalability of EV batteries to improve energy management efficiency.
% form combined battery systems (CBS) with stationary ESS, providing significant advantages in energy scheduling for commercial buildings. These benefits include reduced investment in ESS capacity and improved flexibility of the CBS and energy utilization efficiency \cite{8048016TII} \cite{6901266SmartGrid}. Current research has extensively explored the operational mechanisms and control methods of ESDs \cite{8931747} \cite{zhang2024load} \cite{cao2020deep} \cite{thomas2018optimal}, aiming to formulate more economical, efficient, and flexible energy scheduling strategies.

% Benefit of using the heterogeneous battery energy storage system instead of stand-alone two systems. Shortcoming of current research: 1. Degradation characteristics of the two batteries are different; 2. The different degradation means different cost and making the scheduling difficuilt. EV driving range and its transporation importance should be prioritized. 
Compared to individual battery storage systems, EVs combined with stationary ESS form a combined battery system (CBS) that offers clear advantages in energy dispatch within BEMS. These benefits include reducing the initial investment in ESS capacity and enhancing the system's energy throughput in energy arbitrage markets, thereby reducing the overall life-cycle operating costs of the system \cite{6901266SmartGrid}. However, the integration of ESS and EVs within BEMS faces challenges in battery degradation modeling, system model design, and scheduling strategy formulation. Firstly, current research has not accurately modeled the degradation process of heterogeneous battery energy storage systems within the energy management practices of GBs. This oversight leads to deviations in system scheduling strategies and a decline in operational economic efficiency. Battery degradation costs from frequent charging and discharging during energy dispatch represent a key factor in the development of an accurate operational model \cite{cao2020deep}. Some existing studies have overlooked these degradation costs, leading to inaccurate estimates of the system's energy arbitrage capabilities \cite{onile2023energy} \cite{shen2021real}. Increasingly, research is highlighting the importance of battery degradation within energy systems \cite{cao2020deep} \cite{zhang2024load}. However, in energy scheduling modeling, the differing degradation characteristics of ESS and EV batteries are often overlooked. Lithium-ion batteries dominate the energy storage battery market \cite{zhang2024load}. ESS primarily uses lithium iron phosphate (LFP) materials, known for their low cost, high safety, stability, and long lifespan \cite{li2023accelerated} \cite{olmos2021modelling}. In contrast, lithium nickel manganese cobalt oxide (NMC) materials are widely used in EV batteries, offering high energy and power densities that meet the demands for long range and instant high power output \cite{wen2020overview} \cite{lv2024recent}. Compared to ESS, EV batteries generally have shorter lifespans and demonstrate more significant capacity decay, particularly under high temperatures or deep discharge conditions \cite{hu2024review}. To bridge this knowledge gap, it is essential to develop differentiated battery degradation models that account for the distinct chemical properties of heterogeneous batteries within the CBS. This approach will provide a foundational basis for battery health assessment, lifecycle management, and the optimization of charging and discharging strategies.

Due to the varied degradation costs of ESS and EVs, along with the mileage anxiety of EV users, coordinating the joint operation of ESDs within the CBS to improve energy management efficiency becomes challenging \cite{harnischmacher2023two}. The different degradation characteristics of ESS and EVs imply varying scheduling costs. This requires the system to implement more intelligent control strategies within the scheduling window, cautiously using the more expensive batteries to optimize long-term returns while ensuring battery health. On the other hand, EVs, primarily serving as transportation components in BEMS, and their driving range must be prioritized. A decrease in battery capacity can compromise their primary function as passenger vehicles  \cite{zhang2011review}. In most cases, EV users typically prefer to maintain at least the same battery capacity after scheduling as they had at the start, as a reduction would significantly diminish their willingness to participate in scheduling \cite{thomas2018optimal}. Consequently, scheduling plans for EVs must align with users' transportation needs. Faced with the aforementioned challenges, scheduling arrangements must consider EV driving ranges, scheduling windows, and cost disparities between ESS and EVs, facilitating the collaborative operation of energy storage batteries with diverse characteristics.

% Futhermore, the interest of the building managers is to maximize its profit by fully utilized stationary batteries and EV batteries in energy maganment. While the EV users tend to enjoy the potential income while not influencing their daily demand of EV usage. These different demand poses challenges when formulating cost function in designing a optimization problem. cost, reward design, energy allocation and trading. 

The design of financial and energy interaction mechanisms between CBS, which involves multiple stakeholders, and BEMS presents challenges. Building managers and EV users belong to two distinct interest entities \cite{wu2018transactive}. The interest of the building managers is to maximize their profit by fully utilizing stationary ESS and EV batteries in energy management. While the EV users tend to enjoy the potential income without influencing their daily demand for EV usage. These different demands necessitate the construction of a refined objective function design framework in the BEMS, achieving a balance among various stakeholders through clear principles of revenue allocation and cost sharing. Additionally, to optimize local energy consumption and enhance overall system benefits, an energy utilization logic should be established to promote positive interactions between the CBS and BEMS \cite{bu2020}.
% The integration of ESDs into the BEMS framework presents trade-offs between benefits and costs. From the building manager's perspective, energy dispatch aims to minimize operational costs, often overlooking the interests of EV users. For example, frequent scheduling of EVs can negatively impact battery health \cite{zhang2024load}, while BEMS seeks to maximize EV participation for profit. It is essential to clarify the principles of revenue attribution and cost allocation. Based on this understanding, a coherent overall operational strategy should be developed to balance the profits and costs associated with the participation of the CBS in the BEMS scheduling. Furthermore, an energy utilization logic should be established to promote positive interactions between the CBS and BEMS, optimizing local energy consumption and enhancing overall system benefits \cite{thomas2018optimal} \cite{bu2020} \cite{ma2025fuzzy}.

% Interm of designing and implementing control algorithm, the combined energy system introduces more state and action spaces. 

In BEMS that utilize multiple battery storage units, heterogeneous ESDs create complex system state spaces coupled with numerous interdependent control actions, making it challenging to effectively manage the charging and discharging of ESS and EVs to reduce operational costs. Traditional model-based scheduling methods require pre-establishing physical models and defining system parameters \cite{yoon2018multiple}. In contrast, model-free approaches based on deep reinforcement learning (DRL) demonstrate significant advantages in handling dynamic and complex decision-making scenarios \cite{homod2025massive}. In the microgrid environment of GB, data from heterogeneous devices such as PV, EV and ESS, along with dynamic information like energy prices and loads, are interconnected, creating a complex system state space. As a result, mapping the relationship between system states and control actions becomes challenging. The control actions of multi-battery system represent combinatorial arrangements of individual battery actions, leading to a high-dimensional coupled scheduling strategy space. The coordination among control actions of multiple ESDs must consider factors such as costs, scheduling windows, and operational constraints, complicating the optimization of scheduling decisions. Moreover, in the dynamically changing system environment, the ineffective use of training data due to random sampling can adversely impact the quality of scheduling decisions. Therefore, there is an urgent need to design a high-performance DRL framework tailored for multi-battery energy scheduling, addressing the challenges of complex coupling in state and action spaces, as well as the declining efficiency of sample learning. 

To address above challenges, this paper proposes a DRL-based joint real-time energy scheduling method aimed at optimizing the operational costs of BEMS. First, a multi-battery storage system is constructed to enhance energy utilization. Based on this system, an accurate multi-battery degradation model is developed, taking into account the distinct electrochemical characteristics of the ESS and EV, to estimate the actual arbitrage potential of the energy system. By ensuring that the state of charge (SoC) of EVs upon departure is no less than their SoC upon arrival, the travel needs of EV users are explicitly satisfied. Simultaneously, a prediction network is designed to guide the CBS in interacting with the system based on an energy allocation mechanism (EAM) to optimize operational efficiency. Finally, an improved deep Q-network (DQN) algorithm is proposed to solve the energy dispatch problem in a multi-energy storage battery environment. This algorithm combines double networks and a dueling mechanism to optimize Q-value estimation and action selection of DQN for dynamic complex systems, while prioritized experience replay enhances learning efficiency.

The contributions of this work are summarized as follows:
\begin{enumerate}
	
    \item To enhance system flexibility and energy utilization efficiency, a CBS comprising various types of batteries is constructed. Based on the distinct characteristics of ESS and EV batteries, precise aging models tailored for each battery type are investigated, filling the knowledge gap in quantifying degradation costs for heterogeneous battery systems under DRL-based energy management.
    % To the best of our knowledge, this aspect has not been previously explored.
	
    \item We explicitly consider the requirement of EV users while encouraging their participation in the building energy management system where the priority is given to satisfying the EVs range demand. Meanwhile, the energy management fully considers the differentiated degradation costs and scheduling windows of the ESS and EV. It designs degradation scaling coefficients based on battery cost sensitivity to coordinate the operation of multi-battery energy storage system.

    \item The cost allocation and benefit attribution of heterogeneous ESDs participating in BEMS energy management are clarified and incorporated into scheduling decisions, preventing excessive utilization of EV batteries. A prediction network is designed to be embedded in the energy dispatch process, utilizing forecasting information to guide CBS in positively interacting with BEMS based on energy utilization priorities. This approach reduces operational costs while promoting local energy consumption.

    \item The energy management problem is formulated as a Markov Decision Process (MDP) problem, and a DQN-based joint real-time scheduling algorithm is proposed to derive the control strategy for the CBS. To address the challenge of analyzing heterogeneous and coupled state features, double networks are introduced to improve the accuracy of state-action mapping. The dueling mechanism is employed for Q-function approximation, optimizing control action selection under joint scheduling decisions. Additionally, prioritized experience replay enhances the sample learning efficiency in dynamic environments.

    \item The method has been validated in real commercial building environments, demonstrating the effectiveness of the proposed scheduling method across different seasons and energy consumption patterns.
	
\end{enumerate}

% The remainder of this paper is organized as follows. Section 2 reviews related works. In Section 3, we model the operation of each component in the system. Then, Section 4 presents the framework for solving the optimization energy scheduling problem using the DRL-based algorithm. Section 5 provides a detailed description of the implementation steps of the proposed method. In Section 6, numerical experiments are conducted to validate the effectiveness of the proposed approach, followed by an analysis and discussion of the results. Finally, Section 7 summarizes the paper and discusses future research directions.

\renewcommand{\arraystretch}{1.2}
\begin{tcolorbox}[colback=white, colframe=black, title={}, boxrule=0.5pt]

\textbf{Nomenclature}

\vspace{1ex}

\noindent
\begin{tabularx}{\textwidth}{@{}>{\raggedright\arraybackslash}p{0.12\textwidth}X >{\raggedright\arraybackslash}p{0.12\textwidth}X@{}}
\textbf{Abbreviations} & & & \\[0.5ex]
% GB & Green building & DQN & Deep Q-networks \\
% BEMS & Building energy management system & MILP & Mixed-integer linear programming \\
% EVs & Electric vehicles & MDP & Markov Decision Process \\
% ESDs & Energy storage devices & DERs & Distributed energy resources \\
% ESS & Energy storage system & EAM & Energy allocation mechanism \\
% LFP & Lithium iron phosphate & NMC & Nickel manganese cobalt \\
% CBS & Combined battery system & PV & Photovoltaic \\
% DRL & Deep reinforcement learning & SoC & State of charge \\
% DoD & Depth of discharge & V2G & Vehicle-to-grid \\
% RL & Reinforcement learning & SEI & Solid electrolyte interphase \\
% TD & Temporal difference & O-NET & Online network \\
% T-NET & Target network & D2QN & Double DQN \\
% D3QN & Double dueling DQN & ERM & Experience replay mechanism \\
% PER & Prioritized experience replay & RVFL & Random vector functional link \\
% BOA & Bayesian optimization algorithm & MSE & Mean square error \\
% RMSE & Root mean square error & MAE & Mean absolute error \\
% MASE & Mean absolute scaled error & $R^2$ & $R$-square \\
% LSTM & Long short-term memory & SoH & State of health \\
% CSC & Cumulative scheduling cost & CSP & Cumulative scheduling power \\
% RDedRVFL & Ranking-based dynamic ensemble deep RVFL & ARIMA & Auto-regressive integrated moving average\\[1.5ex]

ARIMA & Auto-regressive integrated moving average & LSTM & Long short-term memory \\
BOA & Bayesian optimization algorithm & MAE & Mean absolute error \\
BEMS & Building energy management system & MASE & Mean absolute scaled error \\
CBS & Combined battery system & MDP & Markov Decision Process \\
CSC & Cumulative scheduling cost & MILP & Mixed-integer linear programming \\
CSP & Cumulative scheduling power & MSE & Mean square error \\
D2QN & Double DQN & NMC & Nickel manganese cobalt \\
D3QN & Double dueling DQN & O-NET & Online network \\
DERs & Distributed energy resources & PER & Prioritized experience replay \\
DQN & Deep Q-networks & PV & Photovoltaic \\
DoD & Depth of discharge & RDedRVFL & Ranking-based dynamic ensemble deep RVFL \\
DRL & Deep reinforcement learning & RL & Reinforcement learning \\
EAM & Energy allocation mechanism & RMSE & Root mean square error \\
ERM & Experience replay mechanism & RVFL & Random vector functional link \\
ESDs & Energy storage devices & SEI & Solid electrolyte interphase \\
ESS & Energy storage system & SoC & State of charge \\
EVs & Electric vehicles & SoH & State of health \\
GBs & Green buildings & TD & Temporal difference \\
LFP & Lithium iron phosphate & T-NET & Target network \\
$R^2$ & $R$-square & V2G & Vehicle-to-grid \\[1.5ex]
\multicolumn{4}{l}{\hspace*{-0.01\textwidth}\textbf{Parameters and variables}} \\

${P}^\text{PV}(t)$ & PV generation output at $t$ & $\text{SoC}^\text{EV}(t_\text{arr})$ & The state of charge for all EVs upon arrival \\
${P}^\text{PV}_\text{min}$ & PV minimum output & $\text{SoC}^\text{EV}(t_\text{off})$ & The state of charge for EVs during offline period \\
${P}^\text{PV}_\text{max}$ & PV maximum output & $\text{SoC}^\text{EV}(t_\text{dep})$ & The state of charge for EVs upon departure \\
$P^\text{load}(t)$ & Building load at $t$ & $P^{\text{CBS,ch}}(t)$ & CBS charging power at $t$ \\
${P}^\text{net}(t)$ & Building net load at $t$ & $P^{\text{CBS,dis}}(t)$ & CBS discharging power at $t$ \\
$P^\text{ESS,ch}(t)$ & ESS charging power at $t$ & $P^{\text{ESS,PG}}(t)$ & Power from ESS injected into the grid at $t$ \\
$P^\text{EV,ch}(t)$ & EVs charging power at $t$ & $P^{\text{ESS,build}}(t)$ & Power supplied by ESS to the building at $t$ \\
$P^\text{ESS,dis}(t)$ & ESS discharging power at $t$ & $P^{\text{EV,PG}}(t)$ & Power from EVs injected into the grid at $t$ \\
$P^\text{EV,dis}(t)$ & EVs discharging power at $t$ & $P^{\text{EV,build}}(t)$ & Power supplied by EVs to the building at $t$ \\
$P^\text{ESS,ch}_\text{min}$ & ESS minimum charging power & $P^\text{PG,ESS}(t)$ & Power supplied by the grid for ESS charging at $t$ \\
$P^\text{EV,ch}_\text{min}$ & EVs minimum charging power & $P^\text{PG,EV}(t)$ & Power supplied by the grid for EVs charging at $t$ \\

\end{tabularx}
\end{tcolorbox}

\vspace{1em}

% ---------- 第二页：参数剩余 ----------
\begin{tcolorbox}[colback=white, colframe=black, title={}, boxrule=0.5pt]
\textbf{Nomenclature}

\vspace{1ex}

\noindent
\begin{tabularx}{\textwidth}{@{}>{\raggedright\arraybackslash}p{0.14\textwidth}X >{\raggedright\arraybackslash}p{0.14\textwidth}X@{}}
\multicolumn{4}{l}{\hspace*{-0.01\textwidth}\textbf{Parameters and variables (Part 2)}} \\
$P^\text{ESS,ch}_\text{max}$ & ESS maximum charging power & $c^{b}(t)$ & Energy price purchased from the grid at $t$ \\
$P^\text{EV,ch}_\text{max}$ & EVs maximum charging power & $c^{s}(t)$ & Energy price sold back to the grid at $t$ \\
$P^\text{ESS,dis}_\text{min}$ & ESS minimum discharging power & $\Delta Q_{\text{cal}}$ & Capacity fade of calendar aging \\
$P^\text{EV,dis}_\text{min}$ & EVs minimum discharging power & $\Delta Q_{\text{cycle}}$ & Capacity fade of cycle aging \\
$P^\text{ESS,dis}_\text{max}$ & ESS maximum discharging power & $f_{d,1}$ & Battery degradation rate \\
$P^\text{EV,dis}_\text{max}$ & EVs maximum discharging power & $\alpha ^\text{ESS}_{d,j}$ & Cycle aging degradation coefficient for ESS of episode $j$ \\
$\mu^\text{ESS}$ & Binary variable of charge/discharge mutual exclusivity for ESS & $\alpha ^\text{EV}_{d,j}$ & Cycle aging degradation coefficient for EV of episode $j$ \\
$\mu^\text{EV}$ & Binary variable of charge/discharge mutual exclusivity for EVs & $C^\text{ESS}_\text{cycle}(t)$ & ESS cycle aging cost at $t$ \\
$\text{SoC}^\text{ESS}(t)$ & State of charge for ESS at $t$ & $C^\text{EV}_\text{cycle}(t)$ & EV cycle aging cost at $t$ \\
$\text{SoC}^\text{EV}(t)$ & State of charge for EVs at $t$ & $C^\text{ESS}_{\text{cal},j}$ & ESS calendar aging cost over episode $j$ \\
$\text{SoC}^\text{ESS}(t-1)$ & State of charge for ESS at $t-1$ & $C^\text{EV}_{\text{cal},j}$ & EV calendar aging cost over episode $j$ \\
$\text{SoC}^\text{EV}(t-1)$ & State of charge for EVs at $t-1$ & $C^\text{total}$ & Total battery degradation cost over period $T$ \\
$E^\text{ESS}$ & Energy capacity of battery for ESS & $C^\text{EV}$ & Battery degradation cost borne by EV \\
$E^\text{EV}$ & Energy capacity of battery for EVs & $C^\text{build}$ & Battery degradation cost borne by building \\
$\eta^\text{ESS,ch}$ & ESS charging efficiency & $P^\text{PV,rev}(t)$ & PV selling revenue at $t$ \\
$\eta^\text{EV,ch}$ & EVs charging efficiency & $P^\text{PG,cost}(t)$ & Energy purchase cost from the grid at $t$ \\
$\eta^\text{ESS,dis}$ & ESS discharging efficiency & $\phi^\text{pri}$ & Electricity price ratio \\
$\eta^\text{EV,dis}$ & EVs discharging efficiency & $\phi^\text{net}$ & Net load ratio \\
$\text{SoC}^\text{ESS}_\text{min}$ & Minimum ESS state of charge & $\theta_\text{ESS}$ & ESS aging parameter \\
$\text{SoC}^\text{EV}_\text{min}$ & Minimum EVs state of charge & $\theta_\text{EV}$ & EV aging parameter \\
$\text{SoC}^\text{ESS}_\text{max}$ & Maximum ESS state of charge & $\theta_\text{scale}$ & Scaling  aging parameter \\
$\text{SoC}^\text{EV}_\text{max}$ & Maximum EVs state of charge & $\theta_\text{base}$ & Base aging parameter \\
$\text{SoC}^\text{EV}_k(t_\text{arr})$ & The state of charge for the $k$-th EV upon arrival & & \\

\end{tabularx}
\end{tcolorbox}

\section{Literature review}
\subsection{Degradation cost of ESDs within energy systems}
In energy systems, ESDs are crucial for energy arbitrage, ensuring system flexibility, and maintaining stability. The authors of \cite{onile2023energy} introduce embedded energy storage technologies within the grid, enhanced by reinforcement learning (RL) control and recommendation systems to improve grid reliability and achieve self-consumption and demand response goals. In \cite{shen2021real}, a microgrid is considered, which includes components such as EV stations, combined heat and power systems, and external natural gas stations. In this framework, EVs are utilized as ESDs to regulate system operations. However, these studies neglect to account for the additional degradation costs that arise from frequent charging and discharging during the scheduling process.

The degradation mechanisms of storage batteries have gradually garnered attention \cite{ortega2014optimal} \cite{wan2018model} \cite{cao2020deep}  \cite{xu2016modeling}. In \cite{ortega2014optimal}, a linear degradation model for lithium batteries is proposed, which is utilized in \cite{wan2018model} for evaluating the degradation costs of EVs in home energy management scenarios. This model assumes that degradation costs are only sensitive to the number of cycles. In real-world conditions, lithium battery degradation is influenced by various factors such as temperature and SoC, making it a typical nonlinear evolutionary process. Failing to account for calendar aging and the aforementioned stress factors, and merely describing the relationship between cycle count and battery degradation cost with a simple linear equation will lead to significant estimation errors. Xu et al. \cite{xu2016modeling} considers important factors affecting battery aging, such as SoC, time, depth of discharge (DoD) and temperature, establishing a more accurate semi-empirical battery degradation model. However, the battery aging calculations in \cite{xu2016modeling} can only estimate degradation over a specific time period. This limitation makes it impossible to compute the aging costs associated with each control action in the DRL-based energy management framework, resulting in reward delay issues. To address this, Cao et al. \cite{cao2020deep} proposed a degradation coefficient to estimate degradation costs for individual time steps. Nevertheless, this model does not clearly separate the degradation costs of calendar aging from those of cycle aging, which hinders accurate cost allocation.

In summary, the aforementioned studies have not considered the degradation modeling of multi-battery system within BEMS. Due to the significant differences in the uses and chemical properties of ESS and EV batteries, establishing a heterogeneous battery degradation cost assessment framework is crucial for the formulation of system energy scheduling strategies.

\subsection{Energy scheduling with the participation of EV}
By participating in energy scheduling as flexible energy storage resources, EV significantly enhance system flexibility and boost energy utilization efficiency. A coordinated matching strategy for a multi-energy supply system, combining random EVs, stationary battery storage, PV, and PG, is proposed in \cite{bai2024collaborative}. This study designed a supply-demand matching operation strategy without altering the electricity consumption habits of office building users. Mohammad et al. \cite{mohammad2020transactive} examine an interactive energy management system for a commercial parking lot equipped with V2G functionality and rooftop PV. This system optimizes EV charging station operational costs through a flexible bidding mechanism for bidirectional energy trading. Expanding on V2G, the concept of building-to-vehicle-to-building is discussed in \cite{buonomano2020building}, utilizing EVs as energy carriers to facilitate power exchange between buildings, thus achieving more efficient energy management. Additionally, research by \cite{wang2025coordinated} and \cite{tan2019multi} focuses on response costs and reactive power support issues related to EV participation in grid scheduling. However, these studies do not address the mileage constraints of EVs as primarily transportation-based energy components, which dampens EV users' enthusiasm for scheduling. Furthermore, the collaborative operational mechanisms under the differentiated characteristics of ESS and electric vehicles have not been explored. Given the transportation attributes of EVs, they have more flexible scheduling windows compared to ESS, necessitating that scheduling occurs within reasonable time frames. The scheduling costs for EVs are often higher than for ESS, and they are more sensitive to battery health. Thus, it may be a more prudent approach to schedule EVs only when the overall benefits are optimal.

\subsection{Energy utilization in the energy system}
A reasonable energy utilization mechanism supports local energy consumption and stabilizes the operation of the PG. Thomas et al. \cite{thomas2018optimal} examine the collaborative operation of BEMS that includes various DERs such as ESS and PV. It explores how energy utilization priority factors impact total system costs. In \cite{zhang2024load}, an industrial energy storage management system is established, where the system's electricity demand is primarily met by PV, with any shortfall supplied by the grid. The study emphasizes that the price of purchased electricity should exceed the price for selling back to the PG to reflect market supply and demand dynamics. Ma et al. \cite{ma2025fuzzy} focus on the uncertainties in multi-energy systems and proposes a two-stage scheduling method based on fuzzy logic and model predictive control to optimize energy distribution. A bi-level regional integrated energy system optimization scheduling model is constructed in \cite{liu2025generalized} to enhance the management efficiency of ESS. This model formulates a game strategy to optimize energy utilization processes between gas power plant carbon capture  equipment and storage devices. However, these studies do not consider the energy exchange logic between multiple batteries and the energy system. Given the energy price differential between purchase and sale, prioritizing supply to buildings can yield higher profits, thereby reducing operating costs and promoting local energy consumption. It is crucial to guide the agent to develop optimized energy utilization strategies within the DRL scheduling framework, ensuring operational efficiency of the system.

\subsection{DRL-based energy management method}

Model-based energy scheduling methods require the pre-definition of physical models and system parameters \cite{yoon2018multiple} \cite{kassab2024optimal}. To achieve optimized design and operation of microgrids, \cite{kassab2024optimal} integrates microgrid scale and energy management issues into a single decision-making framework, proposing a joint multi-objective mixed-integer linear programming algorithm to minimize energy costs and lifecycle emissions. An energy management system for multi-driven building is introduced in \cite{yoon2018multiple}. It presents an overall operational scheme based on grid electricity utilization and the distribution of renewable energy, demonstrating the system's effectiveness by analyzing the economic benefits derived from energy management. 

Due to their self-learning capabilities, model-free DRL methods have found widespread application in complex energy scheduling scenarios \cite{yadollahi2024optimal} \cite{homod2025massive} \cite{hamdi2021lora}. In \cite{yadollahi2024optimal}, DRL is employed to arrange the scheduling plan of an energy hub's management system, optimizing the operational costs of combined heat and power generation. To tackle peak energy demand and rising costs, a deep reinforcement clustering adaptive decision policy algorithm is proposed in \cite{homod2025massive}, enhancing the operational efficiency of chiller units by optimizing charge and discharge processes during low cooling demand periods. A DRL-based green LoRa wireless network system powered by a mix of grid and renewable energy is presented in \cite{hamdi2021lora}, capable of handling intermittent energy supply while reducing energy consumption, provided that the system's service quality requirements are met. However, existing works do not address the DRL design framework within multi-storage energy management systems. It is essential to consider the impact of the complex, heterogeneous state space and the high-dimensional coupled action space on DRL scheduling decisions. Additionally, mitigating the decline in learning efficiency due to random sampling is necessary.

\section{Description of system model}

\begin{figure*}[th]
	\centering
	\includegraphics[width=0.95\textwidth]{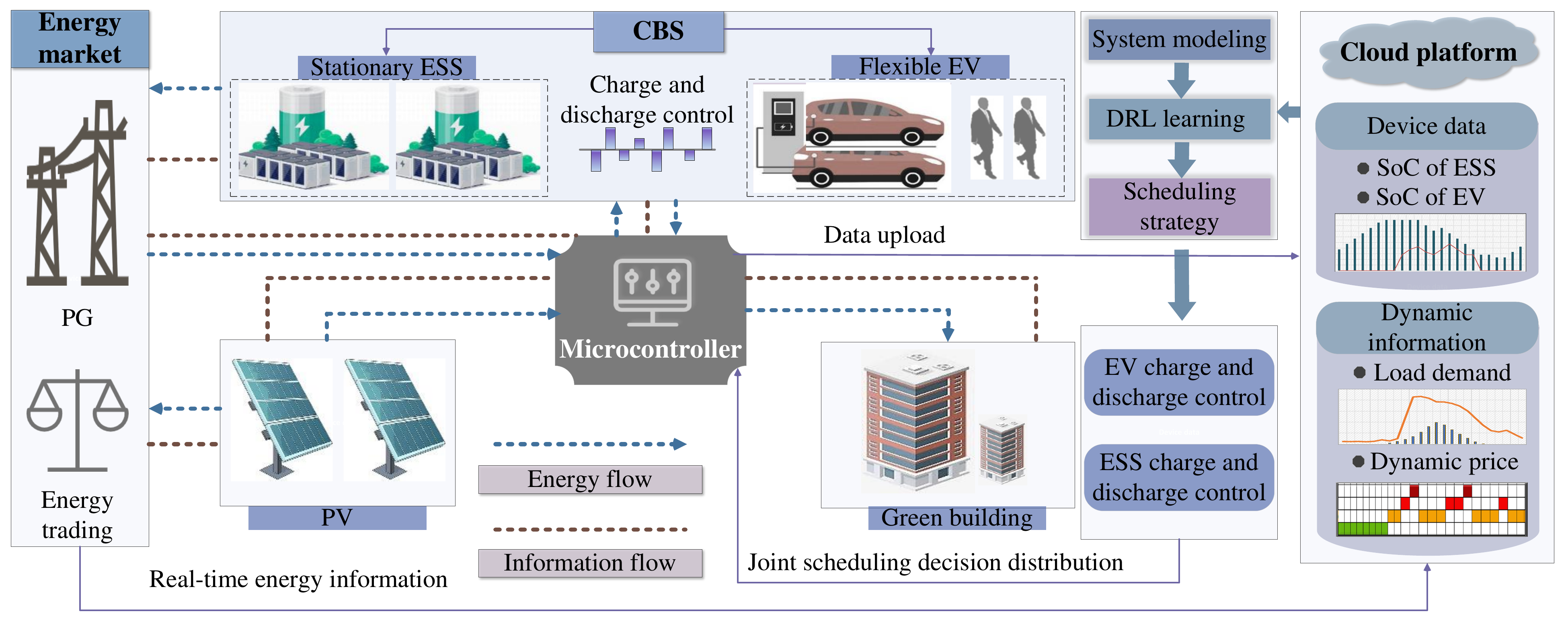}
	\caption{The framework of the system model.}
	\label{system}
\end{figure*}

\subsection{Overview of the joint real-time energy scheduling system model}
In this paper, a joint real-time energy scheduling model is proposed. This model is provided  to investigate the collaborative assessment of multiple DERs in GB, aiming to achieve low-carbon and cost-effective operation of BEMS. The system model we consider includes the PG, PV renewable energy generation, a stationary ESS, and an electric vehicle parking lot equipped with charging stations and V2G capabilities, as depicted in Fig. \ref{system}. The EVs are regarded as flexible energy storage, managed collectively by an EV aggregator, and work in conjunction with the ESS to form a CBS that participates in the building's energy scheduling. To promote local energy consumption, the energy flow from the CBS, PG, and PV follows a specific priority order to meet the building's real-time electricity demand. Specifically, the power generated by the PV has the highest priority and is utilized first to power the building. If the PV output is insufficient to meet the entire load, the CBS is activated to release stored energy as a supplement. If a deficit remains after the CBS discharges energy, electricity is purchased from the PG as a last resort to ensure a stable energy supply. While meeting the building's energy demands, both the PV and CBS can sell excess energy to the grid at market prices, generating revenue and optimizing operational costs. To encourage EV users to participate in BEMS scheduling, the building management will cover the costs associated with battery degradation resulting from their participation in the scheduling process. Furthermore, to address range anxiety, the BEMS ensures that the remaining battery level of EVs upon departure is not less than upon arrival, and any excess energy is not charged for recharging. The BEMS intelligently evaluates the benefits and costs of EV participation in scheduling from the building manager's perspective, balancing operational efficiency and resource allocation. When the EVs are parked, the EMS can control the CBS. When the EVs leave, the EMS focuses solely on managing the charge and discharge of the ESS. The BEMS operates in discrete time steps $t\in\{1,2,...,T\}$, with all control operations executed at each time step $t$. The operational mathematical modeling and constraints of the BEMS are presented in detail in the following subsections.

\subsection{PV renewable energy generation}
The renewable energy generation output ${P}^\text{PV}(t)$ is provided by PV at time step $t$, which satisfies the following constraints:
\begin{align}
{P}^\text{PV}_{\min} \leq {P}^\text{PV}(t) \leq {P}^\text{PV}_{\max},
\end{align}
where ${P}^\text{PV}_\text{min}$ and ${P}^\text{PV}_\text{max}$ represent the minimum and maximum power generation of PV respectively. The ${P}^\text{PV}(t)$ should be prioritized to meet building load demand and the excess will be sold to the PG for revenue \cite{zhang2024load}. This energy supply priority can be expressed as:
\begin{align}
P^{\text{net}}(t) = P^{\text{load}}(t) - P^{\text{PV}}(t),
\end{align}
where $P^{\text{load}}(t)$ and ${P}^\text{net}(t)$ represent the load and net load at time step $t$, respectively. When the PV power generation exceeds the load demand, the ${P}^\text{net}(t)$ is negative, indicating a surplus of PV power. Conversely, when the load demand surpasses the PV output, ${P}^\text{net}(t)$  will be positive.

\subsection{Combined energy storage system}
\subsubsection{ESS operation modeling}
The ESS serves two primary functions in GB. First, it facilitates charge and discharge arbitrage based on market price signals, reducing overall operational costs. Second, as a stationary ESD connected to the BEMS, the ESS remains continuously online, providing power to critical systems such as fire safety and emergency lighting. This ensures an uninterrupted and stable power supply, mitigating risks from grid failures, natural disasters, and other emergencies, thus enhancing the reliability and resilience of the GB microgrid. 

The system implements energy management strategies on a 1-day, 24-hour cycle. Although the SoC of the ESS changes continuously, the SoC at the beginning of the next day is treated as the initial value for the subsequent scheduling cycle, excluding the SoC at the end of the previous day from the next scheduling strategy. The charging/discharging logic of ESS is described in Eqs. \eqref{eq:ESS1}-\eqref{eq:ESS5}, where $\forall t \in T $. The $P^\text{ESS,ch}(t)$ and $P^\text{ESS,dis}(t)$ of ESS at time step $t$ are constrained by the maximum charging/discharging power, which can be expressed as:
\begin{align}
\label{eq:ESS1}
P^\text{ESS,ch}_\text{min} \leq P^\text{ESS,ch}(t) \leq  P^\text{ESS,ch}_\text{max},
\end{align}
\begin{align}
P^\text{ESS,dis}_\text{min}  \leq P^\text{ESS,dis}(t) \leq  P^\text{ESS,dis}_\text{max} ,
\end{align}
According to the physical characteristics of the battery, charging/discharging should be mutually exclusive and are constrained by the following formula:
\begin{align}
\mu^{\text{ESS}}\times P^\text{ESS,ch}(t) +(1-\mu^\text{ESS})\times P^\text{ESS,dis}(t)=0,
\end{align}
The dynamic change of the SoC for ESS at time step $t$ depends on the SoC at time $t-1$ and the charging/discharging actions at time $t$, which is described as follows: 
\begin{align}
\text{SoC}^\text{ESS}(t) = \text{SoC}^\text{ESS}(t-1) + \frac{\eta^\text{ESS,ch}  P^\text{ESS,ch}(t) }{E^\text{ESS}}  - \frac{\eta^\text{ESS,dis} P^\text{ESS,dis}(t)}{ E^\text{ESS}},
\end{align}
The BEMS controls the ESS to perform discharge, charging or standby modes. Note that, the scheduling time interval $\Delta t^\text{sch}$ is 1 hour in this study. For simplicity, it is not explicitly given in the equations for power and energy conversion. Finally, the SoC of ESS is limited between the maximum $\text{SoC}^\text{ESS}_\text{max}$ and the minimum $\text{SoC}^\text{ESS}_\text{min}$, with the value of the boundary depending on the type of battery.
\begin{align}
\label{eq:ESS5}
\text{SoC}^\text{ESS}_\text{min}\leq\text{SoC}^\text{ESS}(t)\leq\text{SoC}^\text{ESS}_\text{max}.
\end{align}

\subsubsection{EVs operation modeling}
EVs, serving as flexible energy resources, are centrally managed by an EV aggregator. When EVs are discharging, they act as power sources supplying energy to buildings. When charging, they function as power loads. The benefits of integrating EVs into the microgrid system of green buildings are evident. On one hand, modern EV batteries have increasingly larger capacities \cite{chen2023techno}, and the additional storage capacity provided by integrating EVs into the into the BEMS enables the building operators to achieve greater profits through energy market arbitrage. On the other hand, as the power system increasingly relies on unstable renewable energy sources as the primary supply, there is an urgent need for active engagement on the consumption side to achieve a dynamic energy balance. The integration of EVs enhances the flexibility of interaction between GB and the grid, ensuring the stable operation of the system. Furthermore, these benefits can be obtained simply by the BEMS making reasonable arrangements for the energy of EVs, without the need for additional expensive upgrades to the ESS capacity.

Based on the research from \cite{bai2024collaborative} and \cite{wu2022coordinated}, EVs arrive gradually during work hours and leave consecutively during off duty hours, with most EVs parked in the parking lot during work hours. This study mainly focuses on the scheduling of EVs during their fixed parking periods. Eqs. \eqref{eq:EV1}-\eqref{eq:EV8} outline the energy scheduling process for EVs, where $\forall t \in T $. Similar to the ESS, constraints Eqs. \eqref{eq:EV1}-\eqref{eq:EV3} limit the charging and discharging power range of EVs. The introduction of the Boolean variable $\mu^\text{EV}$ prevents the physical constraint of simultaneous charging and discharging from being violated. Eqs. \eqref{eq:EV4}-\eqref{eq:EV6} define the start and end states of EVs participating in the scheduling process. The time $t_\text{arr}$ is designated in Eq. \eqref{eq:EV4} as the starting time for EVs to participate in scheduling, treating all EVs as a single entity for scheduling, where ${k \in \mathcal{K}}$. Prior to time $t_\text{arr}$, arriving EVs do not connect to the BEMS. Eq. \eqref{eq:EV5} describes the regular scheduling process for EVs during their parking period. The current SoC of EVs depends on the remaining energy from the previous time and the charging/discharging actions at the current time. Specifically, during charging, $\mathrm{SoC}^\text{EV}(t)$ equals $\mathrm{SoC}^\text{EV}(t-1)$ plus the energy injected into the battery. During discharging, $\mathrm{SoC}^\text{EV}(t)$ equals $\mathrm{SoC}^\text{EV}(t-1)$ minus the energy released from the battery. Energy conversion losses are taken into account both during charging and discharging. In standby mode, no standby losses are considered, and $\mathrm{SoC}^\text{EV}(t)$ equals $\mathrm{SoC}^\text{EV}(t-1)$. Eq. \eqref{eq:EV6} depicts the state of EVs during the period they disconnect from BEMS scheduling, where $ t_\text{off} \in (t_\text{dep}, t_\text{arr})$. The SoC at time $t_\text{dep}$ can still be expressed using Eq. \eqref{eq:EV5}, but the EVs disconnect from the BEMS, and their SoC no longer updates. At time $t_\text{off}$, the SoC of the EVs is set to 0, indicating that the EVs are in a non-scheduled period. Constraint Eq. \eqref{eq:EV7} restricts the boundaries of the SoC. Typically, to preserve battery life, the boundary values $\text{SoC}^\text{EV}_\text{min}$ and $\text{SoC}^\text{EV}_\text{max}$ should be greater than 0 and less than 1.  

The purpose of Eq. \eqref{eq:EV8} is to alleviate range anxiety among EV users. Despite promises of economic incentives from building managers to encourage EV participation in scheduling, the driving range of EVs remains a primary consideration as they serve as the main mode of transportation. Most EV users prefer that the SoC upon departure is no less than the SoC upon arrival \cite{thomas2018optimal}, otherwise their willingness to participate in scheduling decreases. Therefore, it is necessary to ensure the rationality of EV participation in coordinated energy scheduling through Eq. \eqref{eq:EV8}. 
% It should be noted that the BEMS communicates with the EVs aggregator rather than individual EV. The building manager only ensures that the SoC of the EV aggregator upon departure is greater than that upon arrival, without directly participating in the energy distribution between the aggregator and each EV.

\begin{align}
\label{eq:EV1}
P^\text{EV,ch}_\text{min} \leq P^\text{EV,ch}(t) \leq  P^\text{EV,ch}_\text{max},
\end{align}
\begin{align}
\label{eq:EV2}
P^\text{EV,dis}_\text{min} \leq P^\text{EV,dis}(t) \leq  P^\text{EV,dis}_\text{max},
\end{align}
\begin{align}
\label{eq:EV3}
\mu^\text{EV}\times P^\text{EV,ch}(t)+(1-\mu^\text{EV})\times P^\text{EV, dis}(t)=0,
\end{align}
\begin{align}
\label{eq:EV4}
\text{SoC}^\text{EV}(t_\text{arr})=\sum_{k \in \mathcal{K}}\text{SoC}^\text{EV}_{k}(t_\text{arr}),
\end{align}
% \begin{align}
% \mathrm{SoC}^{EV}_{t+1} = 
% \begin{cases}
% {SoC}^{EV}_{t_{arr}} + \frac{\eta^{EV,ch} \cdot P^{EV,ch}_{t} \cdot \Delta t}{E^{EV}}  - \frac{ P^{EV,dis}_{t} \cdot \Delta t}{\eta^{EV,dis} \cdot E^{EV}}  & t = t_{arr} \\
% {SoC}^{EV}_{t} + \frac{\eta^{EV,ch} \cdot P^{EV,ch}_{t} \cdot \Delta t}{E^{EV}}  - \frac{ P^{EV,dis}_{t} \cdot \Delta t}{\eta^{EV,dis} \cdot E^{EV}}  & t  \neq  t_{arr}
% % \mathrm{SoC}^{EVs}_{t_{arr}} + \frac{1}{E^{EVs}} \cdot \eta^{EVs,ch} \cdot \int_{t_{arr}}^{t_{arr}+1} P^{EVs,ch}_{t} \, dt - \frac{1}{E^{EVs}} \cdot \frac{1}{\eta^{EVs,dis}} \cdot \int_{t_{arr}}^{t_{arr}+1} P^{EVs,dis}_{t} \,dt & t = t_{arr} \\
% %  \mathrm{SoC}^{EVs}_{t} + \frac{1}{E^{EVs}} \cdot \eta^{EVs,ch} \cdot \int_{t}^{t+1} P^{EVs,ch}_{t} \, dt - \frac{1}{E^{EVs}} \cdot \frac{1}{\eta^{EVs,dis}} \cdot \int_{t}^{t+1} P^{EVs,dis}_{t} \,dt & t  \neq  t_{arr} 
% \end{cases}
% \end{align}
\begin{align}
\label{eq:EV5}
\mathrm{SoC}^\text{EV}(t) = \mathrm{SoC}^\text{EV}(t-1) + \frac{\eta^\text{EV,ch} P^\text{EV,ch}(t)}{E^\text{EV}}  - \frac{\eta^\text{EV,dis} P^\text{EV,dis}(t)}{ E^\text{EV}},
\end{align}
\begin{align}
\label{eq:EV6}
\text{SoC}^\text{EV}(t_\text{off})=0,
\end{align}
\begin{align}
\label{eq:EV7}
\text{SoC}^\text{EV}_\text{min}\leq\text{SoC}^\text{EV}(t)\leq\text{SoC}^\text{EV}_\text{max},
\end{align}
\begin{align}
\label{eq:EV8}
\text{SoC}^\text{EV}_{}(t_\text{arr}) \leq \text{SoC}^\text{EV}(t_\text{dep}).
\end{align}
% The EV aggregator and ESS form the CBS, with the dynamic modeling of its SoC changes as follows:
% \begin{align}
% {SoC}^{All}(t) = {SoC}^{ESS}(t) + {SoC}^{EV}(t)
% \end{align}

\subsection{Energy allocation and trading models}
The energy allocation rules of CBS are defined in this subsection. Eqs. \eqref{eq:allocation1}-\eqref{eq:allocation16} describe the energy allocation logic for CBS discharging, under the precondition that the system have issued discharging instructions to both the ESS and EV based on electricity prices and load signals. The energy released by CBS can be utilized for power supply or sold back to the PG for profit. To promote local energy consumption and reduce transmission losses, CBS supplies $P^\text{load}(t)$ with the highest priority, and the surplus energy will be sold back to the PG for arbitrage. The energy allocation logic described above can be summarized as follows.
\begin{align}
\label{eq:allocation1}
P^{\text{CBS,ch}}(t) = P^{\text{ESS,ch}}(t) + P^{\text{EV,ch}}(t),
\end{align}
\begin{align}
\label{eq:allocation2}
P^{\text{CBS,dis}}(t) = P^{\text{ESS,dis}}(t) + P^{\text{EV,dis}}(t),
\end{align}
\begin{align}
\label{eq:allocation3}
P^{\text{ESS,PG}}(t) + P^{\text{ESS,build}}(t) = P^{\text{ESS,dis}}(t),
\end{align}
\begin{align}
\label{eq:allocation4}
P^{\text{EV,PG}}(t) + P^{\text{EV,build}}(t) = P^{\text{EV,dis}}(t),
\end{align}
where Eqs. \eqref{eq:allocation1} and \eqref{eq:allocation2} define the charging/discharging power from the CBS, which is the sum of the charging/discharging power of both the ESS and EVs. The specific allocation of energy discharged by the ESS and EVs is described by Eqs. \eqref{eq:allocation3} and \eqref{eq:allocation4}. The energy distribution logic is divided into several scenarios based on the different values of ${P}^\text{net}(t)$ and $P^\text{{CBS,dis}}(t)$. When the ${P}^\text{net}(t)$ is positive and $P^{\text{CBS,dis}}(t) \leq  P^\text{net}(t)$, the energy released by CBS takes precedence in meeting the building’s demand, and no energy is injected into the PG, which is defined as:
\begin{align}
\label{eq:allocation5}
P^{\text{ESS,build}}(t) = P^{\text{ESS,dis}}(t),
\end{align}
\begin{align}
\label{eq:allocation6}
P^{\text{EV,build}}(t) = P^{\text{EV,dis}}(t),
\end{align}
\begin{align}
\label{eq:allocation7}
P^{\text{ESS,PG}}(t) = 0,
\end{align}
\begin{align}
\label{eq:allocation8}
P^{\text{EV,PG}}(t) = 0,
\end{align}
Eqs. \eqref{eq:allocation9}-\eqref{eq:allocation12} describe the energy allocation logic when ${P}^\text{net}(t)$ is positive and $P^{\text{CBS,dis}}(t) > P^\text{net}(t)$. In this scenario, where there is an energy surplus released by the CBS, the ESS and EV supply power to the building according to their original discharge ratios, with the remaining portion being sold back to the PG. The energy distribution process is formulated as follows:
\begin{align}
\label{eq:allocation9}
P^{\text{ESS,build}}(t) = P^{\text{ESS,dis}}(t) - P^{\text{ESS,PG}},
\end{align}
\begin{align}
\label{eq:allocation10}
P^{\text{EV,build}}(t) = P^{\text{EV,dis}}(t) - P^{\text{EV,PG}},
\end{align}
\begin{align}
\label{eq:allocation11}
P^{\text{ESS,PG}}(t) = \frac{P^{\text{ESS,dis}}(t)}{P^{\text{CBS,dis}}(t)} \left(P^{\text{CBS,dis}}(t) - P^{\text{net}}(t)\right),
\end{align}
\begin{align}
\label{eq:allocation12}
P^{\text{EV,PG}}(t) = \frac{P^{\text{EV,dis}}(t)}{P^{\text{CBS,dis}}(t)} \left(P^{\text{CBS,dis}}(t) - P^{\text{net}}(t)\right),
\end{align}
For the last scenario, where ${P}^\text{net}(t)$ is zero or negative, indicating that ${P}^\text{PV}(t)$ meets the entire building load. The discharging power of the CBS is entirely injected into the PG for arbitrage. This process can be described as:
\begin{align}
\label{eq:allocation13}
P^{\text{ESS,build}}(t) = 0,
\end{align}
\begin{align}
\label{eq:allocation14}
P^{\text{EV,build}}(t) = 0,
\end{align}
\begin{align}
\label{eq:allocation15}
P^{\text{ESS,PG}}(t) = P^{\text{ESS,dis}}(t),
\end{align}
\begin{align}
\label{eq:allocation16}
P^{\text{EV,PG}}(t) = P^{\text{EV,dis}}(t),
\end{align}
The energy allocation logic for CBS charging is defined by Eqs. \eqref{eq:allocation17}-\eqref{eq:allocation18}, on the premise that charging control actions have been issued to both the ESS and EV within the CBS. The energy acquired during the CBS charging process is sourced from the grid, satisfying:
\begin{align}
\label{eq:allocation17}
P^\text{PG,ESS}(t) = P^{\text{ESS,ch}}(t) ,
\end{align}
\begin{align}
\label{eq:allocation18}
P^\text{PG,EV}(t) = P^{\text{EV,ch}}(t).
\end{align}

The electricity price at which the BEMS purchases energy from the PG at time step $t$ is denoted as $c^{b}(t)$. This purchase price is typically published by the utility market in advance, allowing the BEMS to plan its energy procurement strategies for ESS, EVs and GB. According to \cite{zhang2024load}, the price $c^{s}(t)$ at which the BEMS sells energy back to the PG should be less than $c^{b}(t)$, reflecting market incentives and operational constraints. Utility companies typically set a slightly lower buyback rate to account for transmission losses, operational profit margins, and to promote local energy consumption. Mathematically, this pricing relationship can be expressed as:
\begin{align}
c^{s}(t) = \beta_\text{pri} c^{b}(t).
\end{align}
where $\beta_\text{pri}$ is the price coefficient, with $0 < \beta_\text{pri} < 1$, fostering local energy use and enhancing the operational stability of the PG.

\subsection{Battery degradation model}
During the entire battery usage cycle, degradation occurs due to external environmental conditions and operational behaviors. In the joint real-time energy management process, additional charge/discharge cycles in CBS accelerate battery degradation, affecting its lifespan. To effectively balance the operational profits and costs of BEMS, it is essential to establish precise battery degradation models to accurately estimate the costs associated with degradation.

Based on a comprehensive consideration of aspects such as cost, safety, service life, and energy density, lithium battery energy storage devices dominate the market \cite{zhang2024load} \cite{farzin2016practical} \cite{ wu2021strategies}. The aging of lithium batteries involves coupled aging reactions among various internal components, including anode and cathode materials, separators, and electrolytes \cite{redondo2018efficiency} \cite{xiong2025multi}. This manifests as a loss of usable capacity, a decline in power capability, and other related issues. Aging can be primarily categorized into two types: cycle aging and calendar aging \cite{cao2020deep} \cite{kassem2013postmortem}. Cycle aging refers to the lifespan lost in each charge-discharge cycle, primarily related to battery usage behavior and operational frequency. Calendar aging, on the other hand, signifies the degradation of batteries over time, even when not in use, with performance decreasing gradually over time based on storage conditions.

In the joint energy scheduling task of ESS and EV batteries, the BEMS makes control decisions based on various information such as load, PV generation, and price. EVs participate in scheduling by leveraging their mobile energy storage characteristics to enhance the energy utilization efficiency and flexibility of the system.
Generally, the needs of stationary energy storage batteries for buildings meet the requirements of long service life, low cost, high safety, long-term stable operation \cite{li2023accelerated}, and adapt to the frequency modulation power fluctuation needs of the grid \cite{krieger2013comparison}. The LFP, with the chemical formula LiFePO4, meets the aforementioned requirements and is currently the primary material for stationary energy storage batteries \cite{yang2023life}. The mainstream material for EV batteries is ternary lithium, typically referring to typically referring to Li(NiMnCo)O2 (NMC) \cite{wen2020overview}. NMC batteries, serving as the primary power source for passenger vehicles, offer high instantaneous power and energy density to meet the demands for acceleration and range \cite{lv2024recent}. For life and cost, fixed energy storage devices pay more attention to the number of cycles and calendar life, and pursue the full life cycle economy. The power lithium batteries focus more on cycling stability and power density. Compared to energy storage batteries, they typically have a lower service life and higher costs \cite{li2023accelerated} \cite{hu2024review}.

Given the significant differences between the two types of batteries, it is urgently necessary to establish accurate degradation models for them respectively. However, in previous studies, the aging characteristics of different energy storage batteries and their corresponding degradation costs have been overlooked \cite{onile2023energy} \cite{cao2020deep} \cite{wan2018model}. This is not only unrealistic but also directly impacts the formulation of coordinated operation and energy scheduling strategies for specific multi-battery systems, leading to unreasonable optimization results. To address the challenge of lacking precise multi-battery degradation models for coordinated operation of CBS and efficient energy management strategy arrangement, this paper builds accurate battery degradation models based on the differing chemical properties of ESS and EV.

In addition to the factor of time, there are mainly two variables that have a significant impact on calendar aging: temperature and SoC \cite{xu2016modeling} \cite{najera2023semi}. The specific ways in which these parameters affect calendar aging depend on the chemical composition of the battery. Calendar aging follows the Arrhenius exponential expression and can be used to estimate the correlation between the above-mentioned variables and calendar aging, which is manifested as the decay of battery capacity over time:
\begin{align}
\label{eq:cal}
 \Delta Q_{\text{cal}} (\Delta t)= e^{k_\alpha \cdot e^{k_\beta \cdot \sigma} \cdot e^{\frac{k_\gamma}{T}} \cdot \Delta t^{k_z}},
\end{align}
where $\sigma$ and $T$ represent the average SoC and battery temperature, $\Delta t \triangleq [t_{s}, t_{e}]$ is the duration of the calendar aging test, $t_{s}$ and $t_{e}$ are the start and end times of the calendar aging test, $\Delta Q_{\text{cal}}$ represents the capacity fade that undergoes $\Delta t$  under the above stress factors. $k_z$ is a power law factor whose value is generally in the range of 0 to 1. This paper follows the setting of \cite{najera2023semi} and sets it to 0.5, indicating that the capacity decay is proportional to the square root of time. It is worth noting that three constants, $k_\alpha$, $k_\beta$, and $k_\gamma$, are used to adjust the model to adapt to the characteristics of different batteries. Specifically, $k_\alpha$ affects the overall decay rate. The larger the $k_\alpha$ value, the faster the capacity decay. $k_\beta$ is an exponential coefficient related to SoC, and the larger the value of $k_\beta$, the greater the influence of SoC on attenuation. $k_\gamma$ represents the exponential coefficient related to temperature, and the larger the value of $k_\gamma$, the more significant the influence of temperature on attenuation. According to Eq. \eqref{eq:cal}, it is evident that the SoC and temperature stress factors significantly impact battery performance. Under the condition that $k_\gamma$ is negative, high temperatures and elevated SoC can lead to exponential degradation of battery capacity, which aligns with the findings in \cite{najera2023semi} \cite{wang2011cycle}.

The cycle aging of a battery not only depends on the number of cycles, but also is usually related to factors such as SoC, battery temperature, and DoD. The stress factors in each independent battery cycle are different, so the entire battery cycle degradation can be regarded as the sum of all independent cycle degradation. The battery degradation rate $f_{d,1}$ and battery capacity fade $Q_{\text{cycle}}$ in a cycle can be expressed as:
\begin{align}
\label{cycle1}
\ Q_{\text{cycle}}(N)= 1 - \alpha_{\text{SEI}} e^{-N \beta_{\text{SEI}} f_{d,1}} - \left(1 - \alpha_{\text{SEI}}\right) e^{-N f_{d,1}},
\end{align}
\begin{align}
\label{cycle2}
f_{d,1}(\delta,\Delta t,\sigma,T) = \left(S_{\delta}(\delta) + S_{t}(\Delta t)\right) S_{\sigma}(\sigma) S_{T}(T),
\end{align}
where $\alpha_{\text{SEI}}$ and $\beta_{\text{SEI}}$ are battery solid electrolyte interphase (SEI) parameters, $N$ represents the number of cycles identified from irregular cycles using the rainflow cycle-counting method \cite{xu2016modeling} \cite{xu2017factoring} \cite{xu2018optimal}.   $S_{\Delta t}(\Delta t)$, $S_{\sigma}(\sigma)$, and $S_{T}(T)$ represent the stress factor equations for time ($\Delta t$), average SoC ($\sigma$) and battery temperature ($T$) of each identified cycle, respectively. Their formulas are as follows:
\begin{align}
S({T})=e^{k_T(T-T_{\mathrm{ref}})\frac{T_{\mathrm{ref}}}{T}},
\end{align}
\begin{align}
S({\sigma})=e^{k_\sigma(\sigma-\sigma_{\mathrm{ref}})},
\end{align}
\begin{align}
S(\Delta t)={k_t}\Delta t,
\end{align}
where $k_T$, $k_\sigma$ are temperature and SoC stress coefficients, ${k_t}$ is time stress coefficient; $T_{\mathrm{ref}}$, $\sigma_{\mathrm{ref}}$ are the reference values of temperature and SoC, respectively. It should be noted that the degradation characteristics of LFP and NMC batteries are reflected based on different DoD stress models, which can be expressed as:
\begin{align}
S^\mathrm{LFP}_{{\delta}} (\delta) = k_{\delta 1} \delta  e^{k_{\delta 2}  \delta},
\end{align}
\begin{align}
S^\mathrm{NMC}_{{\delta}} (\delta) = \left( k_{\delta 1} \delta^{k_{\delta 2}} + k_{\delta 3} \right)^{-1},
\end{align}
where $\delta$ represents the DoD value for the identified cycle, $k_{\delta 1}$, $k_{\delta 2}$ and $k_{\delta 3}$ are DoD model parameters respectively. The differing degradation characteristics of LFP and NMC arise from variations in the chemical stability of their cathode materials \cite{li2023accelerated}. For LFP, capacity decay increases exponentially with DoD. Specifically, $k_{\delta 1}$ is the baseline aging coefficient, controlling the rate of linear degradation at low DoD, while $k_{\delta 2}$ is the exponential coefficient representing the nonlinear impact of DoD on the degradation rate. This reflects the increased internal stress in LFP during deep discharge, accelerated electrolyte decomposition, and the loss of lithium inventory due to the thickening of the SEI \cite{redondo2018efficiency} \cite{kassem2013postmortem}. In contrast, NMC aging is primarily due to the repeated intercalation and de-intercalation of $\mathrm{Li}^+$, resulting in the dissolution of transition metals and structural collapse, alongside increased internal side reactions \cite{zhao2018high} \cite{yang2022extreme}. Consequently, the aging of NMC follows a power function growth with DoD but is constrained by the baseline aging constant $k_{\delta 3}$, eventually stabilizing its capacity decay. Therefore, it is essential to balance energy demand and lifespan based on the distinct degradation characteristics of LFP and NMC to achieve efficient energy management. The above calendar and cycle degradation parameters are derived from theoretical analyses and empirical fitting of actual battery degradation experiments based on material differences \cite{xu2016modeling} \cite{najera2023semi}.

\begin{figure*}[hbt]
	\centering
	\subfloat[]{\includegraphics[height=2.0in]{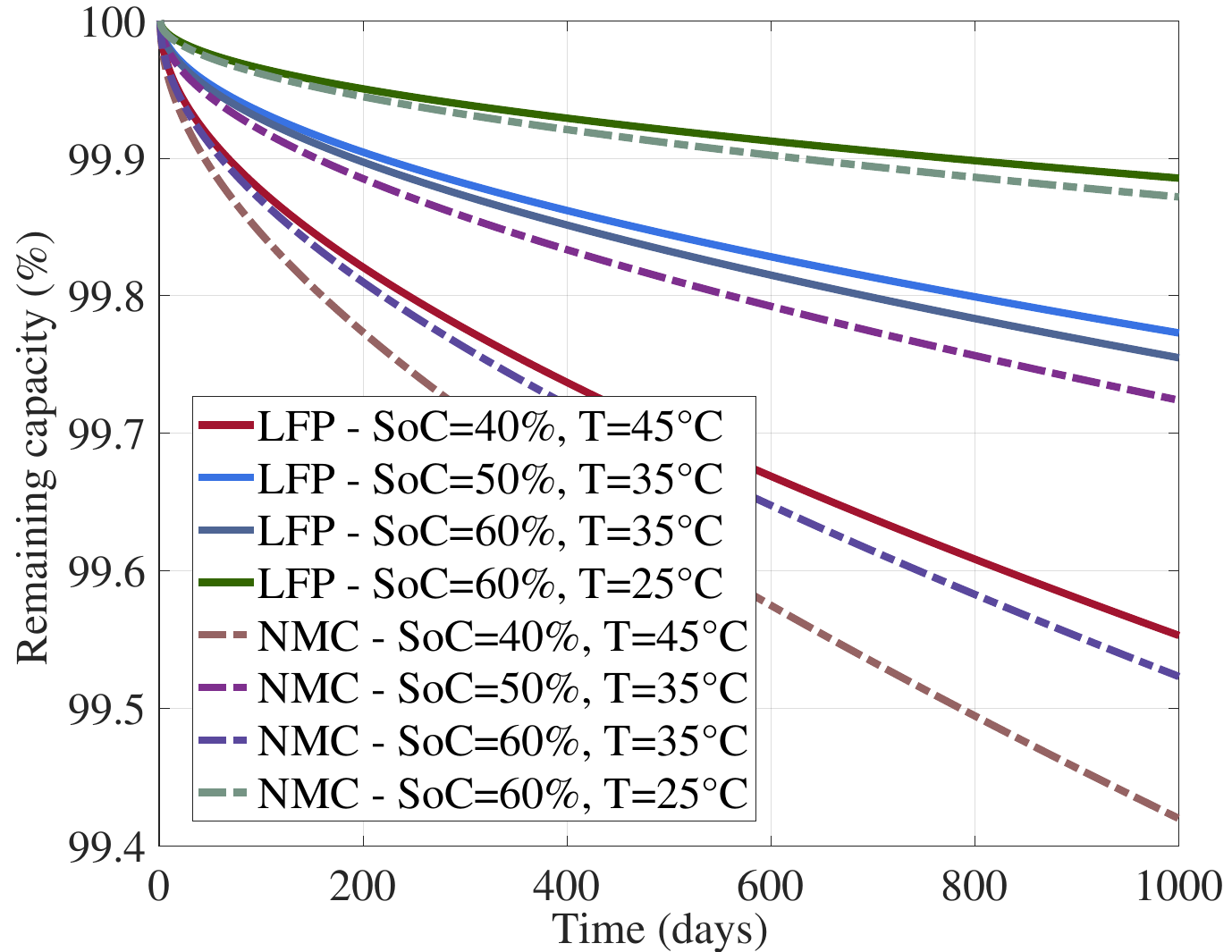}%
		\label{cal}}
	\hfil
	\subfloat[]{\includegraphics[height=2.0in]{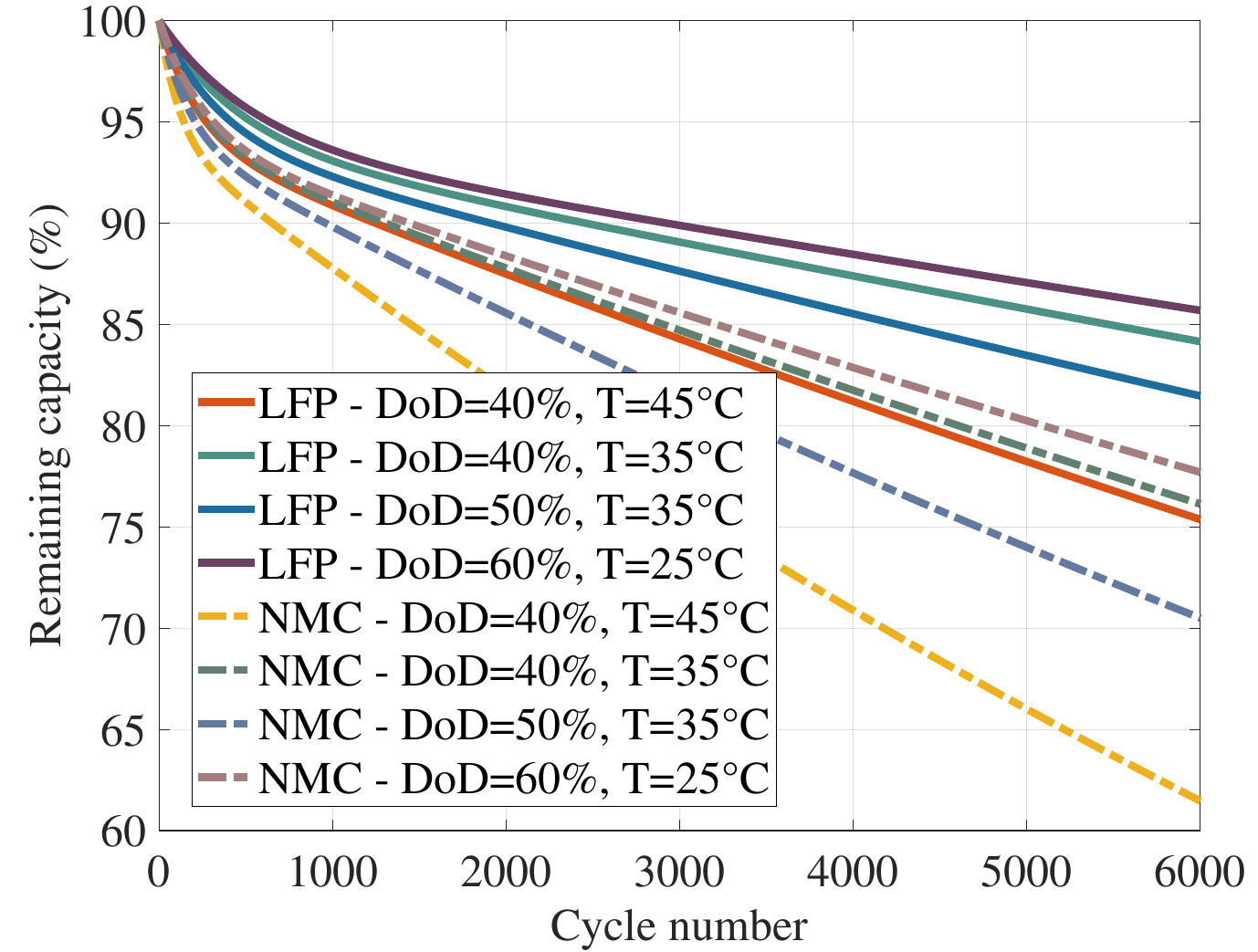}%
		\label{cycle}}
	\hfil
	\caption{The degradation characteristics of LFP and NMC lithium-ion batteries under various working conditions. (a) Comparison of calendar aging characteristics of LFP and NMC at different SoC and temperatures; (b) Comparison of the cycle aging characteristics of LFP and NMC at 50\% SoC and different DoD and temperatures.}
	\label{degline}
\end{figure*}

Based on the lithium battery capacity degradation modeling analysis mentioned above, we further illustrate the calendar aging and cycle aging trends of LFP and NMC under different working conditions in Fig. \ref{degline}. This is to demonstrate the necessity of establishing battery loss models for ESS and EV in energy scheduling. Fig. \ref{cal} compares the aging characteristics of LFP and NMC batteries under varying temperatures and SoC. It is evident that high SoC and high temperature conditions accelerate the aging process of both battery types, with the latter's impact being more pronounced. High SoC leads to elevated battery voltage, causing significant mechanical stress on materials and resulting in unwanted byproducts, such as the SEI layer on electrode surfaces, thereby accelerating calendar aging. High temperatures primarily harm battery life by accelerating internal chemical reactions. 
Due to differences in material properties, NMC exhibits a faster calendar aging process compared to LFP. After 1,000 days at 25°C, 35°C, and 45°C with corresponding SoC levels, NMC experiences, on average, a capacity loss that is 39.46\% greater than that of LFP. These results are generally consistent with the findings from actual battery tests reported in \cite{najera2023semi}. For the cycle aging presented in Fig. \ref{cycle}, NMC consistently exhibits a faster capacity loss than LFP under the same conditions, and this trend is significantly influenced by the temperature. Specifically, after 6,000 cycles at a constant 50\% SoC across varying temperatures and DoD, NMC experienced an average capacity decline of 55.51\% more than LFP. Notably, under 40\% DoD at 45°C, NMC's remaining capacity dropped to 61.46\%, reflecting an 56.42\% greater decline compared to LFP. As shown in Fig. \ref{degline}, it is clear that LFP and NMC batteries exhibit significantly different degradation characteristics. In subsequent sections, we will optimize their coordinated scheduling and energy management strategies based on these distinct degradation profiles.

\subsection{Battery aging cost assessment}
In this subsection, we will elaborate on how to integrate the CBS composed of batteries with different chemical characteristics into the DRL-based energy real-time scheduling framework. Building on this, we will design a profit-cost coordination mechanism for energy management that addresses the needs of various stakeholders, facilitating the collaborative optimization of multiple energy storage batteries within the BEMS.

\begin{figure*}[th]
	\centering
	\includegraphics[width=0.85\textwidth]{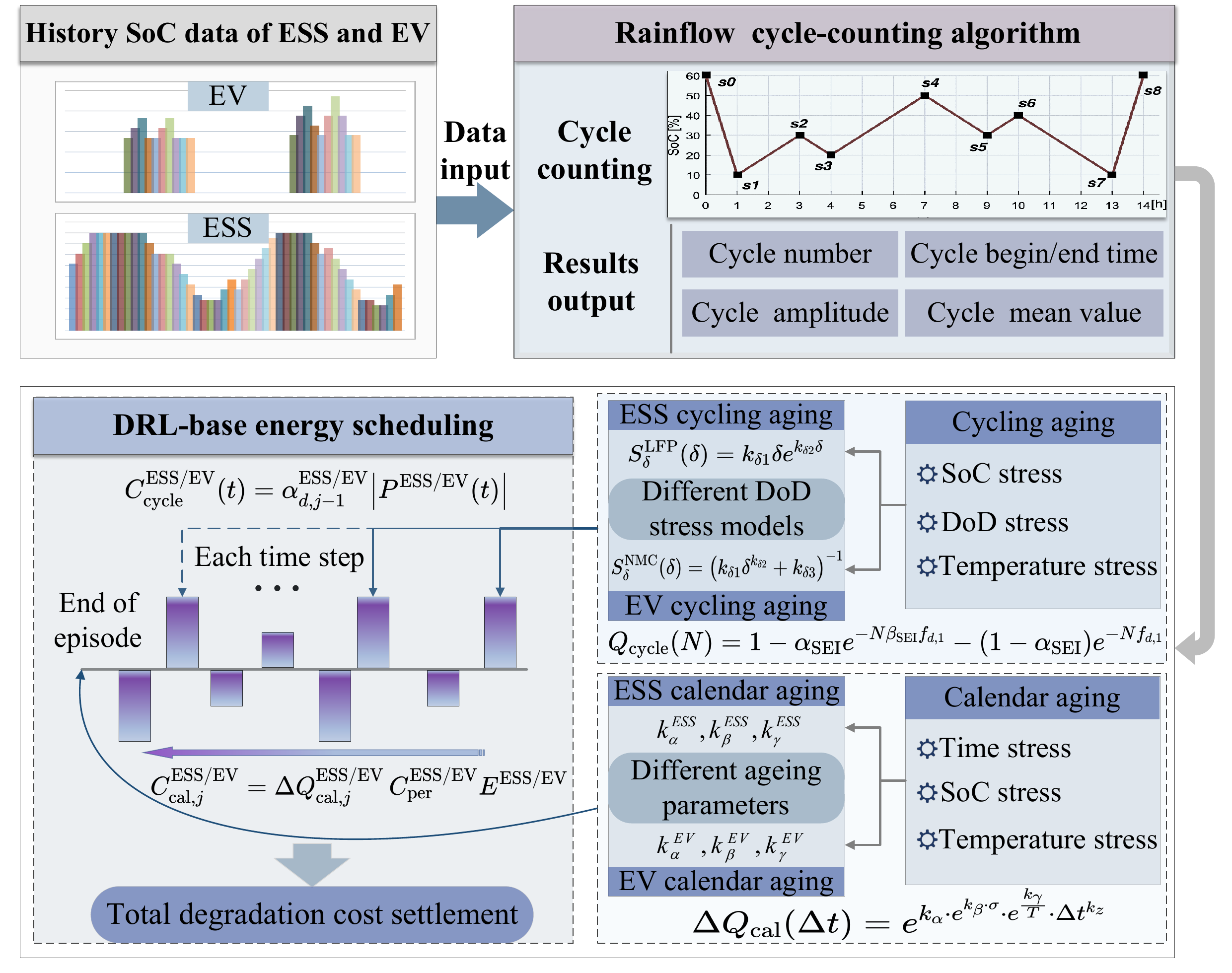}
	\caption{Multi-battery degradation assessment framework in DRL-based real-time energy scheduling.}
	\label{Deg_Framework}
\end{figure*}

After incorporating ESDs with different degradation costs into the energy management process of a green building microgrid, a trade-off between operational profits and battery degradation costs arises. ESS and EVs belong to different stakeholders, but the scheduling strategy is designed with the aim of minimizing the operating costs of the BEMS. Based on the objective function, BEMS tends to over-utilize the CBS to increase revenue, which is unacceptable for controlling battery degradation costs and maintaining battery health. If the allocation of revenue and the distribution of costs are not clearly defined, it will be impossible to formulate an optimized scheduling strategy. Previous studies either used overly simplified degradation models or failed to accurately calculate the costs of calendar and cycle aging in a multi-battery environment environment, making cost-sharing calculations in this context unfeasible. Therefore, it is essential to investigate how to integrate ESS and EVs with different battery aging and operational characteristics into the DRL-based real-time energy scheduling framework. This is crucial for designing a reasonable energy scheduling plan in a microgrid environment that incorporates various DERs.

To address these challenges, we propose a multi-battery degradation assessment framework within an  DRL-based scheduling environment, as illustrated in Fig. \ref{Deg_Framework}. First, we gather historical SoC data from ESS and EVs to support our calculations. The rainflow algorithm is employed to calculate degradation parameters for irregular battery operating cycles, using the historical SoC profile from the previous DRL episode. This provides cycle amplitude, mean value, count, and start and end times, which are converted into stress factors for the degradation model, such as DoD and average SoC. Next, we establish precise degradation models for ESS and EVs based on LFP and NMC lithium-ion battery chemistries, explicitly modeling the processes of calendar aging and cycle aging to facilitate subsequent cost-sharing calculations. We then compute the costs associated with both calendar and cycle aging. Degradation coefficients are used to determine cycle aging costs at each DRL time step, which are reflected in the reward function to guide the agent in making informed scheduling decisions. Calendar aging costs are calculated at the end of each episode for accuracy. Finally, we summarize the system's revenue and costs for each episode, accumulating them into the total operating costs. Detailed allocations of battery degradation and system revenue costs are provided below.

In the DRL-based energy scheduling framework, we need to pay special attention to the additional battery cycle aging caused by the energy scheduling operations of the BEMS. To reflect the instantaneous degradation costs in the agent's learning process, inspired by \cite{cao2020deep}, we use the cycle aging degradation coefficient $\alpha_{d}$ to describe the impact of each charge/discharge action on battery degradation. This coefficient can be calculated as follows:
\begin{align}
\alpha^{\text{ESS}}_{d,j} = \frac{C^{\text{ESS}}_{\text{per}} \Delta Q^{\text{ESS}}_{\text{cycle},j} E^{\text{ESS}}}{\sum_{t=1}^T \left|P^{\text{ESS}}(t)\right|},
\end{align}
\begin{align}
\alpha^{\text{EV}}_{d,j} = \frac{C^{\text{EV}}_{\text{per}} \Delta Q^{\text{EV}}_{\text{cycle},j} E^{\text{EV}}}{\sum_{t=1}^T \left|P^{\text{EV}}(t)\right|},
\end{align}
where the superscripts ESS and EV of the symbols in the formulas respectively refer to the names of the two types of ESDs in this subsection. $C_\text{per}^\text{ESS}$ and $C_\text{per}^\text{EV}$ represent the battery cost per kWh, $P(t)$ denotes the charge/discharge power at time step $t$, and $T$ indicates the time step length within an episode. $E^\text{ESS}$ and $E^\text{EV}$ represent the energy capacity. Due to the nonlinear relationship between battery cycle aging and $N$, the amount of capacity degradation varies across each episode. Therefore, we utilize $\Delta Q^\text{ESS}_{\text{cycle},j}$ and $\Delta Q^\text{EV}_{\text{cycle},j}$ to track this nonlinear capacity degradation process, defined as the difference between the initial remaining capacity and the end remaining capacity of the battery in the $j$-th episode. $\Delta Q^\text{ESS}_{\text{cycle},j}$ and $\Delta Q^\text{EV}_{\text{cycle},j}$ are continuously updated throughout the battery's cycle aging process and are refreshed when the battery reaches the termination condition of 80\% remaining life \cite{zhang2024load} \cite{xu2016modeling}. The $\alpha^\text{ESS}_{d}$ and $\alpha^\text{EV}_{d}$ are updated at the end of each episode, being sensitive to the frequency of battery control actions. The $\alpha_{d}$  not only calculates the cycling degradation cost of the ESS but also directly reflects the degradation costs incurred by EV participation in building energy scheduling. This provides important insights for the joint operation and scheduling of multi-battery systems involving different stakeholders. Then, the battery cycle aging cost $C^\text{ESS}_\text{cycle}(t)$ and $C^\text{EV}_\text{cycle}(t)$ to be paid for dispatching $P^\text{ESS}$ and $P^\text{EV}$ at time step $t$ during episode $j$ are:
\begin{align}
C^{\text{ESS}}_{\text{cycle}}(t) = \alpha^{\text{ESS}}_{d,j-1} \left|P^{\text{ESS}}(t)\right|,
\end{align}
\begin{align}
C^{\text{EV}}_{\text{cycle}}(t) = \alpha^{\text{EV}}_{d,j-1} \left|P^{\text{EV}}(t)\right|.
\end{align}

Calendar aging is an inherent chemical reaction process in batteries, and the associated costs are not directly linked to the fees incurred by control actions at each time step in the DRL-based method. Therefore, calendar aging costs can be calculated more accurately than approximated on the basis of $\alpha_{d}$. We compute the battery calendar aging costs at the end of each episode, represented as follows:
\begin{align}
\label{ESScal}
C^{\text{ESS}}_{\text{cal},j} &= \Delta Q^{\text{ESS}}_{\text{cal},j} \, C^{\text{ESS}}_{\text{per}} E^{\text{ESS}},
\end{align}
\begin{align}
C^{\text{EV}}_{\text{cal},j} &= \Delta Q^{\text{EV}}_{\text{cal},j} \, C^{\text{EV}}_{\text{per}} E^{\text{EV}},
\end{align}
where $C^\text{ESS}_{\text{cal},j}$ and $C^\text{EV}_{\text{cal},j}$ are batteries calendar aging cost during episode $j$. The battery degradation of EVs during the scheduling process primarily stems from cycle aging, and this cost should be borne by the building manager to compensate for EV users' losses. The calendar aging of EVs batteries is inherent to the battery's usage cycle and should not be included in the operational costs of joint energy scheduling from the perspective of building managers. Ultimately, for joint energy scheduling over a time period $T$, or during one episode, the total battery degradation cost $C^\text{total}$ consists of four components, which can be expressed as:
\begin{align}
C^{\text{total}} = \sum_{t=1}^T \left[C^{\text{ESS}}_{\text{cycle}}(t) + C^{\text{EV}}_{\text{cycle}}(t)\right] + C^{\text{ESS}}_{\text{cal},j} + C^{\text{EV}}_{\text{cal},j},
\end{align}
From the perspectives of building managers and EV users, $C^\text{total}$ can be further divided into two components:
\begin{align}
C^{\text{EV}} = C^{\text{EV}}_{\text{cal},j},
\end{align}
\begin{align}
C^{\text{build}} = \sum_{t=1}^T \left[C^{\text{ESS}}_{\text{cycle}}(t) + C^{\text{EV}}_{\text{cycle}}(t)\right] + C^{\text{ESS}}_{\text{cal},j},
\end{align}
where $C^\text{EV}$ and $C^\text{build}$ represent the battery degradation costs should be borne by the EV users and building managers, respectively, during the $j$-th episode.

\subsection{Energy supply and demand balance}
The power balance constraint, as defined in Eq. \eqref{eq:balance}, ensures that the input and output power of the BEMS remain equal at each time step.
\begin{align}
\label{eq:balance}
P^{\text{PG}}(t) + P^{\text{CBS,dis}}(t) = P^{\text{net}}(t) + P^{\text{CBS,ch}}(t),
\end{align}
Eq. \eqref{eq:balance} illustrates that the total system load comprises the net load of the building and the CBS charging load, while the system's power supply comes from the PG and the CBS discharging power. Both ${P}^\text{net}(t)$ and ${P}^\text{PG}(t)$ can be either positive or negative, depending on the system's power supply and demand status. The CBS charging and discharging powers are absolute values of the scheduling actions, hence positive.
Eq. \eqref{eq:balance} describes the instantaneous power equilibrium at a macroscopic level, encompassing energy allocation variables such as $P^{\text{ESS,build}}(t)$ and $P^{\text{EV,build}}(t)$ within specific terms, thus eliminating the necessity for explicit enumeration.

\subsection{Optimization problem}
Based on the model mentioned above, we can establish the optimization objective of this study, as shown in Eq. \eqref{obj}.
\begin{align}
\label{obj}
\mathcal{P}_{1}: \quad &\min_{\pi} \left( \sum_{t=1}^{T} \left[ c^{\text{s}}(t) \left( P^{\text{PV,rev}}(t) + P^{\text{ESS,PG}}(t) + P^{\text{EV,PG}}(t) \right) - c^{\text{b}}(t) \left( P^{\text{PG,cost}}(t) + P^{\text{CBS,ch}}(t) \right) \right] - C^{\text{build}} \right) \\
& \text{s.t.} \ (1) - (52) \nonumber,
\end{align}
\begin{align}
\label{obj1}
P^{\text{PV,rev}}(t) = \max\{-P^{\text{net}}(t), 0\},
\end{align}
\begin{align}
\label{obj2}
P^{\text{PG,cost}}(t) = \max\{P^{\text{net}}(t) - P^{\text{ESS,build}}(t) - P^{\text{EV,build}}(t), 0\}.
\end{align}
The objective of $\mathcal{P}_{1}$ is to minimize the total operational cost of the system by effectively coordinating the joint real-time charging/discharging actions of the ESS/EVs within the scheduling period $T$. The system's revenue consists of two components. First, it includes proceeds from the ESS and EVs selling energy to the power grid, provided they meet the building's net load. Second, as depicted in Eq. \eqref{obj1}, when the net load is negative, it encompasses the profits obtained from selling $P^\text{PV,rev}(t)$ from PV. The system's costs can be categorized into the cost of purchasing electricity and the incurred battery degradation costs. For the former, when $P^\text{net}(t)$ is positive, as shown in Eq. \eqref{obj2}, if the energy released by the CBS is insufficient to cover $P^\text{net}(t)$, electricity needs to be purchased from the PG. On the other hand, all the charging energy of the CBS must be obtained from the PG. Regarding degradation costs, the system calculates the expenses based on the revenue-cost coordination mechanism outlined in subsection 3.6.

The challenges of $\mathcal{P}_{1}$ can be summarized as follows. (1)  The significantly different degradation costs for ESS and EVs require BEMS to carefully schedule the CBS. This schedule must achieve a precise balance between battery loss costs and energy arbitrage revenue to achieve $\mathcal{P}_{1}$. (2) As transportation-oriented energy components, scheduling strategies for EVs need to prioritize their mileage and range constraints. Additionally, the differing scheduling windows and cost constraints of ESS and EVs narrow the profit margin of system. (3) The $\mathcal{P}_{1}$ expects that the CBS interacts with BEMS based on optimized energy utilization logic, leveraging the differences between $c^{s}(t)$ and $c^{b}(t)$ to further optimize costs. However, the trade-off between long-term returns and short-term gains is challenging to learn. (4) Unlike most single-battery storage systems, multi-battery systems exhibit complex heterogeneous state characteristics and a high-dimensional coupled action selection space, making it difficult to find energy scheduling decisions that minimize costs. To address these challenges, we transform the joint optimization problem of charge-balancing strategies into a MDP problem, deriving a progressively optimal solution for $\mathcal{P}_{1}$ using a DRL-based algorithm.

\section{Problem formulation}
\subsection{Formulation of MDP}
In this study, we develop a real-time energy joint dispatch strategy for the CBS in BEMS to optimize the operation of GB. The SoC of CBS at time step $t$ depends solely on the previous state and action, independent of any states or actions prior to time step $t-1$, satisfying the Markov property. Thus, this problem can be formulated as a finite MDP problem with discrete time steps. The sequential decision-making process of the joint energy dispatch MDP in the BEMS is as follows: within the BEMS, the agent acts as a controller, formulating real-time strategies for the CBS at hourly intervals. At time step $t$, the agent observes the system state $s_t \in \mathcal{S}$, which includes information on the battery states of the ESS and EVs, the real-time net load of the GB, predictions of net load for upcoming time steps, as well as energy price series. Based on this information, the agent selects a composite charging/discharging action $a_t \in \mathcal{A}$ for CBS according to policy $\pi$. Upon executing action $a_t$, a reward score $R_{t}$ is generated, and the system transitions to a new state $s_{t+1}$ based on the state transition probability ${P}(s_{t + 1} \mid s_t, a_t)$. The objective of the proposed scheduling method is to determine an optimal energy joint dispatch strategy to minimize the expected long-term operational costs of the GB. The detailed MDP modeling process is provided in the following sections.

\subsection{Description of state space}
The system state $s_t$ is typically designed in practice to be a set that captures key features reflecting the real system state. In the study, $s_t \in \mathcal{S}$ is defined as a 51-dimensional vector, $s_t=(c^{b}(t), ... , c^{b}(t+22), c^{b}(t+23), {P}^\text{net}(t), ... , {P}^\text{net}(t+22), {P}^\text{net}(t+23), \xi(t), \text{SoC}^\text{ESS}(t), \text{SoC}^\text{EV}(t))$. Here, $c(t), ..., c(t+22), c(t+23)$ is the series of prices for the 24 hours published by PG in advance. The ${P}^\text{net}(t), ... , {P}^\text{net}(t+22), {P}^\text{net}(t+23)$  includes the current net load and the predicted net loads for the next 23 hours. The two sets of feature series reflect future price and building demand trends, respectively, aiding the agent in forming long-term strategic decisions. The variable $\xi(t)$  is a Boolean identifier indicating whether the EV battery is within the scheduled period, being true during that time and false otherwise. By embedding $\xi(t)$ into $s_t$, we guide the agent to learn the travel patterns of EVs, ensuring that it issues charging/discharging commands only within the scheduling period. Lastly, $\text{SoC}^\text{ESS}(t)$ and $\text{SoC}^\text{EV}(t)$ provide crucial battery state information for the CBS, directly influencing the charging/discharging potential of the batteries.

\subsection{Description of action space}
Based on the observed $s_t$, the agent can choose charging or discharging actions. The control actions for ESS and EV are defined as discrete decision variables \cite{wan2018model} \cite{chics2016}:
\begin{align}
\mathcal{A}^{\text{ESS}} = \left[ P^{\text{ESS,ch}}_{\text{min}}, 0.5P^{\text{ESS,ch}}_{\text{min}}, 0, 0.5P^{\text{ESS,dis}}_{\text{max}}, P^{\text{ESS,dis}}_{\text{max}} \right],
\end{align}
\begin{align}
\mathcal{A}^{\text{EV}} = \left[ P^{\text{EV,ch}}_{\text{min}}, 0.5P^{\text{EV,ch}}_{\text{min}}, 0, 0.5P^{\text{EV,dis}}_{\text{max}}, P^{\text{EV,dis}}_{\text{max}} \right],
\end{align}
The action space of CBS, denoted as $\mathcal{A}$, can be defined as the Cartesian product of individual ESS and EV action spaces: 
\begin{align}
 \label{eq:action_space}
\mathcal{A} = \mathcal{A}^{\text{ESS}} \times \mathcal{A}^{\text{EV}} = \left\{ \left( a_t^{\text{ESS},i}, a_t^{\text{EV},j} \right) \mid a_t^{\text{ESS},i} \in \mathcal{A}^{\text{ESS}}, a_t^{\text{EV},j} \in \mathcal{A}^{\text{EV}} \right\},
\end{align}
Specifically, with 5 discrete actions available for both the ESS and EVs, the CBS action space $\mathcal{A}$ contains 25 discrete control options, where each action coordinates ESS and EV charging/discharging power exchange simultaneously. The CBS action $a_t$ can be denoted as $a_t = (a_t^{\text{ESS},i}, a_t^{\text{EV},j})$. When the agent selects a control action, CBS maps the action index to the corresponding ESS and EV actions, subsequently converting them into charging and discharging power for further calculations. Considering the constraints of charging and discharging efficiency as well as the SoC boundaries, the control actions must satisfy the following constraints:
\begin{align}
\left( \text{SoC}^{\text{ESS}}(t) - \text{SoC}^{\text{ESS}}_{\text{max}} \right) E^{\text{ESS}} \eta^{\text{ESS},m} \leq a_t^{\text{ESS}} \leq \left( \text{SoC}^{\text{ESS}}(t) - \text{SoC}^{\text{ESS}}_{\text{min}} \right) E^{\text{ESS}} \eta^{\text{ESS},m},
\end{align}
\begin{align}
\left( \text{SoC}^{\text{EV}}(t) - \text{SoC}^{\text{EV}}_{\text{max}} \right) E^{\text{EV}} \eta^{\text{EV},m} \leq a_t^{\text{EV}} \leq \left( \text{SoC}^{\text{EV}}(t) - \text{SoC}^{\text{EV}}_{\text{min}} \right) E^{\text{EV}} \eta^{\text{EV},m},
\end{align}
where $ m \in \{\text{ch, dis}\} $ indicates that the charging/discharging of ESS/EV are constrained by the mandatory SoC boundary limits.
% \begin{align}
% {({SoC}^{EV}(t)-SoC^{EV}_{max}) E^{EV}}{\eta^{EV,ch}}\leq P^{EV,ch}(t)\leq{({SoC}^{EV}(t)-SoC^{EV}_{min}) E^{EV}}{\eta^{EV,ch}}
% \end{align}
% \begin{align}
% {({SoC}^{EV}(t)-SoC^{EV}_{max}) E^{EV}}{\eta^{EV,dis}}\leq P^{EV,dis}(t)\leq{({SoC}^{EV}(t)-SoC^{EV}_{min}) E^{EV}}{\eta^{EV,dis}}
% \end{align}

\subsection{Design of reward function}
The design of the reward function $R_t$ is crucial for guiding the agent's decisions. In this paper, the reward function considers three main aspects. First, $R_t$ incentivizes the agent to maximize energy arbitrage profits based on fluctuating price and net load signals. Second, it accounts for the energy interaction logic between the CBS and the building, prioritizing the supply of energy to the building over energy trading. Finally, it takes into account the precise degradation costs of different ESDs, optimizing the collaborative operation among various types of storage batteries within the CBS. The reward function $R_t$ can be represented as:
\begin{align}
\label{reward}
R_t = w^{\text{dis}} P^{\text{CBS,dis}}(t) - w^{\text{ch}} P^{\text{CBS,ch}}(t) - w^{\text{ESS}} C^{\text{ESS}}_{\text{cycle}}(t) - w^{\text{EV}} C^{\text{EV}}_{\text{cycle}}(t) - w^{\text{pen}} \text{SoC}^{\text{pen}}(t),
\end{align}
where $w^\text{dis}$ and $w^\text{ch}$ are scaling coefficients for discharging and charging, respectively. $w^\text{ESS}$ and $w^\text{EV}$ are scaling coefficients for the degradation costs of ESS and EV batteries. $w^\text{pen}$ represents the penalty coefficient for SoC violations.
\begin{figure*}[hbt]
	\centering
	\subfloat[]{\includegraphics[height=2.2in]{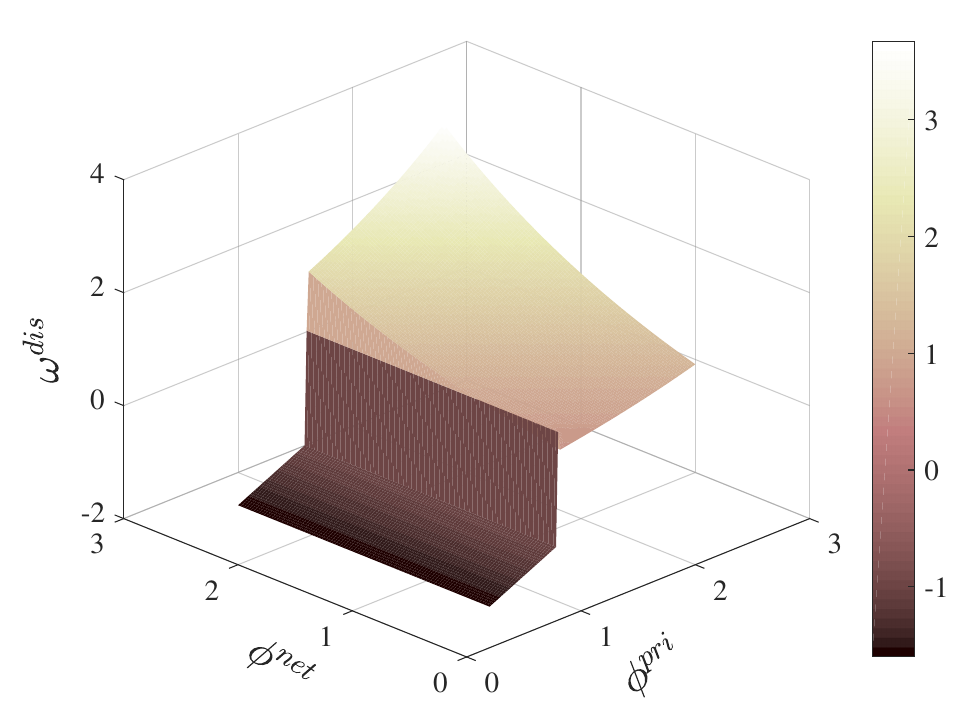}%
		\label{dis}}
	\hfil
	\subfloat[]{\includegraphics[height=2.2in]{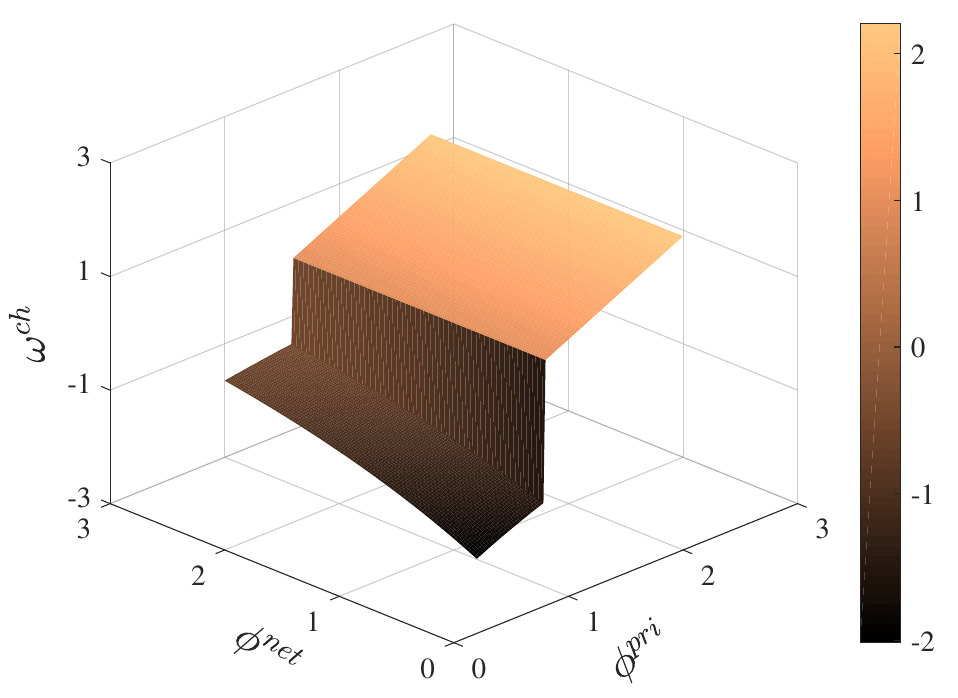}%
		\label{ch}}
	\hfil
	\caption{The design principles of scaling coefficients for discharging and charging. (a) Discharging scaling coefficient; (b) Charging scaling coefficient.}
	\label{disch}
\end{figure*}

\begin{align}
w^{\text{dis}} &= 
\begin{cases} 
e^{((\phi^{\text{pri}} + \phi^{\text{net}})/2 - 1)} & \text{if } \phi^{\text{pri}} > 1 \\
\phi^{\text{pri}} - 2 & \text{if } \phi^{\text{pri}} < 1 \\
1 & \text{otherwise}
\end{cases}, \\
\end{align}
\begin{align}
w^{\text{ch}} &= 
\begin{cases} 
-e^{(1 - (\phi^{\text{pri}} + \phi^{\text{net}})/2)} & \text{if } \phi^{\text{pri}} < 1 \\
\phi^{\text{pri}} & \text{if } \phi^{\text{pri}} > 1 \\
1 & \text{otherwise}
\end{cases},
\end{align}
\begin{align}
\phi^{\text{pri}} = \frac{c^{b}(t)}{c^{\text{avg}}(t)},
\end{align}
\begin{align}
\phi^{\text{net}} = \frac{P^{\text{net}}(t)}{P^{\text{net,avg}}(t)},
\end{align}
where $\phi^\text{pri}$ is the ratio of $c^b(t)$ to the daily average electricity price $c^\text{avg}(t)$. $\phi^\text{net}$ represents the ratio of the current net load $P^\text{net}(t)$ to the average net load $P^\text{net,avg}$ over the past 48-hour window. The value of $P^\text{net,avg}$ is continuously updated as the scheduling progresses. To guide the agent in learning the appropriate control actions for energy storage during valley tariff and discharging during peak tariff, we devise scaling coefficients $w^\text{dis}$ and $w^\text{ch}$. Fig. \ref{disch} clarifies the design principles behind them.  When $\phi^\text{pri}$ is greater than 1 and the current net load is also relatively high, the reward for CBS discharging is significantly increased through an exponential function, while charging actions are penalized. Conversely, when $\phi^\text{pri}$ is less than 1 and net load is low, the reward for CBS charging is exponentially amplified, while discharging actions are penalized. When the ratio equals 1, it indicates an average price period. In this case, no scaling of charging or discharging costs occurs, maintaining the fundamental logic that discharging generates revenue while charging incurs costs. The variable $\phi^\text{net}$  serves as an indicator of net load levels, incentivizing the agent to supply energy to buildings when net load is high and to store energy when it is low. The ranges of variation for $\phi^\text{pri}$ and $\phi^\text{net}$ are restricted to $[0.4, 2.2]$ and $[0.2, 2.4]$, respectively. The broader range for $\phi^\text{net}$ encourages the agent to prioritize net load when formulating scheduling strategies. Thus, rather than just responding to high electricity prices for discharging, the agent analyzes predicted net load trends and prioritizes power supply to the building during high electricity prices and high net load periods, optimizing the operational costs of the BEMS.

The rationale for exponentially amplifying rewards for correct actions while applying linear penalties for incorrect actions lies in the complex control logic inherent in the CBS energy scheduling. For instance, during nighttime, the electricity price may be slightly below the flat tariff but higher than the valley tariff. If an exponential penalty is imposed on discharging actions during this period, it could discourage the CBS from emptying stored energy at the flat tariff. This would diminish the incentive to charge at high power during the valley tariff. Ultimately, it would affect the operational benefits of the BEMS. Conversely, exponential rewards for correct actions provide a clear and strong incentive for the desired behavior of the system. This specifically guides the agent to charge/discharge during valley/peak tariff periods.
\begin{figure*}[th]
	\centering
	\includegraphics[width=0.45\textwidth]{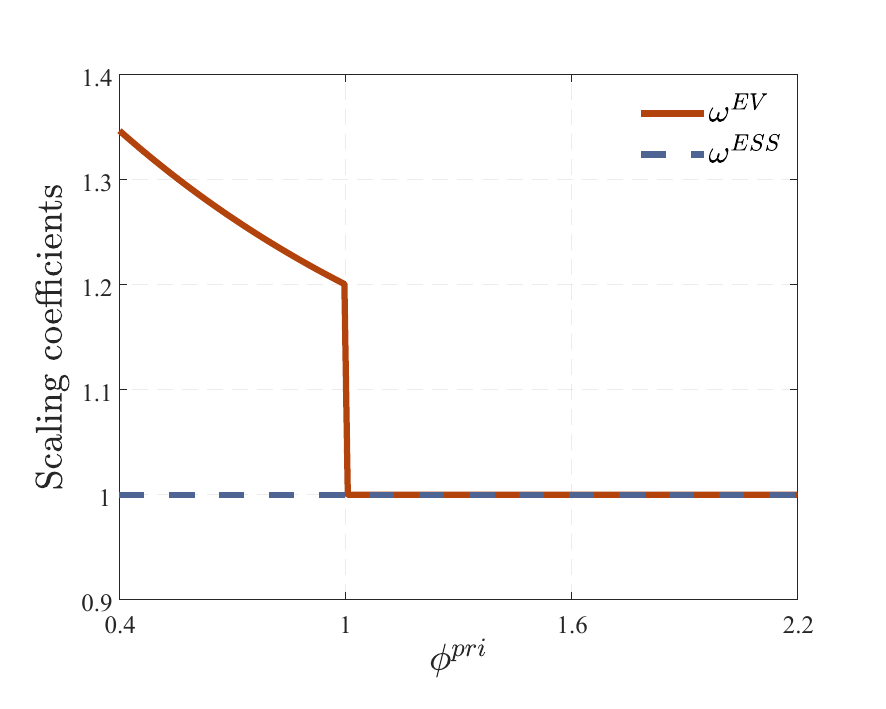}
	\caption{The design principles of scaling coefficients for the degradation cost of ESS and EV.}
	\label{ESSEV_scaling}
\end{figure*}

\begin{align}
w^{\text{ESS}} = \frac{\theta_{\text{ESS}}}{\theta_{\text{base}}},
\end{align}
\begin{align}
w^{\text{EV}} = \frac{\theta_{\text{EV}}}{\theta_{\text{base}}},
\end{align}
\begin{align}
\theta_{\text{EV}} &= 
\begin{cases} 
\theta_{\text{ESS}} \left(1 + 0.5 \cdot e^{-\theta_{\text{scale}} \cdot \phi^{\text{pri}}}\right) & \text{if } \phi^{\text{pri}} \leq 1 \\
1 & \text{otherwise}
\end{cases}, \\
\end{align}
where $\theta_\text{ESS}=\theta_\text{base}=1$, $\theta_\text{scale}=0.916$. The design of $w^\text{ESS}$ and $w^\text{EV}$ aims to balance battery degradation costs and arbitrage profits. Specifically, these coefficients ensure the maximization of arbitrage gains while preventing excessive degradation of the EV battery during the energy scheduling process. The functioning of $w^\text{ESS}$ and $w^\text{EV}$ is illustrated in Fig. \ref{ESSEV_scaling}. The $C^\text{ESS}_\text{cycle}(t)$ serves as a baseline, while $C^\text{EV}_\text{cycle}(t)$ is sensitive to electricity prices. When the electricity price is at or below the flat tariff level, an exponential decay function is employed to smoothly amplify the aging cost of the EV battery, thereby restraining the agent's scheduling actions on the EV battery to maintain its health. The $\theta^\text{scale}$ controls the sensitivity of the ESS to different price ratios. The lower the electricity price, the higher the scheduling cost for the EV. Conversely, when electricity prices are higher, the amplification of the EV battery's degradation cost is reduced to prioritize profit-seeking. The maximum value of $w^\text{EV}$ does not exceed 1, ensuring the fundamental logic that the scheduling cost of EVs is always higher than that of the ESS.

In practice, the design of $w^\text{EV}$ incorporates expert knowledge. Since EV scheduling primarily occurs during the day, making it difficult to charge at low prices to store energy. Thus, $w^\text{EV}$ is primarily geared towards penalizing discharge actions during valley tariff periods. Peak tariff periods typically overlap with EV scheduling times, allowing the agent to autonomously learn arbitrage strategies during high-price periods through interactions with the environment. This design enables the agent to optimize decisions that balance arbitrage profits with battery degradation costs. The BEMS is likely to prioritize scheduling the ESS with a lower cyclic aging lifespan to maximize earnings. When profits significantly exceed battery degradation costs, EVs will also participate in the energy scheduling process, ultimately maximizing the system's operational returns.
\begin{align}
\text{SoC}^{\text{pen}}(t) = \left( P^{\text{ESS,bey}}(t) + P^{\text{EV,bey}}(t) \right),
\end{align}
where $P^\text{ESS,bey}(t)$ and $P^\text{EV,bey}(t)$ represent the power by which the charging and discharging actions of the ESS and EV respectively exceed the SoC boundary constraints. The greater the violation of these constraints, the higher the penalty imposed. Excessive charging or discharging of lithium batteries significantly impacts their lifespan. By imposing SoC boundary penalties, the goal is to compel the agent to ensure battery health during energy management.

\section{Methodology} 
\subsection{DRL-based method}
\subsubsection{Preliminary of RL and DQN}
RL is a powerful decision optimization technique based on trial-and-error learning \cite{Sutton1998}. An agent interacts with the environment, receiving rewards to guide its behavior. As a crucial branch of machine learning, RL focuses on enabling intelligent agents to take actions that maximize cumulative rewards through an optimal policy $\pi^*$. For the joint real-time energy scheduling problem in this paper, due to the complexity of coordinating system arbitrage revenue and operating costs, as well as the uncertainty of future building net loads, determining $\pi^*$ is challenging. RL iterates and optimizes the action-value function based on the Bellman equation. This core mechanism effectively captures the dynamic relationship between current actions and future rewards, providing a feasible framework for addressing the optimal scheduling problem.

\begin{figure*}[]
	\centering
	\includegraphics[width=0.7\textwidth]{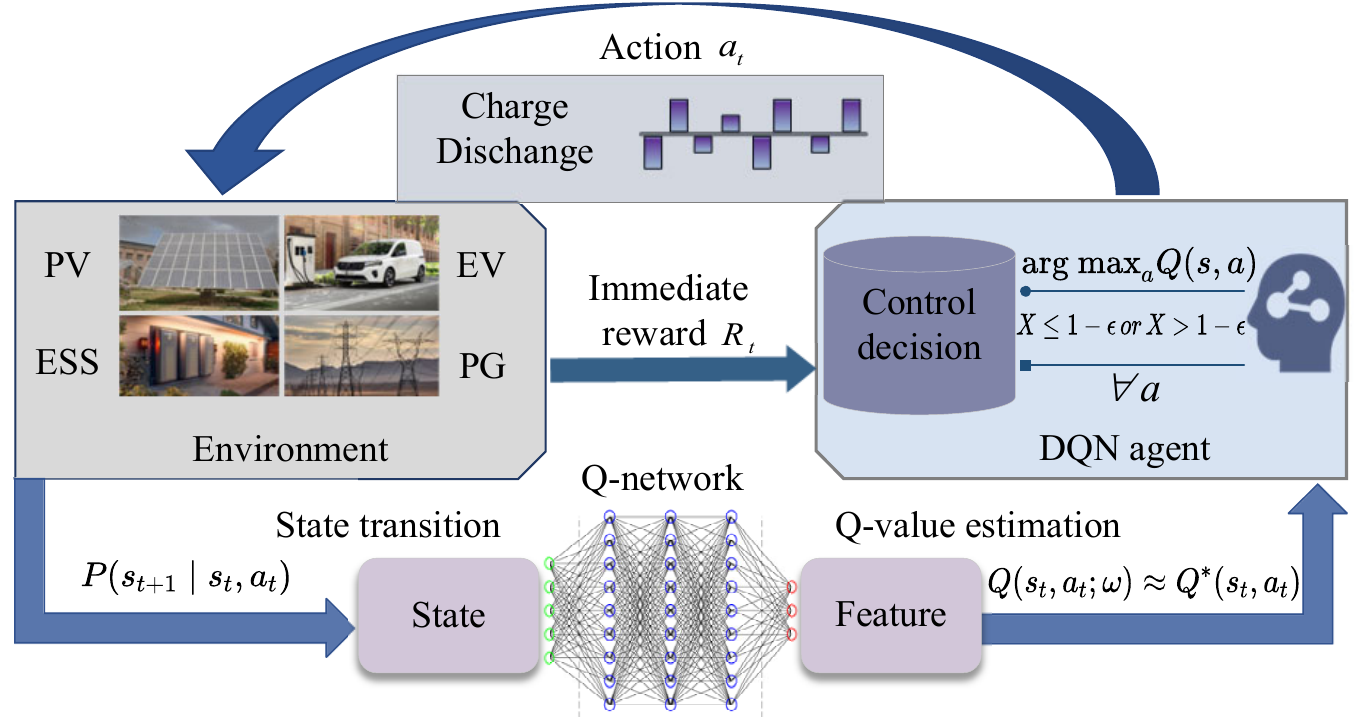}
	\caption{Energy scheduling decision-making based on DQN.}
	\label{DQN}
\end{figure*}

In scheduling decisions, the agent first perceives the current state of the environment, which encapsulates all relevant information that may influence the agent's decision-making. The value of the control action $a$ under state $s$ with policy $\pi$ is calculated by the action-value function $Q_\pi (s,a)$, also known as the Q-value. It is defined through the Bellman equation as follows:
\begin{align}
Q_\pi(s,a)=\mathbb{E}_\pi\left[\sum_{k=0}^{\infty}\gamma^k\cdot R_{t+k}|s_t=s,a_t=a\right],
\end{align}
where $\gamma \in (0,1)$ represents the discount factor, which is used to balance the importance of immediate rewards and future rewards. $k$ represents the index of time steps and is used to accumulate the rewards $R_{t+k}$ for different time steps in the future. The policy $\pi$ maps system states to specific charging and discharging actions of the CBS. The agent constantly explores the environment based on the $\epsilon$-greedy policy as followed:
\begin{align}
\label{greedy}
a = \begin{cases}
\underset{a}{\operatorname{argmax}} Q(s,a), & X \leq 1 - \epsilon \\
\forall a, & X > 1 - \epsilon
\end{cases},
\end{align}
where $X$ represents a randomly generated number between 0 and 1. Once the action is carried out, the agent receives a reward from the environment. The reward serves as a signal that indicates the quality of the action taken in relation to the agent's objective of maximizing long-term rewards. Then the Bellman equation will be iteratively updated to calculate Q-value, as follows:
\begin{align}
Q(s_t,a_t) &\leftarrow Q(s_t,a_t)+\lambda\left[R_t+\gamma\max_aQ(s_{t + 1},a)-Q(s_t,a_t)\right],
\end{align}
where $\lambda \in (0,1)$ represents learning rate, which is used to control the magnitude of updates to the Q-value. As the number of iterations $i$ increases, $i\rightarrow\infty$, $Q_\pi (s,a)$ will converge to the optimal action-value function $Q^* (s,a)$. At this point, the optimal scheduling strategy is determined.  
\begin{align}
a^*=\underset{a\in\mathcal{A}}{\operatorname*{\operatorname*{arg max}}}Q^*(s,a).
\end{align}

DQN mark a significant advancement in RL, efficiently addressing complex Markov decision problems \cite{mnih2015human}. The principle of energy scheduling decision-making based on DQN is illustrated in Fig. \ref{DQN}. Traditional RL algorithms often struggle with high-dimensional state and action spaces of energy scheduling problems. In these cases, the complexity of representing and learning the Q-function becomes unmanageable. DQN overcomes this problem by utilizing the feature extraction and approximation capabilities of deep neural networks. DQN is trained using the Bellman equation and the concept of temporal difference (TD) learning. By minimizing the loss between predicted Q-values and target Q-value, DQN gradually learns the optimal policy. The Q-function represented by DQN with weights $\omega$ is expressed as:
\begin{align}
Q(s_t,a_t;\omega)\approx Q^*(s_t,a_t),
\end{align}
By calculating the mean squared error (MSE) loss function $L(\omega)$ between Q-values and the TD at each training step, the parameters of the neural network are updated. This can be expressed as:
\begin{align}
\label{Q}
L(\omega)=\mathbb{E}\left[(y_t-Q(s_t,a_t;\omega))^2\right],
\end{align}
\begin{align}
y_t=R_{t}+\gamma \underset{a}{\mathrm{max}}\hat{Q}(s_{t+1},a_{t+1};\omega^{-}),
\end{align}
where $y_t$ is the TD target, $Q$ is the online network (O-NET), and $\hat Q$ is the target network (T-NET). $\omega$ and $\omega^{-}$ represent the weights of the O-NET and the T-NET, respectively. The $\omega$ of the O-NET are continuously updated by receiving reward signals and state transition information from the environment. Optimization algorithms, such as gradient descent, minimize the loss function, improving Q-value estimation over time. The primary role of the T-NET is to provide a stable target for the learning of the O-NET. Periodically, $\hat Q$ copies the parameters from the $Q$ and remains fixed for a set duration, not interacting with the environment or learning in real-time. This approach helps prevent fluctuations in learning objectives caused by the continuous changes in Q-value estimates, thereby enhancing the stability and convergence of the learning process.

\subsubsection{Double dueling DQN with with prioritized experience replay}
In the heterogeneous energy storage battery environment of this study, applying the DQN algorithm to address online real-time energy scheduling issues faces challenges related to declining Q-value estimation accuracy and difficulties in action selection. Firstly, the system involves dynamically changing PV and load information, along with diverse DERs such as fixed ESS and mobile EVs. These devices exhibit significant heterogeneity and interact with one another, coordinating charging and discharging between ESS and EVs, as well as managing hierarchical energy utilization among CBS, GB, and the PG. This complexity results in a high-dimensional and intricate state space, making it difficult for the agent to accurately extract the features reflecting the system state, thereby impacting the precision of Q-value estimation. Secondly, the system requires coordinated operation of ESS and EVs within the CBS, with the action space encompassing all possible combinations of their charging and discharging operations. The scheduling strategy for ESS and EVs should not solely rely on electricity price signals but also consider the differential degradation costs arising from their distinct chemical properties. This necessitates that the agent balances long-term returns with short-term economic benefits. Furthermore, energy scheduling decisions must adhere to constraints such as SoC boundaries and the EV scheduling availability windows. This leads to a coupled and multi-constrained complex action space. The agent's ability to select control actions is hindered by the aforementioned issues, making it challenging to effectively learn and devise scheduling strategies that contribute to long-term cumulative returns.

To address the issues of Q-value estimation bias and inefficient action selection from high-dimensional and complex state and action spaces, we constructed the double dueling DQN (D3QN) architecture. This approach helps stabilize and efficiently learn optimized scheduling strategies, and its detailed description is as follows. In the traditional DQN framework, the TD learning process can lead to overestimation of $\underset{a} \max Q(s,a;\omega)$ in Eq. \eqref{Q}, which negatively impacts the agent's learning. By introducing a double network structure, we decouple action selection from target Q-value generation, forming double DQN (D2QN), aiming to stabilize the agent's learning process. In D2QN, the best action in the next state is selected using the current DQN network weights $\omega$. The target Q-value for that action is evaluated using the weights $\omega^{-}$ of the previous T-NET. Thus, the TD target $y_t^{D}$ for D2QN can be expressed as:
\begin{align}
y_t^{D} = R_{t} + \gamma \hat{Q}\left(s_{t + 1}, \underset{a_{}}{\operatorname{argmax}} Q(s_{t + 1}, a_{t + 1}; \omega), \omega^-\right).
\end{align}

To optimize action selection in high-dimensional coupled multi-battery environments, we introduced a dueling mechanism to construct the D3QN based on the D2QN framework. D3QN decomposes the Q-values into the action-independent value function $V(s,v)$ and the advantage function $A(s,a,\omega)$, allowing the network to reduce redundant learning in the high-dimensional CBS action space and analyze the coupling relationships between ESS and EVs. Specifically, D3QN adds two sub-networks before the output layer of the main network to separately learn $V$ and $A$, with network weights denoted as $v$ and $\omega$. Subsequently, D3QN combines $V$ and $A$ to generate a more accurate Q-function:
\begin{align}
\label{Dueling}
Q(s, a) = V(s, v) + A(s, a, \omega),
\end{align}
The D3QN model explicitly distinguishes between the overall value of the current state and the relative advantages of different actions, aiding the agent in comprehending the comprehensive value and long-term returns across various states. This capability facilitates making more optimal coordinated scheduling decisions.

The real-time joint energy scheduling system operates within a dynamic environment. Energy supply and demand can fluctuate significantly due to factors such as sudden changes in renewable energy sources and the stochastic nature of user consumption behavior. The agent must continuously adapt to these dynamics and learn from a vast number of state-action pairs to develop robust control strategies. In traditional DQN architectures, the training of network is advanced through the introduction of experience replay mechanism (ERM). Initially, an experience replay buffer $\mathcal{D}$ with a capacity of $\mathcal{I}$ is established to store training samples $(s_t,a_t,R_t,s_{t+1})$ from each time step. As training progresses and the buffer fills, the oldest experiences are gradually replaced by new ones while maintaining the capacity $\mathcal{I}$. Given the aim of developing a general real-time energy scheduling framework that includes multiple energy components, training data encompass system state features under different seasons, weather conditions, etc., resulting in an environment with high diversity. Traditional ERM randomly selects samples from $\mathcal{D}$ for learning. This may result in highly relevant samples for improving the current policy not being fully utilized, making it difficult to adapt to the system's dynamic changes. Moreover, in a diverse environment, the uniform sampling approach of traditional ERM may struggle to learn effectively from infrequently occurring yet crucial training samples. This limitation can negatively impact the algorithm's optimization efficiency.

To address these issues, prioritized experience replay (PER) is introduced to establish D3QNPER algorithm, facilitating effective learning for the agent. The core idea of PER is to assign a priority to each sample in the experience buffer. Samples with higher priorities have a greater probability of being selected for training. The priority of a sample is determined based on its TD-error. A higher TD-error indicates greater importance for improving the current policy, resulting in a higher priority, which can be expressed as:
\begin{align}
\label{per1}
L^{TD}(\omega)=\left|y_t^{D}-Q(s_t,a_t;\omega)\right|,
\end{align}
For each sample, a probability  $p_i = L^{TD}(\omega) + \psi$ is calculated based on Eq. \eqref{per1}, which reflects the magnitude of the current Q-value prediction error. $\psi$ is an extremely small positive number, which gives low-priority samples a chance to be selected, thereby maintaining sampling diversity. Samples with high contributions during dynamic changes in BEMS will receive high priority. This allows the agent to adjust its strategy dynamically to adapt to the latest system changes. Based on the proportional prioritization method, the sampling probability $P(i)$ of the $i$-th sample can be calculated according to $p_i$:
\begin{align}
\label{per2}
P(i)=\frac{p_i^{\alpha^\text{per}}}{\underset{{i\in \mathcal{I}}}{\sum} {{p_i^{\alpha^\text{per}}}}},
\end{align}
where $\alpha^\text{per}\in (0,1)$ controls the intensity of priority sampling. As it increases from 0 to 1, the algorithm gradually changes from uniform sampling to sampling completely based on priority. Due to the introduction of the priority concept, samples with low occurrence frequency but significant influence on decision-making will have a higher selection probability, thereby alleviating the problem of sample imbalance. The excessive reuse of high-priority samples will introduce biases, which need to be corrected through important sampling weight $\vartheta$ as followed:
\begin{align}
\label{per3}
\vartheta_i=\left(\frac{1}{\mathcal{I}\cdot P(i)}\right)^{\alpha^\text{dev}},
\end{align}
where ${\alpha^\text{dev}}\in (0,1)$ controls the intensity of deviation correction. It increases from the initial value of 0.4 to 0.99 as the training process progresses, balancing early instability with later deviation correction. After the error is corrected, the loss function with PER can be expressed as:
\begin{align}
\label{per4}
L_i(\omega_i)=\mathbb{E}\left[\left(\vartheta_{i}(y^{D}_t-Q(s_t,a_t;\omega_i))\right)^2\right].
\end{align}

\subsection{Electricity load and PV generation prediction with RDedRVFL}
The real-time energy scheduling of BEMS based on the DQN algorithm relies on prediction trend information of PV generation and building load. This guides the agent in accurately assessing long-term benefits and avoiding short-term decision-making. However, the achievement of accurate predictions for PV generation and building load presents significant challenges. On one hand, PV generation is highly correlated with factors such as weather and seasons, which significantly impact the accuracy of PV output predictions \cite{xu2025double}. On the other hand, in addition to the aforementioned factors, building load exhibit greater irregularity and volatility due to the random energy consumption habits of occupants, making it difficult for predictive models to accurately capture the changing patterns of building load \cite{liu2025building}.

In response to the high volatility and time-varying characteristics of new energy output and building load, we construct the prediction model based on ranking-based dynamic ensemble deep random vector functional link, namely RDedRVFL \cite{liu2025building}. The RDedRVFL effectively integrates deep learning and ensemble learning techniques, and it incorporates a dynamic ensemble strategy to enhance the network's learning capability. To cope with the time-varying nature of PV output and building load, RDedRVFL stacks multiple random vector functional link (RVFL) modules to construct a deep network that extracts temporal features from the time series data layer by layer. Meanwhile, the  dynamic combination weights based on the latest accuracy is designed to adapt to the dynamic fluctuations of the time series data. To mitigate the decline in prediction accuracy caused by the high volatility of PV output and building load, RDedRVFL employs ensemble learning techniques and a diversity measurement mechanism to fuse decision-level features from different enhancement layers, thereby ensuring the stability of predictions. 

RDedRVFL is divided into two stages: edRVFL learning and decision-level feature fusion. Algorithm \ref{alg1} outlines the specific learning process of the edRVFL, which consists of $L$ enhancement layers, with $N$ samples and $m$ feature dimensions. In Algorithm \ref{alg1}, $U$ represents the number of neurons in the enhancement layer, $+$ denotes the Moore–Penrose pseudo-inverse, $D=[H^{l},X]$, and $I$ represents the identity matrix. In edRVFL, a direct link is established between each enhancement layer and the output layer. This setup enables multi-scale feature extraction from load and PV data through the features of each RVFL module, ultimately resulting in a collection of prediction outputs of each layer, denoted as $\hat{Y}=[\hat{y}_{1}, \hat{y}_{2},\dots,\hat{y}_{L}]^\mathrm{T}$.

\begin{algorithm}
\caption{The edRVFL learning}
\label{alg1}
\begin{algorithmic}[1]
\State \textbf{Input}: Input data $X$, target output $Y$, number of enhancement layers $L$, non-linear activation function $g(\cdot)$, regularization parameter $\lambda^\text{re}$
\State \textbf{Output}: Optimal output layer weight vector $\beta = [\beta_{1},\beta_{2},\ldots,\beta_{L}]^{\mathrm{T}}$, the predicted value $\hat{Y}=[\hat{y}_{1}, \hat{y}_{2},\dots,\hat{y}_{L}]^\mathrm{T}$
\State \textbf{Initialize} the hidden layer weights $B_1,B_2,\ldots,B_l$
\For{$l = 1$ to $L$}
    \If{$l = 1$}
        \State Calculate the enhancement features of the first layer based on ${H}^{1}=g({X}{B}_{1})$ 
        \State Calculate the output of the first layer based on ${Y}^{1}=[{H}^{1}, {X}]\beta_{1}$ 
    \Else
        \State Calculate the enhancement features of the $l$-th layer based on $H^{l}=g([H^{l-1},X] B_{l})$
        \State Calculate the output of the $l$-th layer based on $\hat{Y}_{l}=[ H^{l}, {X}]\beta_{l}$
    \EndIf
    \State Calculate the loss function of the $l$-th layer based on $Loss^{l}_\text{edRVFL}=\|[ H^{l}, {X}]\beta_{l}-Y\|^2+\lambda^\text{re}\|\beta_{l}\|^2$
    \If{$m + U \leq N$}
        \State $\beta_{l} = (D^{\mathrm{T}}D + \lambda^\text{re} I)^{-1}D^{\mathrm{T}}Y$; 
    \Else
        \State $\beta_{l} = D^{\mathrm{T}}(DD^{\mathrm{T}} + \lambda^\text{re} I)^{-1}Y$; 
    \EndIf
\EndFor
\State \Return $\beta = [\beta_{1},\beta_{2},\ldots,\beta_{L}]^{\mathrm{T}}$, $\hat{Y}=[\hat{y}_{1},\hat{y}_{2},\dots,\hat{y}_{L}]^\mathrm{T}$
\end{algorithmic}
\end{algorithm}

The decision-level feature fusion stage is detailed in Algorithm \ref{alg2}. In this algorithm, the RDedRVFL employs a ranking-based dynamic strategy to integrate the outputs from all prediction layers, effectively addressing the challenges posed by irregular fluctuations, time-varying characteristics, and outlier predictions in building loads and renewable energy generation. First, the accuracy and diversity metrics are defined based on the latest errors and the forecast values distribution in Euclidean space, respectively. Then, a ranking-based strategy is used to quantify the contribution of each layer, which is then transformed into combination weights. Finally, the output of each prediction layer is used to calculate the final combination forecast $\hat{Y}_c (t)$ based on combination weights.

\begin{algorithm}
\caption{Decision-level feature fusion}
\label{alg2}
\begin{algorithmic}[1] % 使用1开始进行编号
\State \textbf{Input}: Prediction output of the edRVFL $\hat{Y}=[\hat{y}_{1}, \hat{y}_{2},\dots,\hat{y}_{L}]^\mathrm{T}$, actual values $y$, trade-off parameter $\alpha^\text{trade}$
\State \textbf{Output}: Combination weights $w^{R}_i$ for each layer, final combination forecast $\hat{Y}_c (t)$

% 计算基于最新精度的贡献指标
\For{$l = 1$ to $L$}
    \State Define contribution indicators based on the latest error by $f_{l}={1}  /  {(\hat{y}_{l}(t - 1)-y(t - 1))^2}$
    % 对基于最新精度的贡献指标进行排序
    \State Sort $f_{l}$ in descending order, $f_{1}\ge f_{2}\ge\cdots\ge f_{l}\ge\cdots\ge f_{L}$, obtain the sorting index $I^{f}_{l}$
    \State Calculate the ranking of latest accuracy contribution by $F_{l}=L-I^{f}_{l}+1$ 
    % 计算基于预测分布的多样性贡献
    \State Calculate the diversity contribution based on Euclidean distance by $E_{l}=\sum_{i=1}^L\sqrt{(\hat{y}_{l}(t)-\hat{y}_{i}(t))^2}$
    % 对基于预测分布的多样性贡献进行排序
    \State Sort $E_{l}$ in descending order, $E_{1}\ge E_{2}\ge\cdots\ge E_{l}\ge\cdots\ge E_{L}$, obtain the sorting index ${I}^{d}_{l}$
    \State Calculate the diversity ranking based on prediction distribution by ${D}^{d}_{l}=L-{I}^{d}_{l}+1$ 
    %%%%%%%%%%%%%%%%%%%%%
    % 计算基于预测性能的多样性贡献
    \State Find the middle layer index by $F_{l}=L/2$, get the middle layer's contribution indicator ${f}_{mid}$
    \State Calculate the deviation from ${f}_{mid}$ by $f^{d}_{l}=|f_{l}-{f}_{mid}|$
    % 对基于预测性能的偏差进行排序
    \State Sort $f^{d}_{l}$ in descending order, $f^{d}_{1}\ge f^{d}_{2}\ge\cdots\ge f^{d}_{l}\ge\cdots\ge f^{d}_{L}$, obtain the sorting index ${I}^{p}_{l}$
    \State Calculate the diversity ranking based on prediction performance by ${D}^{p}_{l}=L-{I}^{p}_{l}+1$ 
    % 计算整体多样性贡献排名
    \State Define contribution indicator based on the overall diversity measurement by $d_{l}={({D}^{d}_{l}+{D}^{p}_{l})} / {\sum_{i=1}^L({D}^{d}_{i}+{D}^{p}_{i})}$
    \State Sort $d_{l}$ in descending order, ${d}_{1}\ge {d}_{2}\ge\cdots\ge {d}_{l}\ge\cdots\ge {d}_{L}$, obtain the sorting index $I^{o}_{l}$
    \State Calculate the overall diversity ranking by ${D}_{l}=L-I^{o}_{l}+1$ 
    %%%%%%%%%%%%%%%%%%%%%
    % 计算最终贡献指标
    \State Define the final contribution indicator based on accuracy and diversity by $r_{l}=\alpha^\text{trade} F_{l}+(1-\alpha^\text{trade})D_{l}$
    %%%%%%%%%%%%%%%%%%%%%
    % 基于排序策略计算组合权重
    \State Sort $r_{l}$ in descending order, $r_{1}\ge r_{2}\ge\cdots\ge r_{l}\ge\cdots\ge r_{L}$, obtain the sorting index $I^{r}_{l}$
    \State Calculate the final ranking by $R_{l}=L-I^{r}_{l}+1$ 
    \State Calculate the combination weight based on ranking value by $w^{R}_{l}={R_l}/{\sum_{i=1}^LR_{i}}$
    %%%%%%%%%%%%%%%%%%%%%
    % 计算组合预测值
    \State Calculate the combination forecast by $\hat{Y}_c (t)=\sum_{i=1}^L w^{R}_i\hat{y}_{i}{(t)}$
\EndFor
\State \Return $w^{R}_i$, $\hat{Y}_c (t)$
\end{algorithmic}
\end{algorithm}

\subsection{The proposed joint real-time energy scheduling method}

\begin{figure*}[th]
	\centering
	\includegraphics[width=0.96\textwidth]{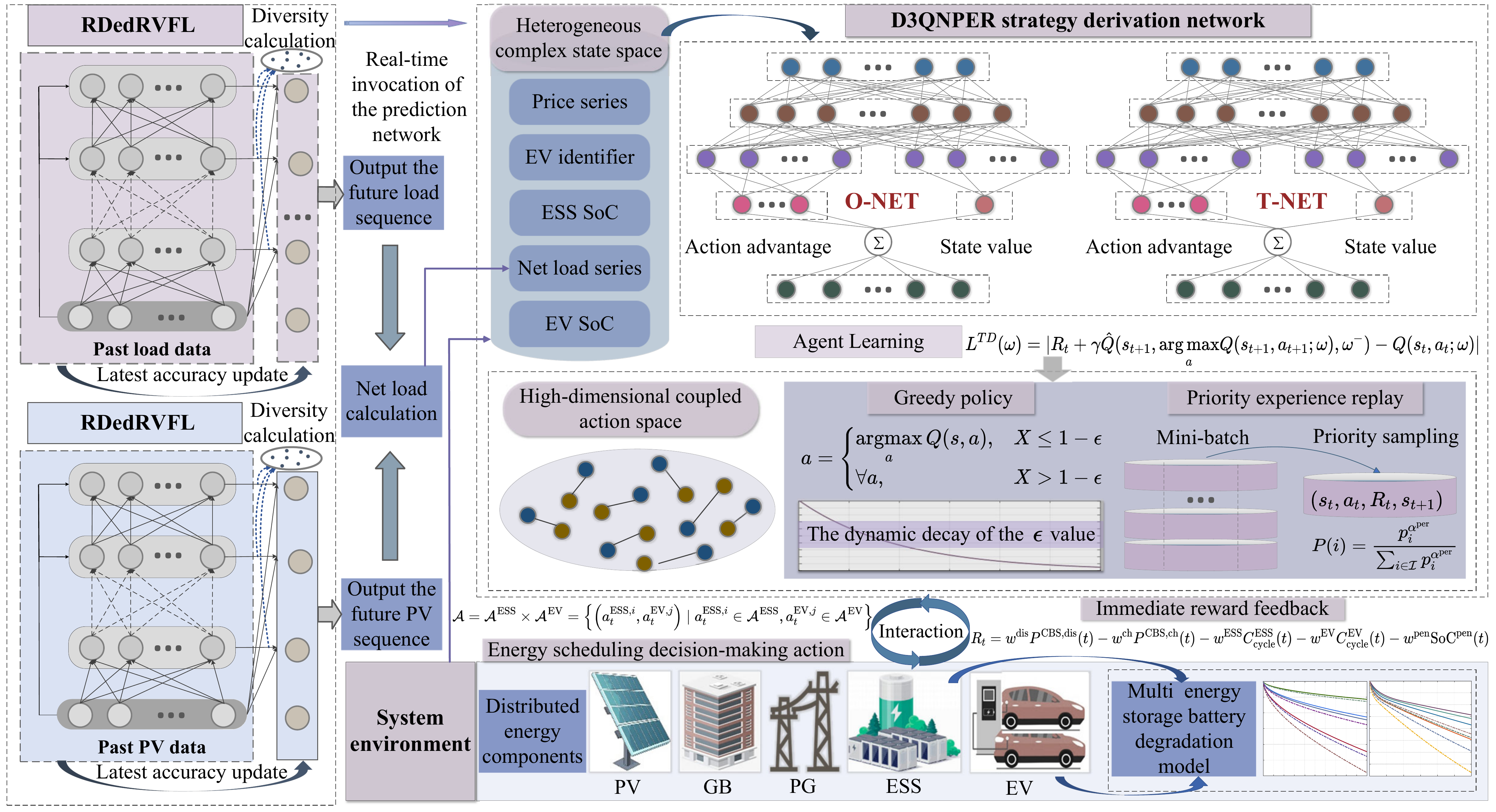}
	\caption{The framework of proposed energy scheduling method.}
	\label{proposed2}
\end{figure*}

In this subsection, the proposed joint real-time energy scheduling method, namely RDedRVFL-D3QNPER, is elaborated upon. The framework of RDedRVFL-D3QNPER is illustrated in Fig. \ref{proposed2}. The online prediction network, constructed based on RDedRVFL, captures future changes in net load, guiding the agent to prioritize building demands for optimizing long-term returns. The output of the prediction network is incorporated as part of the system environment and integrated into the strategy derivation network D3QNPER. To address biased Q-value estimation and inefficient action selection in complex state and action spaces, D3QNPER employs a double dueling network architecture. The two-stream fully connected layer structure decouples the action selection from the generation of the target Q-value. This structure distinguishes between the overall state value and the relative advantages of different actions, optimizing both Q-function computation and action selection. Moreover, D3QNPER leverages the PER strategy to tackle the challenge of declining sample learning efficiency in dynamic environments. Based on above design, the agent continuously interacts with the system environment in real-time, exploring various energy coordination schemes guided by the reward function to achieve the overall optimization of the multi-energy system.

\begin{algorithm}
\caption{The proposed joint real-time energy scheduling method}
\label{alg3}
\begin{algorithmic}[1] % 使用1开始进行编号
\State \textbf{Input}:{Electricity price data $X_c$, historical building load data $X_\text{load}$, historical PV power generation data $X_\text{PV}$, the T-NET $\hat{Q}$ update parameter $\mathcal{C}$, $\xi(t_0)$, $\text{SoC}^\text{ESS}({t_0})$, $\text{SoC}^\text{EV}({t_0})$, $\alpha ^\text{ESS}_{d}$ and $\alpha ^\text{EV}_{d}$, $\epsilon$, $\gamma$}, $\lambda$
\State \textbf{Output}:{The optimal D3QNPER network weights $\omega^*$ for joint real-time energy scheduling}
\State \textbf{Initialize} The replay memory $\mathcal{D}$ and the mini-batch size, the O-NET $Q$ with weights $\omega$, the T-NET $\hat{Q}$ with weights $\omega^- = \omega$
%%%%%%%%%%%%%%%%%%%%%%%
\State Preprocess the historical building load and PV data, $X_\text{load}$ and $X_\text{PV}$
\State Execute Algorithms \ref{alg1} and \ref{alg2} to obtain the predicted PV power generation and building load data series 
\State Calculate $({P}^\text{net}({t_0}) \cdots {P}^\text{net}({t_0}+22), {P}^\text{net}({t_0}+23))$ based on the predicted PV and load data series
\State Initialize the system environment and generate the initial system state $s_{t_0}$
%%%%%%%%%%%%%%%%%%%%%%%
\For{Episode = $1$ to $M$}  
    \State Reset the system environment, and update the $\xi$, $\alpha ^\text{ESS}_{d}$ and $\alpha ^\text{EV}_{d}$, etc, for the new episode
    \For{Time step $t$ = $1$ to $T$}
        \State Estimate and record the Q-value based on Eq. \eqref{Dueling}
        \State Select a combined action $a_t$ of ESS and EV based on $\epsilon$-greedy policy via  Eq. \eqref{greedy}
        \State Decompose $a_t$ into $a^\text{ESS}_t$ , $a^\text{EV}_t$, and map them to the specific charging/discharging powers
        \State Calculate the battery cycle aging costs $C^\text{ESS}_\text{cycle}(t)$ and $C^\text{EV}_\text{cycle}(t)$ of ESS and EV via $\alpha ^\text{ESS}_{d,j-1}$ and $\alpha ^\text{EV}_{d,j-1}$
        \State Calculate the operation cost of the BEMS at time step $t$
        \State Calculate the reward score $R_t$ according to Eq. \eqref{reward}
        \State Execute the online prediction steps in line 5, and update the remaining state features
        \State Obtain the system observation state $s_{t + 1}$ for the next time step $t+1$
        \State Store transition $(s_t, a_t, R_t, s_{t + 1})$ as a training sample in $\mathcal{D}$
        \State Calculate the priority sampling probability $P$ of each sample according to Eqs. \eqref{per1}-\eqref{per2}
        \State Execute mini-batch sampling from $\mathcal{D}$ based on the PER strategy.
        \State Estimate the TD target $y_t^{D} = R_{t} + \gamma \hat{Q}(S_{t + 1}, {\arg\max}_{a} Q(S_{t + 1}, a; \omega), \omega^-)$
        \State Calculate the loss function based on the gradient descent algorithm, $L_i(\omega_i)=\mathbb{E}\left[\left(\vartheta_{i}(y^{D}_t-Q(s_t,a_t;\omega_i))\right)^2\right]$
        \If{mod(Episode, $\mathcal{C}$)=1}
            \State $\omega^- = \omega$;
        \Else
            \State $\omega^-$ remain still;
        \EndIf
    \EndFor
    \State Calculate the ESS calendar aging cost $C^\text{ESS}_{\text{cal},j}$  of this episode via Eq. \eqref{ESScal}
    \State Calculate the objective function $\mathcal{P}_1$ via Eqs. \eqref{obj}-\eqref{obj2} that minimizes the system comprehensive operation cost.
\EndFor
\State \Return The optimal D3QNPER network weights $\omega^*$
\end{algorithmic}
\end{algorithm}

The implementation steps of the proposed method are summarized in Algorithm \ref{alg3}. First of all, the proposed method is initialized to generate the initial network structure (Line 3). In the data preparation stage (Line 4), historical building load and PV sequences are preprocessed to remove outliers and fill missing values.  The prediction procedure is executed to construct the initial state space $s_{t_0}$ (Lines 5-7). The outer loop for episodes begins at Line 8, resetting system state information at the start of each episode (Line 9). The inner loop iterates the algorithm at each time step starting from Line 10. Calculate the Q-value and select the control action (Lines 11-12). Lines 13-18 execute control action mappings, calculate battery cycle aging costs, compute the reward function, and update the system state at each time step.  The prediction module is invoked during the system state update step at each time step to generate the next state observation $s_{t + 1}$. Lines 19-27 execute the agent's learning process, with the T-NET updated based on the conditions in Lines 24-27. Notably, since the calendar aging of lithium batteries is not directly linked to real-time DRL scheduling actions, we prioritize the accuracy of calendar aging cost calculations. Instead of approximating the ESS cycle aging cost $C^\text{ESS}_\text{cycle}(t)$ with 
$\alpha ^\text{ESS}_{d}$ (Line 14), we calculate the ESS calendar aging cost precisely at the end of each episode $C^\text{ESS}_{\text{cal},j}$ (Lines 28). The algorithm computes the objective function $\mathcal{P}_1$ at Line 29 and iterates until the termination condition is met. The cumulative reward score is utilized to assess the convergence of the algorithm.

\section{Experimental results}
\subsection{Experiment setup}

\begin{table}[h]
    \centering
    \begin{threeparttable}
        \caption{Descriptive statistics of load and PV.}
        \label{load_pv_data}
        \begin{tabular}{lccc ccc} % 去掉竖线，调整列格式
            \toprule % 顶线
            \multirow{2}{*}{Season} & \multicolumn{3}{c}{Load} & \multicolumn{3}{c}{PV} \\
            \cmidrule(r){2-4} \cmidrule(l){5-7} % 中间横线，左右分别调整
            & Max & Min & Std & Max & Min & Std \\
            \midrule % 中线
            Summer-2023 & 298.77 & 1.17 & 85.22 & 113.99 & 0.00 & 31.20 \\
            Winter-2023 & 170.65 & 0.79 & 38.03 & 76.94 & 0.00 & 21.31 \\
            Summer-2024 & 300.18 & 1.89 & 79.27 & 145.24 & 0.00 & 41.64 \\
            Winter-2024 & 186.42 & 0.14 & 42.38 & 99.93 & 0.00 & 29.47 \\
            \bottomrule % 底线
        \end{tabular}
    \end{threeparttable}       
\end{table}

\begin{figure*}[hbt]
    \centering
    \subfloat[]{\includegraphics[width=0.8\textwidth]{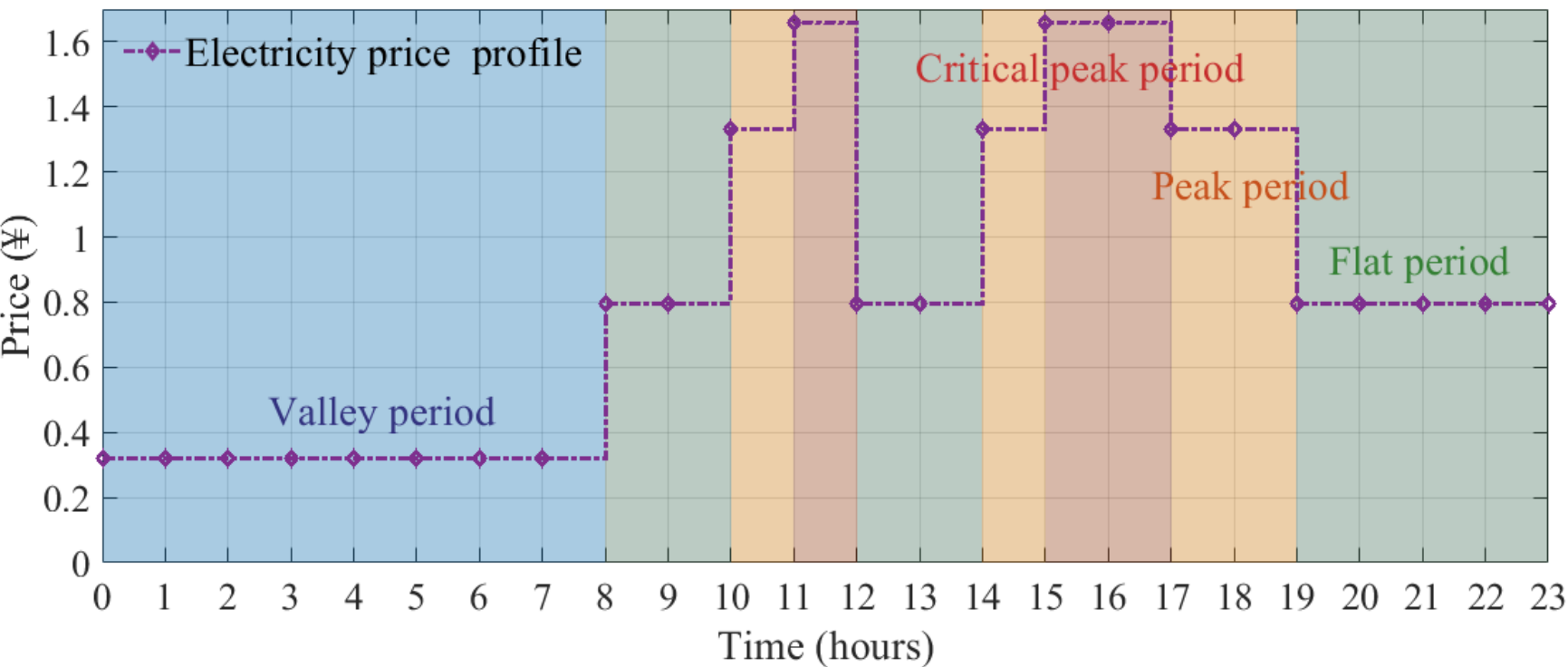}%
        \label{summerprice}}
    \hfil
    \subfloat[]{\includegraphics[width=0.8\textwidth]{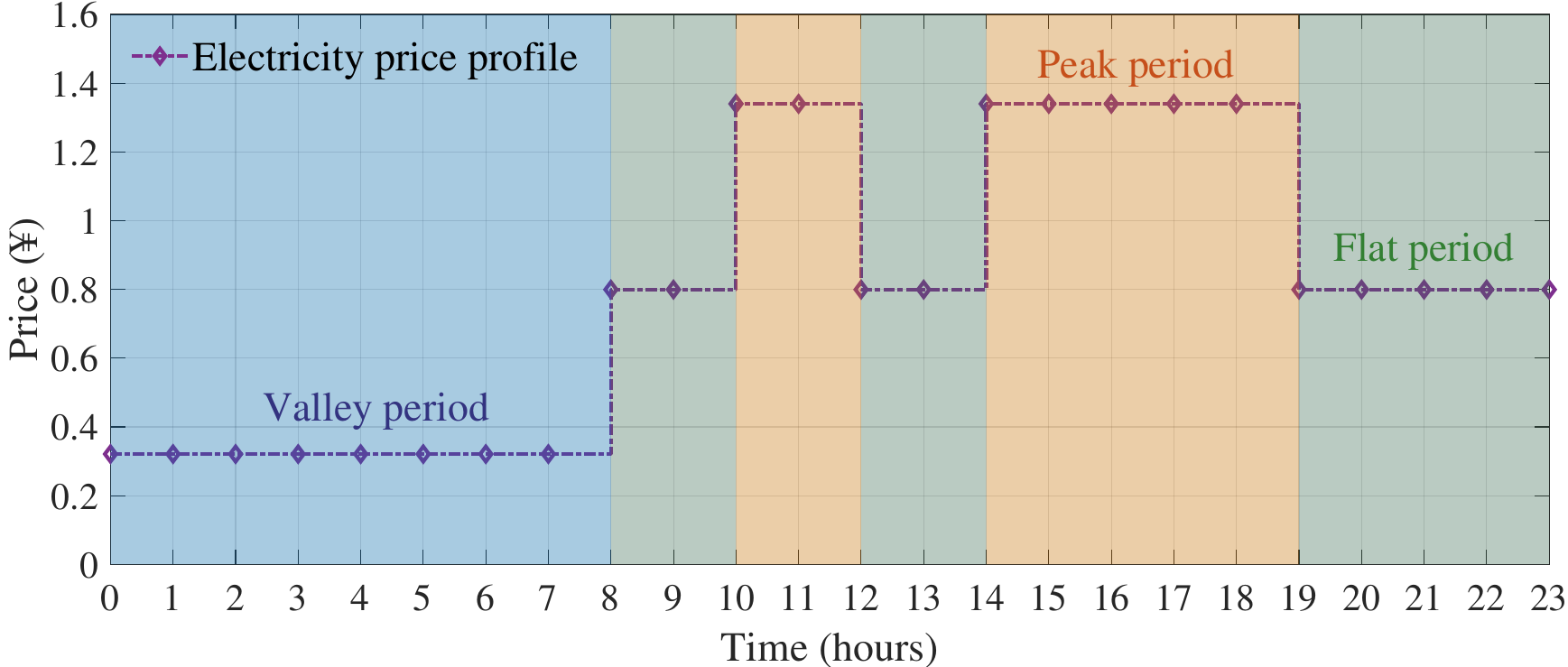}%
        \label{winterprice}}
    \caption{Electricity prices in different seasons. (a) Summer; (b) Winter.}
    \label{fig:seasonalprice}
\end{figure*}

\begin{table}[h]
    \centering
    \begin{threeparttable}
        \caption{Summary of DRL training settings and model parameters.}
        \label{setting}
        \begin{tabular}{clcl}
            \toprule % 绘制表格顶部的横线
            Parameter & Value & Parameter & Value \\
            \midrule % 绘制第一行下方的横线
            Layers & 3 & Discount factor $\gamma$ & 0.99 \\
            Hidden nodes & 128 & Initial $\alpha^\text{ESS}_d$ & 0.35 \\
            Activation function & Relu & Initial $\alpha^\text{EV}_d$ & 0.45 \\
            Learning rate & 0.00025 & Initial $\text{SoC}^\text{ESS}$ & 0.5 \\
            Optimizer & Adam & Initial $\text{SoC}^\text{EV}$ & 0  \\
            Batch size & 32 & $C^\text{ESS}_\text{per}$ & \textyen\,910 \\
            Replay buffer size & 10000 & $C^\text{EV}_\text{per}$ & \textyen\,1092  \\
            Network update & 16 & $\eta^\text{ESS/EV,ch}$ & 0.95 \\
            Epsilon start & 1 & $\eta^\text{ESS/EV,dis}$ & 0.95 \\
            Epsilon end & 0.05 & Price coefficient $\beta_\text{pri}$ & 0.9 \\
            Epsilon decay & 0.99 & Priority sampling coefficient $\alpha^\text{per}$ & 0.95 \\
            Train episodes & 10000 & Deviation correction coefficient $\alpha^\text{dev}$ & (0.4, 0.99) \\
            Trade-off parameter $\alpha^\text{trade}$ & 0.6   & $\theta_\text{ESS}$ & 1\\
            Penalty coefficient $w^\text{pen}$ & 1            & $\theta_\text{base}$ & 1\\
            Small positive number $\psi$ & 0.001          &$\theta_\text{scale}$ & 0.916\\
            \bottomrule % 绘制表格底部的横线
        \end{tabular}
    \end{threeparttable}
\end{table}

% \begin{table}[h]
%     \centering
%     \begin{threeparttable}
%         \caption{Degradation parameters of LFP and NMC.}
%         \label{lfp_nmc_parameters}
%         \begin{tabular}{lllll}
%             \toprule % 绘制表格顶部的横线
%             Parameter & LFP & NMC & Shared parameter & Value \\
%             \midrule % 绘制第一行下方的横线
%             $k_\alpha$ & 5.98E+06 & 1.14E+12 & $k_z$ & 5.00E-01 \\
%             $k_\beta$  & 6.90E-01 & 4.70E+00 & $\alpha_{sei}$ & 5.75E-02 \\
%             $k_\gamma$ & -6.46E+03 & -1.08E+04 & $\beta_{sei}$ & 1.21E+02 \\
%             $k_{\delta1}$ & 9.05E-06 & 1.47E+04 & $T_{ref}$ & 298K \\
%             $k_{\delta2}$ & 1.40E+00 & -1.65E+00 & $\sigma_{ref}$ & 5.00E-01 \\
%             $k_{\delta3}$ & 0.00E+00 & 3.61E+02 & $k_\sigma$ & 1.04E+00 \\
%             &  &  & $k_T$ & 6.93E-02 \\
%             &  &  & $k_t$ & 4.14E-10 \\
%             \bottomrule % 绘制表格底部的横线
%         \end{tabular}
%     \end{threeparttable}    
% \end{table}

\begin{table}[h]
    \centering
    \begin{threeparttable}
        \caption{Degradation parameters of LFP and NMC.}
        \label{lfp_nmc_parameters}
        \begin{tabular}{lllll}
            \toprule % 绘制表格顶部的横线
            Parameter & LFP & NMC & Shared parameter & Value \\
            \midrule % 绘制第一行下方的横线
            $k_\alpha$ & $5.98 \times 10^{6}$ & $1.14 \times 10^{12}$ & $k_z$ & $5.00 \times 10^{-1}$ \\
            $k_\beta$ & $6.90 \times 10^{-1}$ & $4.70 $ & $\alpha_\text{SEI}$ & $5.75 \times 10^{-2}$ \\
            $k_\gamma$ & $-6.46 \times 10^{3}$ & $-1.08 \times 10^{4}$ & $\beta_\text{SEI}$ & $1.21 \times 10^{2}$ \\
            $k_{\delta1}$ & $9.05 \times 10^{-6}$ & $1.47 \times 10^{4}$ & $T_\text{ref}$ & $298 \text{K}$ \\
            $k_{\delta2}$ & $1.40 $ & $-1.65 $ & $\sigma_\text{ref}$ & $5.00 \times 10^{-1}$ \\
            $k_{\delta3}$ & $0 $ & $3.61 \times 10^{2}$ & $k_\sigma$ & $1.04 $ \\
            & & & $k_T$ & $6.93 \times 10^{-2}$ \\
            & & & $k_t$ & $4.14 \times 10^{-10}$ \\
            \bottomrule % 绘制表格底部的横线
        \end{tabular}
    \end{threeparttable}    
\end{table}

\begin{table}[h]
\centering
\begin{threeparttable}
    \caption{Hyper-parameter search space for the prediction models.}
    \label{hyper}
    \begin{tabular}{@{}llllllllll@{}}
    \hline
    \multicolumn{1}{l}{Model}                   & Parameter                & Value               \\\cline{1-3}
    \multirow{3}{*}{ARIMA}                      & Moving average degree          & {[}1, 2, 3{]}         \\
                                            & Auto-regression degree             & {[}30, 40, 50{]}  \\     
                                                & Differentiation order          & {[}1, 2, 3{]}        \\\cline{1-3}
    \multirow{4}{*}{LSTM}                        & Hidden nodes             & {[}16, 24, 32{]}  \\
                                                & Layers                   & {[}2, 3, 4{]}      \\
                                                & Optimizer                & Adam                 \\
                                                & Batch size               & {[}16, 32, 64{]}      \\
                                                & Learning rate            & {[}0.001, 0.01, 0.1{]}         \\
                                                & Activation function      & \textit{Sigmoid,Relu,Tanh}        \\\cline{1-3}
    \multirow{5}{*}{RDedRVFL}                   & Enhancement nodes        & {[}100, 150, 200{]}        \\
                                                & Enhancement Layers        & {[}5, 10, 15{]}        \\
                                                & Regularization parameter & {[}0, 1{]}            \\
                                                & Input scalings           & {[}0, 1{]}            \\
                                                & Batch size               & {[}16, 32, 64{]}      \\
                                                & Learning rate            & {[}0.001, 0.01, 0.1{]}         \\
                                                & Activation function      & \textit{Sigmoid,Relu,Tanh}       \\\cline{1-3}
    \end{tabular}
\end{threeparttable}
\end{table}

In this subsection, we validate the effectiveness of the proposed method in actual energy management scenarios. The load and PV data utilized in this study are collected from a commercial building equipped with BEMS in a coastal province of China during 2023-2024. The data are sampled hourly, yielding 24 data points per day. The time-of-use electricity price data are released in advance by the  local power department. To assess the performance of the proposed method under different seasonal and energy consumption patterns, we deliberately utilized the months of June/July and November/December to construct the summer and winter datasets. Detailed descriptions of the datasets are provided in Table \ref{load_pv_data} and Fig. \ref{fig:seasonalprice}.

The D3QNPER model is trained on the summer and winter datasets respectively, with a training-test split of approximately 7:3. The test set is configured as multiples of 24 hours to align with the model's scheduling cycle.  Specifically, in each season, the first 1,176 hours are used to train the scheduling model, while the last 288 hours serve as the test set to validate the model's performance. Training settings and model parameters are summarized in Table \ref{setting} \cite{liu2025building} \cite{zhang2024load} \cite{cao2020deep} \cite{chen2023techno}. Five LFP batteries with 200kWh capacity compose stationary ESS. EVs in the underground parking lot with V2G capabilities are limited to 10 vehicles per day, each equipped with a 100kWh NMC battery. The SoC of each EV upon arrival follows a normal distribution, expressed as $\text{SoC} \sim N(0.35, 0.1^2)$, while adhering to range and boundary constraints. This study employs semi-empirical modeling parameters for lithium batteries to evaluate the degradation characteristics of lithium-based ESDs in the DRL energy scheduling framework. The parameters for the degradation modeling of LFP and NMC batteries are presented in Table \ref{lfp_nmc_parameters}, combining theoretical analysis with actual battery experimental data to ensure a reasonable level of accuracy \cite{xu2016modeling} \cite{najera2023semi}. The scheduling actions of ESS and EV are discretized into five levels [-50kW, -100kW, 0kW, 50kW, 100kW], where positive values indicate discharging and negative values indicate charging. The battery thermal management system keeps the CBS operating at a fixed temperature of 35°C \cite{xu2016modeling}. Control actions for the CBS are combinations of the aforementioned charging and discharging actions for the ESS and EVs. It is important to note that the joint real-time scheduling method proposed in this study does not depend on the aforementioned battery settings and the distribution of random variables as prerequisites. Researchers can expand upon the method we propose based on specific scenarios and requirements.
% Mainly influenced by raw material prices, the $C^{ESS}_{per}$ for ESS batteries based on LFP is greater than the $C^{EV}_{per}$ for EV batteries based on NMC, with $C^{ESS}_{per}$ and $C^{EV}_{per}$ being 910 $yuan$/kWh and 1092 $yuan$/kWh, respectively \cite{chen2023techno}.The arrival time, parking duration (time available for scheduling), and departure time of EVs are modeled as random variables following the probability density function, with specific settings detailed in Table 1. 

For building load and PV generation prediction, we employ the prediction method proposed in subsection 5.2, training the prediction models using data from the same months and configurations in 2023. We input historical data of 48 time steps to forecast the next 23 hours, with a sliding window size of 1. The training data is further split into a training set and a validation set in an 8:2 ratio. The hyper-parameters of the prediction models are optimized on the validation set using Bayesian optimization algorithm (BOA). The training and test datasets are completely separate with no overlap. To fairly evaluate the performance of each prediction model, the BOA is set to run for a total of 300 iterations on each model. Table \ref{hyper} provides the basic settings and optimization ranges for prediction models \cite{liu2025building} \cite{zhang2024load} \cite{gao2022random}. Four prediction evaluation metrics \cite{liu2025building} \cite{gao2022random}, namely root mean square error (RMSE), mean absolute error (MAE), mean absolute scaled error (MASE), and $R$-square ($R^2$), are employed to assess the performance of models, defined as follows:
\begin{equation}
\text{RMSE}=\sqrt{\frac1N\sum_{i=1}^N\left(\hat{x}_{i}-{x}_{i}\right)^2},
\end{equation}
\begin{equation}
        \text{MAE}=\frac1N\sum_{i=1}^N\left|{\hat{x}_{i}-{x}_{i}}\right|,
\end{equation}
\begin{equation}
    \text{MASE}=\frac{\frac1N\sum_{i=1}^N|\hat{x}_{i}-{x}_{i}|}{\frac{1}{L-1}\sum_{j=2}^{T}|{x}_{j}-{x}_{j-1}|},
\end{equation}
\begin{equation}
         R^2=1-\frac{\sum_{i=1}^N(\hat{x}_{i}-{x}_{i})^2}{\sum_{i=1}^N({x}_{i}-\bar{x})^2},
\end{equation}
where ${x}_{i}$ and $\hat{x}_{i}$ represent the actual and predicted values, while $\bar{x}$ denotes the mean of the actual data. $N$ and $L$ are the sample sizes of the training and test sets, respectively. Smaller values of RMSE, MAE, and MASE indicate higher prediction accuracy, while a larger $R^2$ value reflects a better fit of the regression function to the actual values. Before executing the prediction program, the input data are normalized as follows:
\begin{equation}
x_\text{nor}=\frac{x-x_\text{min}}{x_\text{max}-x_\text{min}},
\end{equation}
where $x_\text{nor}$, $x_\text{max}$ and $x_\text{min}$ are the normalized, maximum and minimum values in the training data, respectively. The prediction models are trained on the training set, followed by hyperparameter cross-validation on the validation set, with the test set used for evaluation. Finally, the trained model is saved for use in the real-time energy scheduling phase to construct state features.

All simulation computations for the experiments mentioned above are conducted on a 64-bit Windows operating system (CPU: Intel Core i7-12700 @ 3.60 GHz, Memory: 32GB) using Python 3.7.3. Upon completion of the training process for the proposed RDedRVFL-D3QNPER model, it generates scheduling actions within 0.12 seconds. This performance is significantly below the control frequency requirement of one hour, making it suitable for real-time energy scheduling in GB scenarios.

\subsection{Performance evaluation of proposed method}
\subsubsection{Prediction performance evaluation}
The RDedRVFL prediction model is trained using the summer data from June/July and the winter data from November/December in 2023. Two representative time series prediction models, auto-regressive integrated moving average (ARIMA) and long short-term memory (LSTM), alongside RDedRVFL, are included in this study. ARIMA is a traditional statistical model based on linear time series analysis \cite{zhang2024load}. It has clear mathematical interpretations and statistical properties, making it easy to analyze and understand. LSTM, widely used for time series forecasting \cite{zhang2024load} \cite{cao2020deep}, effectively captures dependencies between distant time steps and adapts well to complex time series features.

\begin{figure}[!ht]
    \centering
    \subfloat[]{
        \includegraphics[width=0.8\textwidth]{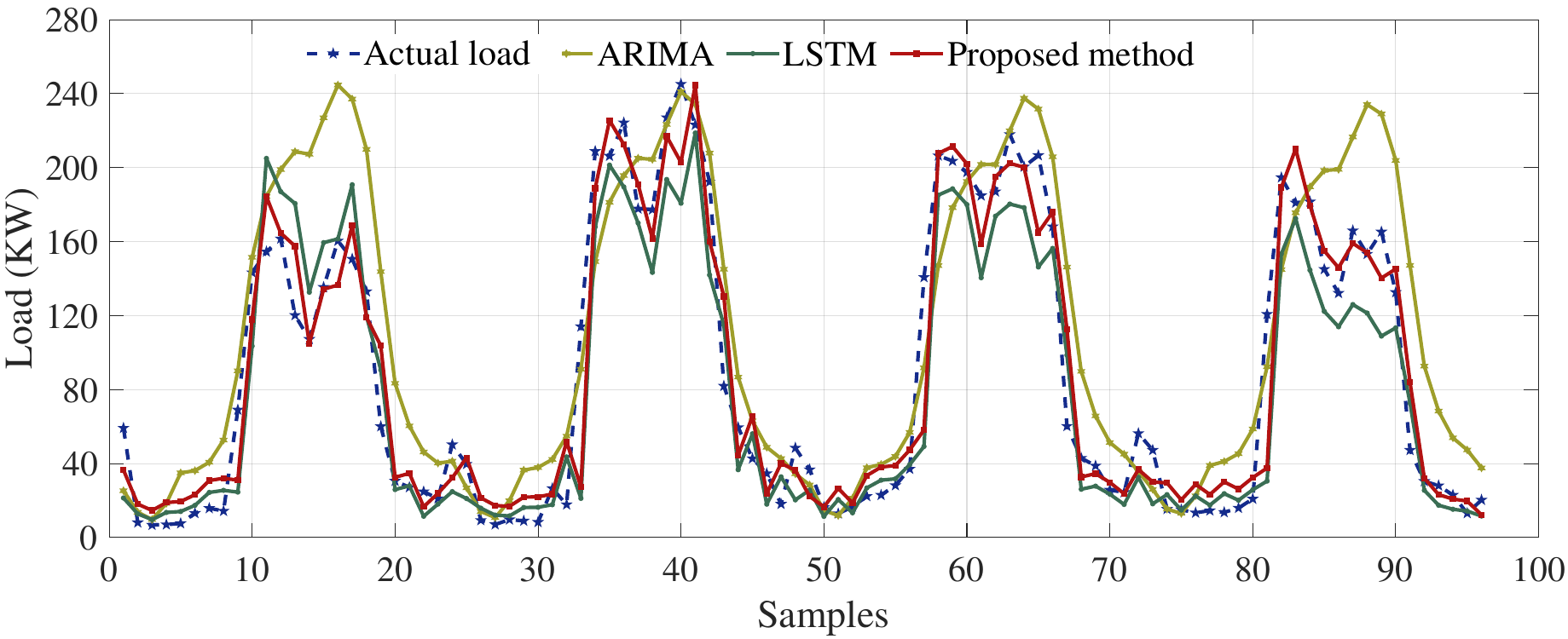}
        \label{summerload}
    }\\
    \subfloat[]{
        \includegraphics[width=0.8\textwidth]{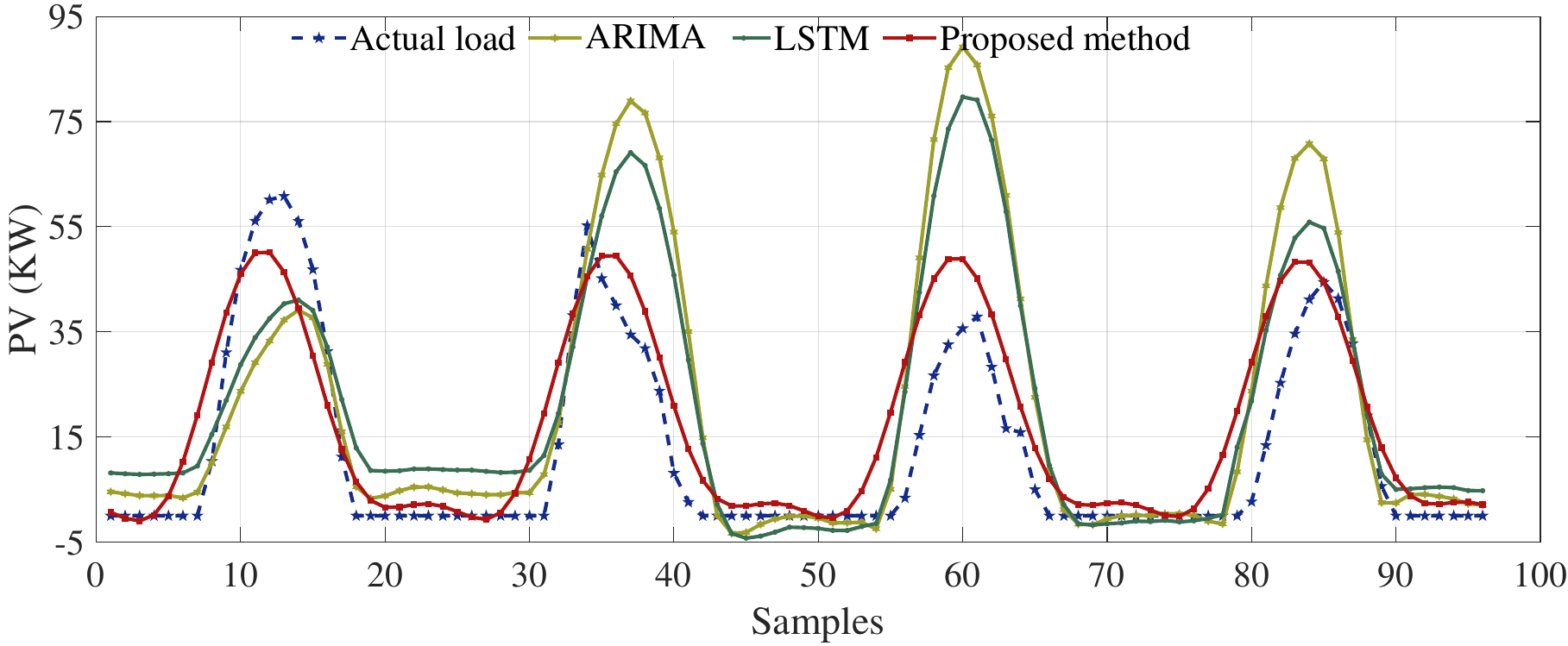}
        \label{summerPV}
    }
    \caption{The prediction results of building load and PV generation. (a) The building load prediction curve; (b) The PV generation prediction curve.}
    \label{fig:summer_load_PV}
\end{figure}

In addition to the performance metrics provided in subsection 6.1, we also record the computing time of different prediction models to evaluate the feasibility of embedding these models into real-time energy scheduling programs. Tables \ref{loadpre} and \ref{PVpre} present the experimental results of all models on different seasonal datasets after achieving optimal parameter configurations. It can be observed that RDedRVFL achieves the best predictive results across all metrics in different seasons. Specifically, RDedRVFL has the smallest RMSE, MAE, and MASE, the highest $R^2$, and the shortest computing time. Taking the seasonally averaged prediction performance in the last five rows of Table \ref{loadpre} as an example, compared to RDedRVFL, LSTM shows increases of 22.81\%, 35.82\%, and 26.12\% in RMSE, MAE, and MASE, respectively, while its $R^2$ decreased by 4.19\%. Notably, LSTM's computing time is 325 seconds longer than that of RDedRVFL, representing an increase of 309.52\%. For PV prediction, RDedRVFL also surpasses LSTM across all four error metrics, with LSTM exceeding RDedRVFL's computing time by 324.06\%. ARIMA shows significant disadvantages in both prediction accuracy and computing time compared to LSTM and RDedRVFL. Fig. \ref{fig:summer_load_PV} illustrates the prediction curves of all models using summer data as examples. It is evident that building loads exhibit significant irregular variations, particularly due to fluctuations occurring at midnight.  Additionally, PV output is greatly affected by weather, resulting in nonlinear fluctuations. Despite these complexities, the RDedRVFL accurately predicts the periodicity and trends of both load and PV generation.

The results indicate that data-driven predictive models based on machine learning generally outperform traditional statistical models in complex time series forecasting. Among the deep learning models, RDedRVFL demonstrates superior overall prediction performance due to its deep ensemble network structure. LSTM's relatively lower accuracy may stem from suboptimal hyperparameter tuning. However, in the context of real-time energy scheduling discussed in this paper, the time-consuming process of BOA optimization for hyperparameter tuning cannot be easily complicated further. In contrast, RDedRVFL, leveraging randomization techniques and non-backpropagation computation, achieves high computational efficiency, making it well-suited for real-time energy scheduling frameworks. The computing times of the comparative models all satisfy the one-hour scheduling frequency. However, considering the complexity of real-world applications and the scalability of scheduling frameworks, RDedRVFL clearly stands out as the optimal option.

\begin{table}[htbp]
    \centering
    \begin{threeparttable}
        \caption{Comparison of load prediction performance among different models.}
        \label{loadpre}
        \begin{tabular}{lcccc}
        \toprule
        \multirow{2}{*}{Season} & \multirow{2}{*}{Metrics} & \multicolumn{3}{c}{Prediction model} \\
        \cmidrule(lr){3-5}
        & & ARIMA & LSTM & Proposed \\
        \midrule
        \multirow{5}{*}{Summer} 
        & RMSE & 38.7920 & 30.0197 & 23.5775 \\
        & MAE & 29.1528 & 24.1260 & 16.5004 \\
        & MASE & 1.3424 & 1.1109 & 0.7499 \\
        & $R^2$ & 0.7829 & 0.8700 & 0.9198 \\
        & Computation time (s) & 552 & 440 & 109 \\
        \midrule
        \multirow{5}{*}{Winter} 
        & RMSE & 24.3548 & 14.2939 & 12.5058 \\
        & MAE & 18.9375 & 10.3643 & 8.8927 \\
        & MASE & 1.2305 & 0.6735 & 0.6649 \\
        & $R^2$ & 0.6652 & 0.8847 & 0.9117 \\
        & Computation time (s) & 507 & 420 & 101 \\
        \midrule
        \multirow{5}{*}{Mean} 
        & RMSE & 31.5734 & 22.1568 & 18.0417 \\
        & MAE & 24.0452 & 17.2452 & 12.6966 \\
        & MASE & 1.2865 & 0.8922 & 0.7074 \\
        & $R^2$ & 0.7241 & 0.8774 & 0.9158 \\
        & Computation time (s) & 529.5 & 430 & 105 \\
        \bottomrule
        \end{tabular}
    \end{threeparttable}
\end{table}

\begin{table}[htbp]
    \centering
    \begin{threeparttable}
        \caption{Comparison of PV prediction performance among different models.}
        \label{PVpre}
        \begin{tabular}{lcccc}
        \toprule
        \multirow{2}{*}{Season} & \multirow{2}{*}{Metrics} & \multicolumn{3}{c}{Prediction model} \\
        \cmidrule(lr){3-5}
        & & ARIMA & LSTM & Proposed \\
        \midrule
        \multirow{5}{*}{Summer} 
        & RMSE & 16.2442 & 10.1320 & 7.6184 \\
        & MAE & 9.6313 & 7.6908 & 5.3266 \\
        & MASE & 1.5834 & 1.2643 & 0.8757 \\
        & $R^2$ & 0.6839 & 0.8770 & 0.9305 \\
        & Computation time (s) & 495 & 415 & 97 \\
        \midrule
        \multirow{5}{*}{Winter} 
        & RMSE & 9.3354 & 6.7956 & 5.4972 \\
        & MAE & 7.1536 & 3.9375 & 3.2725 \\
        & MASE & 1.6288 & 0.8966 & 0.7451 \\
        & $R^2$ & 0.7755 & 0.8810 & 0.9222 \\
        & Computation time (s) & 471 & 378 & 90 \\
        \midrule
        \multirow{5}{*}{Mean} 
        & RMSE & 12.7898 & 8.4638 & 6.5578 \\
        & MAE & 8.3925 & 5.8142 & 4.2996 \\
        & MASE & 1.6061 & 1.0805 & 0.8104 \\
        & $R^2$ & 0.7297 & 0.8790 & 0.9264 \\
        & Computation time (s) & 483 & 396.5 & 93.5 \\
        \bottomrule
        \end{tabular}
    \end{threeparttable}
\end{table}

\subsubsection{Analysis of the convergence process}
\begin{figure*}[hbt]
    \centering
    \subfloat[]{\includegraphics[width=0.45\textwidth]{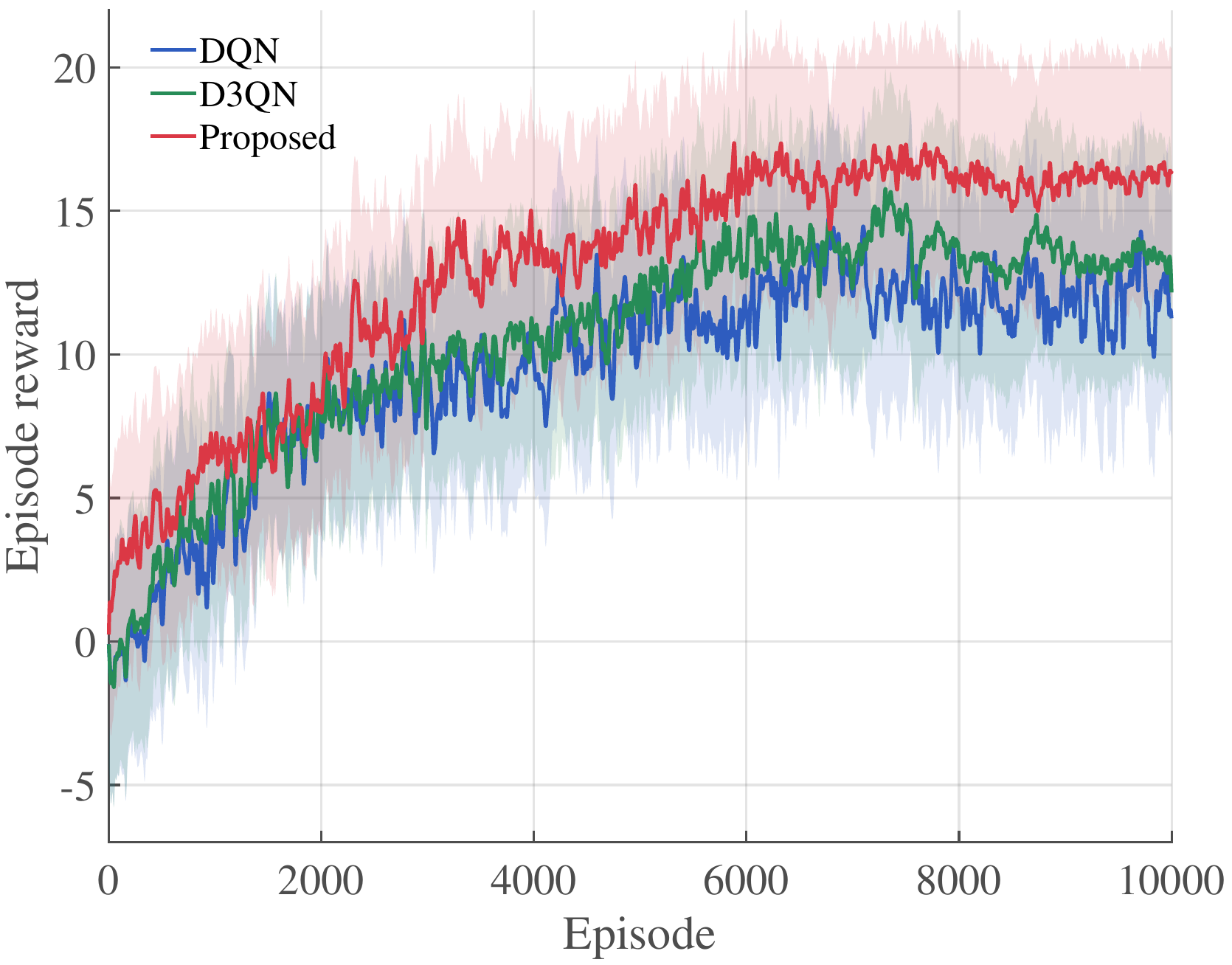}%
        \label{con_summer}}
    \hfil
    \subfloat[]{\includegraphics[width=0.45\textwidth]{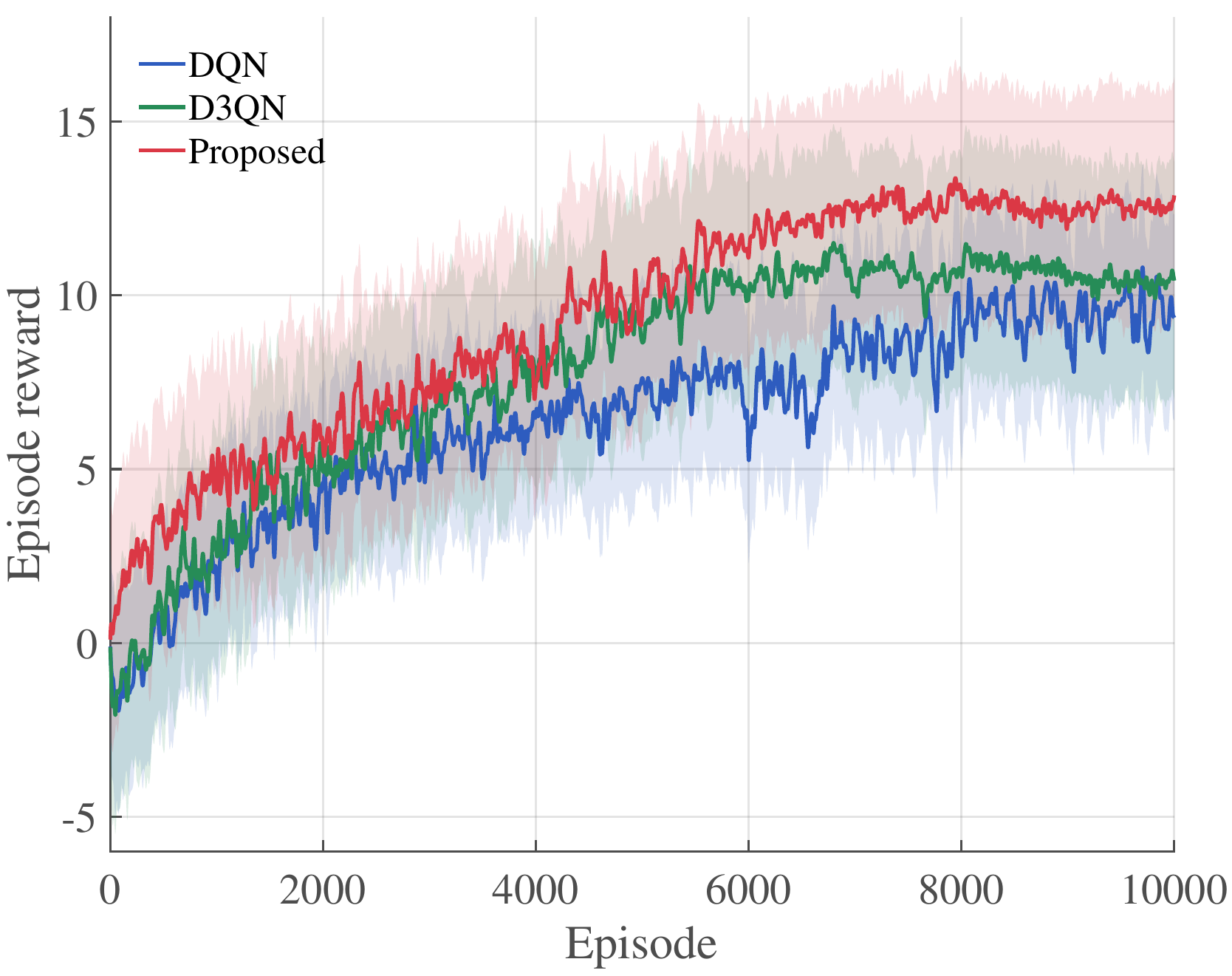}%
        \label{con_winter}}
    \caption{The episode reward during the training process for DRL-based models in different seasons. (a) Summer; (b) Winter.}
    \label{fig:convergence}
\end{figure*}

We evaluate the performance of the proposed joint real-time energy scheduling method is evaluated using the summer data from June/July and the winter data from November/December in 2024. The proposed D3QNPER model and two DRL-based comparative models, DQN and D3QN, are investigated. All scheduling methods are integrated with the RDedRVFL prediction module. DQN and D3QN serve as baseline algorithms to validate the effectiveness of the D3QNPER improvement. To ensure a fair comparison, the basic parameter settings for DQN and D3QN are aligned with those of D3QNPER. 

Fig. \ref{fig:convergence} illustrates the evolution of episode cumulative rewards for the three algorithms during the training process. The proposed algorithm achieves optimal convergence in both summer and winter scenarios. In summer, rewards gradually increase during the first 4,000 episodes, indicating that the agent is continuously exploring and optimizing its scheduling strategy. Between 4,000 and 6,000 episodes, rewards begin to converge, suggesting a reduction in exploratory behavior as the agent accumulates experience through trial and error, stabilizing its learned optimal scheduling strategies. In the final 6000-10000 episodes, the reward evolution trend stabilizes. This indicates that the agent has found an optimized scheduling strategy to maximize cumulative rewards by consistently issuing charging and discharging commands to the CBS. It is worth noting that to prevent the algorithm from missing out on exploring potentially optimal strategies, the proposed algorithm maintains a small exploration probability of $\epsilon =0.05$ during the convergence phase. As a result, the convergence curve fluctuates within a certain range. Among all scheduling algorithms, DQN performs the worst, with rewards fluctuating widely in the later stages of training and failing to identify a higher-reward control strategy. D3QN, benefiting from improved Q-value estimation through double networks and dueling architectures, shows better convergence stability and accuracy compared to DQN. However, due to potential issues like insufficient sample learning efficiency, the rewards for D3QN remain lower than those of the proposed method. In winter, all scheduling algorithms exhibit similar yet season-specific convergence trajectories. They adapt to the dynamic changes in the winter system environment during the first 6,000 episodes, exploring scheduling solutions that align with winter electricity consumption patterns. In the final 2000 episodes, all algorithms tend to converge. The proposed algorithm demonstrates the best convergence effect, followed by D3QN, with DQN in last place.

\subsubsection{Effectiveness analysis of the proposed method}
\begin{figure}[!ht]
    \centering
    \subfloat[]{
        \includegraphics[width=14.1cm]{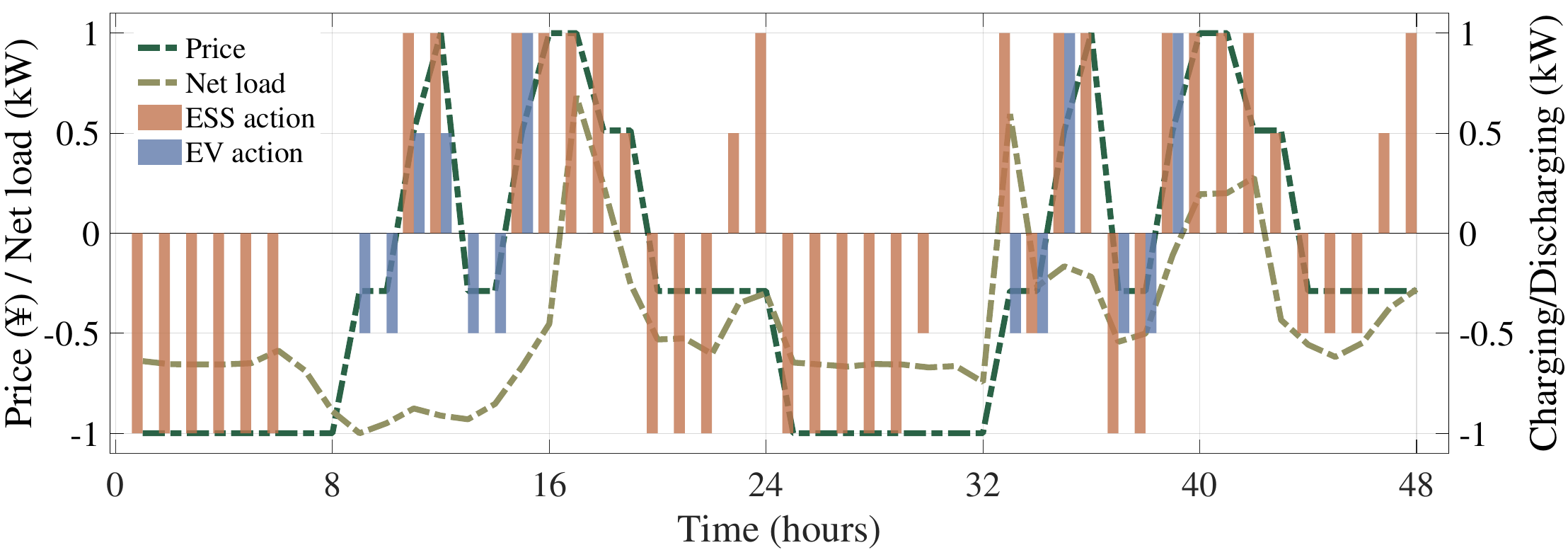}
        \label{action}
    }\\
        \subfloat[]{
        \includegraphics[width=14cm]{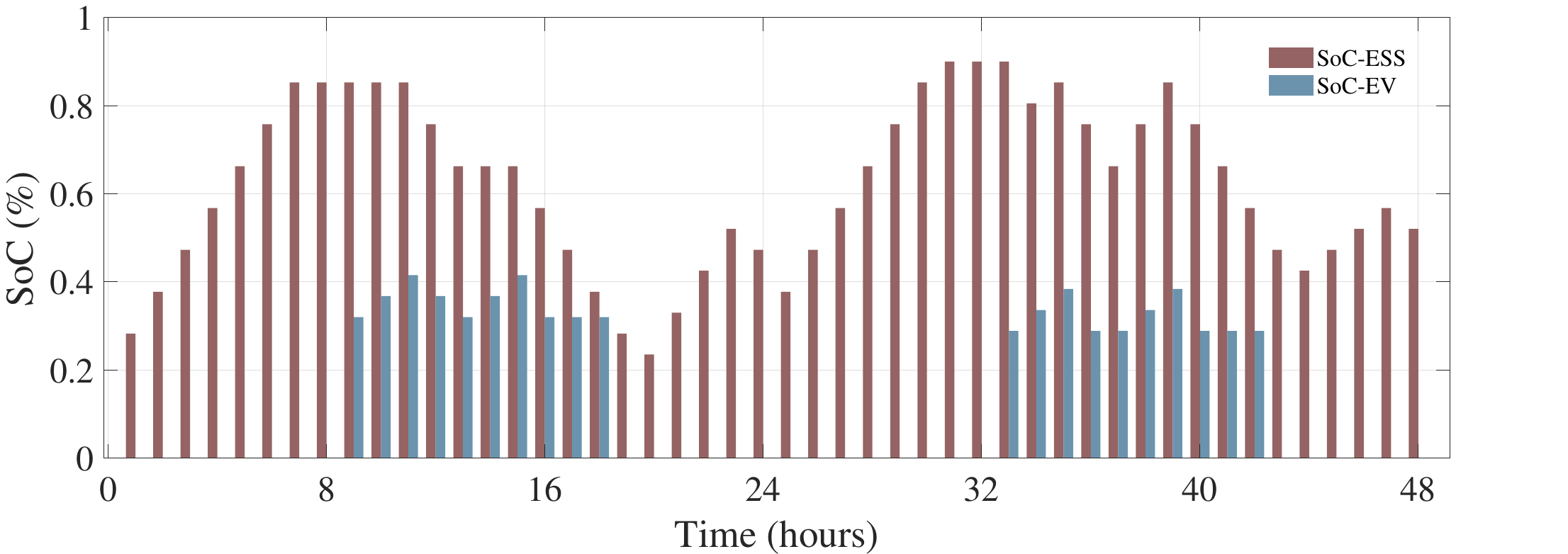}
        \label{soc_summer}
    }\\
    \caption{The charging/discharging control results of the proposed method in summer. (a) The blue bar and orange bar represent the charging(-)/discharging(+) control actions of EV and ESS, the green line and yellow line denote the electricity price and net load. (b) The blue bar and red bar represent the SoC of EV and ESS.}
    \label{fig:action_soc_summer}
\end{figure}

\begin{figure}[!ht]
    \centering
    \subfloat[]{
        \includegraphics[width=14cm]{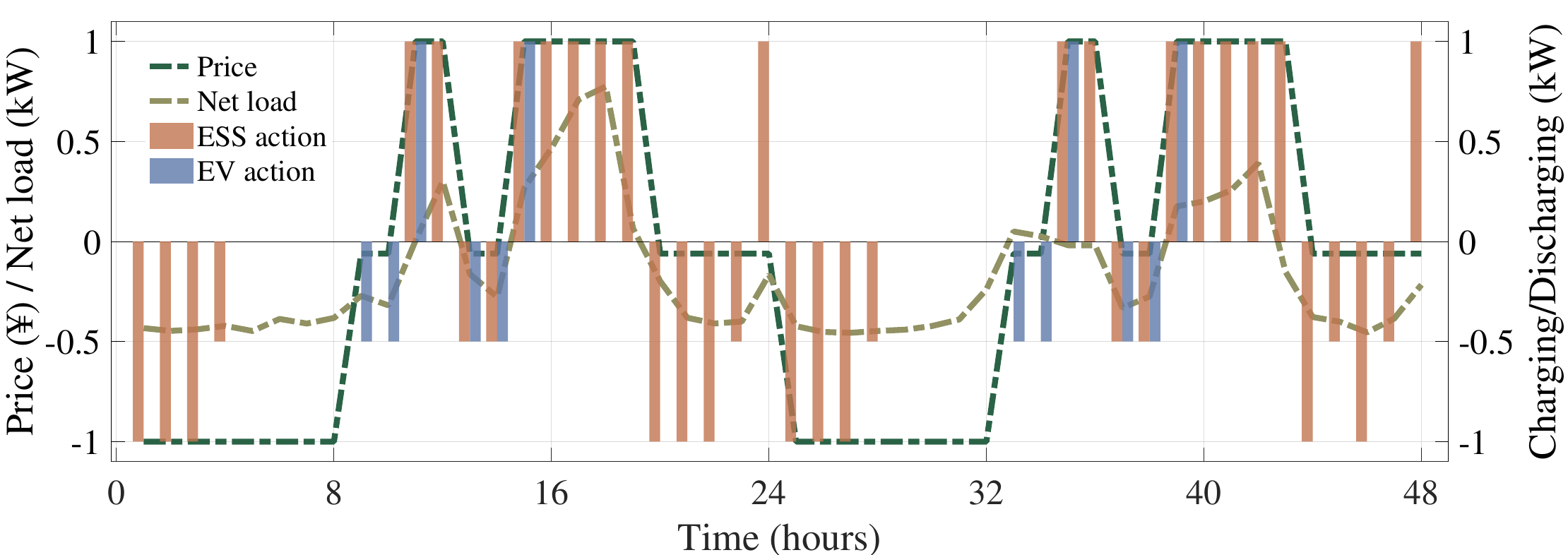}
        \label{action_winter}
    }\\
    \subfloat[]{
        \includegraphics[width=14cm]{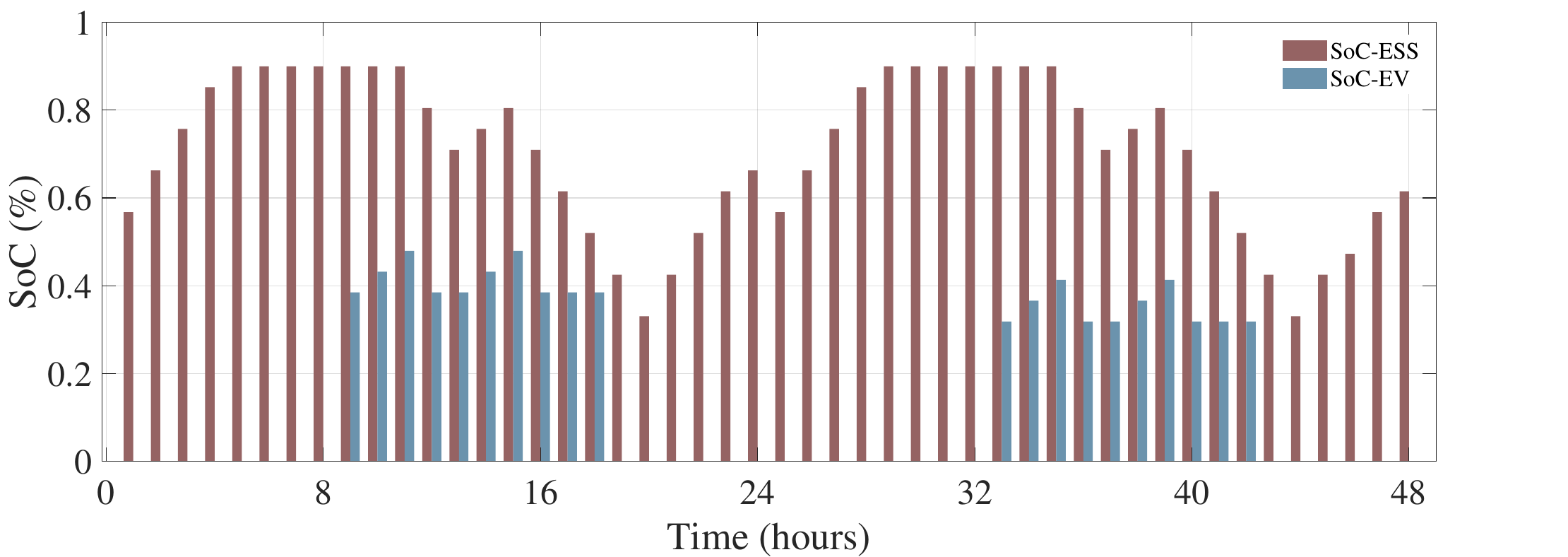}
        \label{soc_winter}
    }\\
    \caption{The charging/discharging control results of the proposed method in winter. (a) The blue bar and orange bar represent the charging(-)/discharging(+) control actions of EV and ESS, the green line and yellow line denote the electricity price and net load. (b) The blue bar and red bar represent the SoC of EV and ESS.}
    % , green line: electricity price, yellow line: net load, blue bar: EV, orange bar: ESS; (b) Remaining battery energy, blue bar: EV, red bar: ESS.
    \label{fig:action_soc_winter}
\end{figure}

After training, the proposed scheduling model is saved to carry out the joint real-time energy scheduling process. In Figs. \ref{fig:action_soc_summer} and \ref{fig:action_soc_winter}, we present the control results of consecutive full episodes under different seasons and system operation modes. The electricity price and net load are represented by green and yellow lines, respectively, while blue and orange bars depict the control actions of EVs and ESS along with their corresponding SoC changes. All data is scaled to the range of -1 to 1 for clarity on the same graph. 

From the figures, it is evident that the proposed method has learned optimal CBS control strategies in dynamic environments across different seasons. Firstly, the method responds promptly to price signals from the energy market, charging during valley tariff and discharging during peak tariff. Secondly, the method considers energy utilization priorities. In commercial buildings, certain distributed energy-consuming devices, such as commercial database servers, shift some non-real-time, delay-tolerant tasks to be processed during midnight to take advantage of surplus electricity. Consequently, fluctuations in load unrelated to occupant energy usage occur during midnight, as shown in Fig. \ref{summerload}. Based on net load forecasting from system state features, the proposed method adjusts charging and discharging strategies according to net load signals. It releases energy during midnight when electricity prices are low and net load is rising, saving on grid purchasing costs and clear the remaining energy of the ESS. Subsequently, during the valley tariff period in the early morning, the agent charges at maximum power to store energy for arbitrage during future peak tariff periods. Last but not least, the proposed method learns a collaborative operation strategy for ESS and EV within the CBS. By leveraging the embedding of $\xi$ in the system state, the agent identifies the EV scheduling cycle. From Figs. \ref{fig:action_soc_summer} and \ref{fig:action_soc_winter}, during collaborative periods, the agent prioritizes scheduling the ESS, allowing the EV to release energy only when prices or net loads are high to maximize profits. This strategy minimizes rapid degradation of EV batteries, which are sensitive to cycling aging, thus maintaining user enthusiasm and reducing scheduling costs. Moreover, the proposed method places a significant emphasis on EV user range anxiety. Since EVs are primarily used for transportation, the agent ensures that the SoC upon departure is not less than upon arrival, guaranteeing the rationality and enthusiasm of EV users to participate in scheduling. Considering the significant differences in peak and valley tariffs, the proposed scheduling method has the potential to reduce the operational costs of BEMS in different system operating environments.

\subsubsection{Performance comparison of different scheduling methods}

\begin{figure*}[hbt]
    \centering
    \subfloat[]{\includegraphics[width=0.45\textwidth]{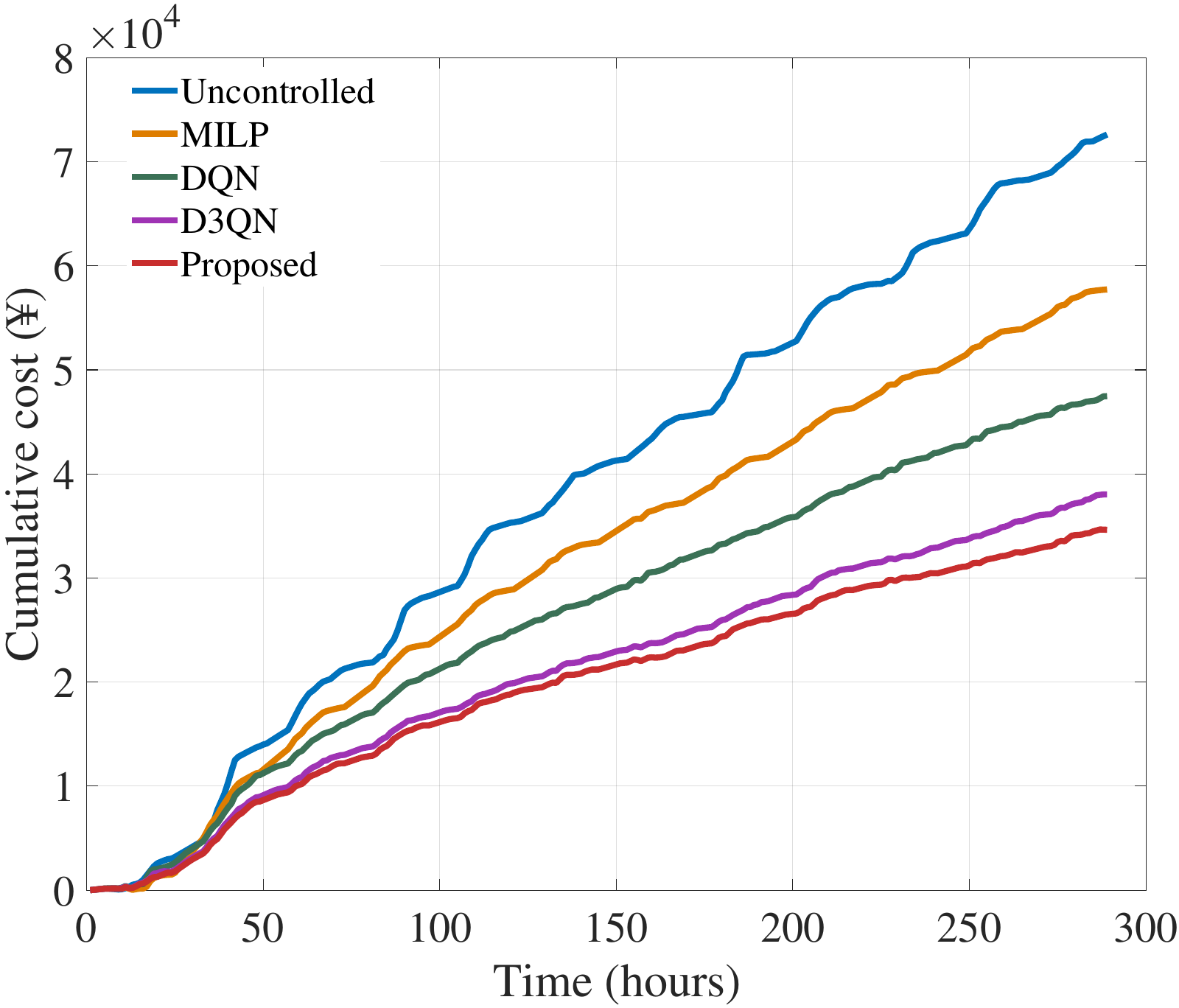}%
        \label{comparison}}
    \hfil
    \subfloat[]{\includegraphics[width=0.45\textwidth]{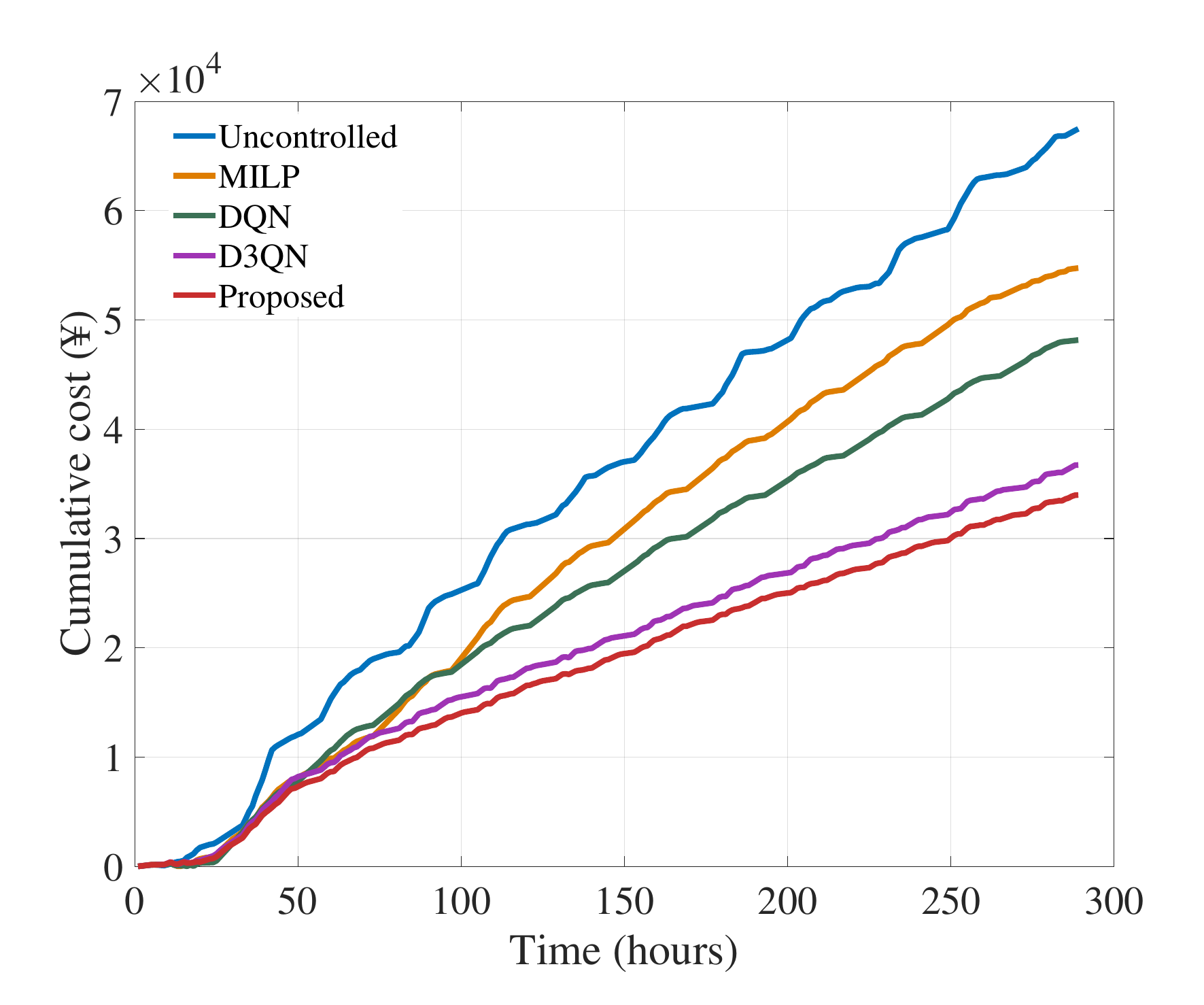}%
        \label{comparison_winter}}
    \caption{Comparison results of cumulative profits for all methods in different seasons. (a) Summer; (b) Winter.}
    \label{fig:comparison}
\end{figure*}

The comparative methods consist of five approaches: DQN, D3QN,  mixed-integer linear programming (MILP), uncontrolled, and the proposed method. MILP is a non-DRL-based scheduling method, which is implemented using CPLEX in Python. The uncontrolled method does not manage the CBS, and the ESS and EV are charged and discharged at maximum power within SoC constraints and scheduling periods \cite{wan2018model}. Figs. \ref{fig:comparison} presents the operational performance of the scheduling methods on the test dataset. It is observed that, regardless of the season, the proposed scheduling method incurs the lowest cumulative operational costs. The operational cost of the scheduling strategy formulated by D3QN ranks second after the proposed method, with a cost difference of no more than 1000 yuan. The operational costs generated by DQN are higher than those of D3QN, exceeding it by 9448.32 yuan in summer and 11429.63 yuan in winter. MILP performs worse than all DRL-based methods, with cost increases of 21.60\% in summer and 13.69\% in winter compared to DQN. Serving as the baseline, the uncontrolled method incurs the highest operational costs, amounting to 72622.65 yuan in summer and 67488.79 yuan in winter, representing increases of 109.68\% and 98.61\%, respectively, compared to the proposed method. These experimental results demonstrate that DRL-based frameworks hold significant promise for real-time energy scheduling. Their model-free advantages and learning-based approaches enable the optimization of control strategies tailored to specific system environments, offering more effective energy management solutions than traditional methods like MILP.

\subsection{Analysis of prediction module and energy allocation}
\subsubsection{Ablation study of prediction module}

\begin{figure*}[hbt]
    \centering
    \subfloat[]{\includegraphics[width=0.45\textwidth]{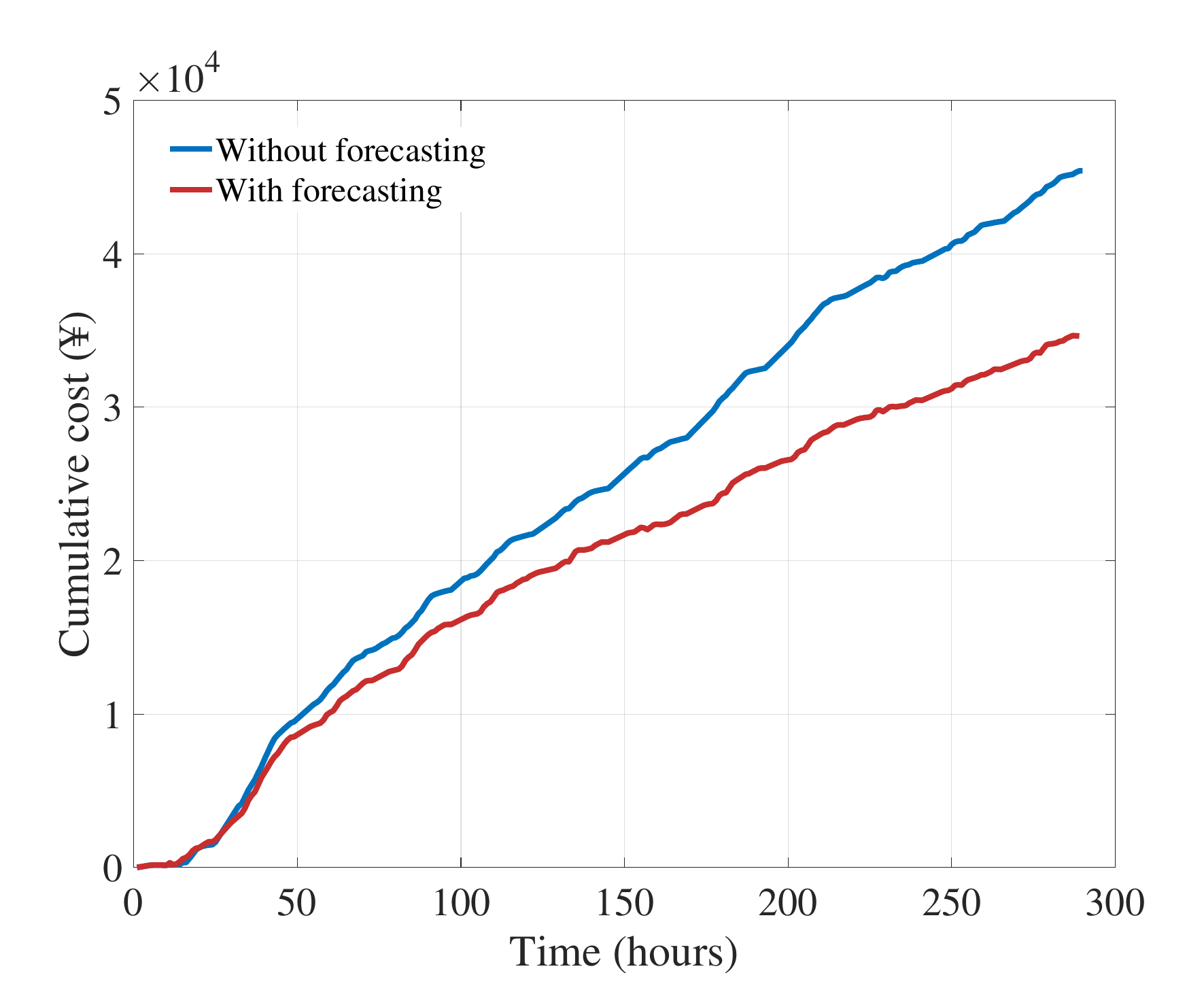}%
        \label{pre_com}}
    \hfil
    \subfloat[]{\includegraphics[width=0.45\textwidth]{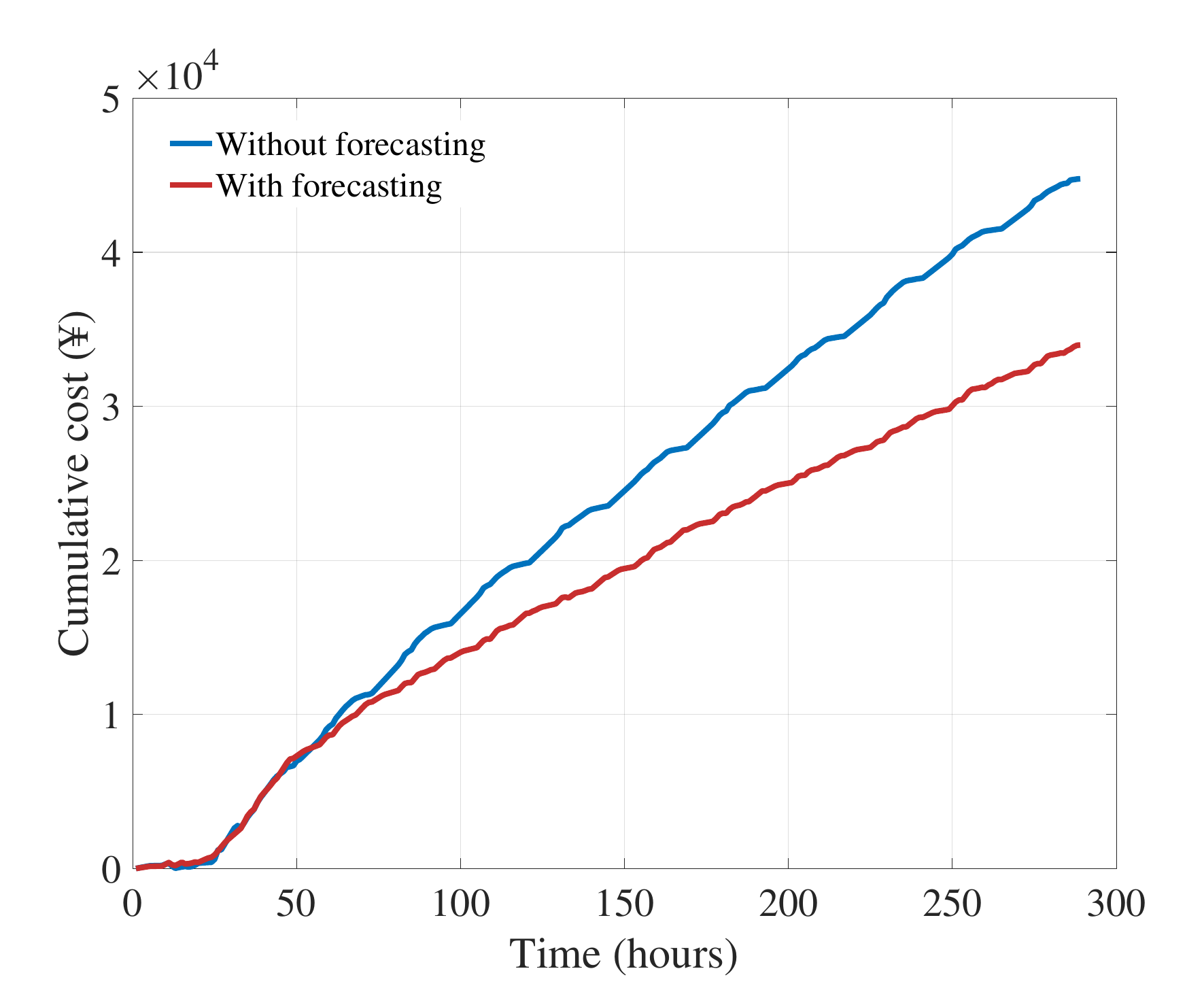}%
        \label{pre_com_winter}}
    \caption{Comparison of operating costs without prediction mechanism in different seasons. (a) Summer; (b) Winter.}
    \label{fig:pre_comparison}
\end{figure*}

\begin{figure}[!ht]
    \centering
    \subfloat[]{
        \includegraphics[width=14cm]{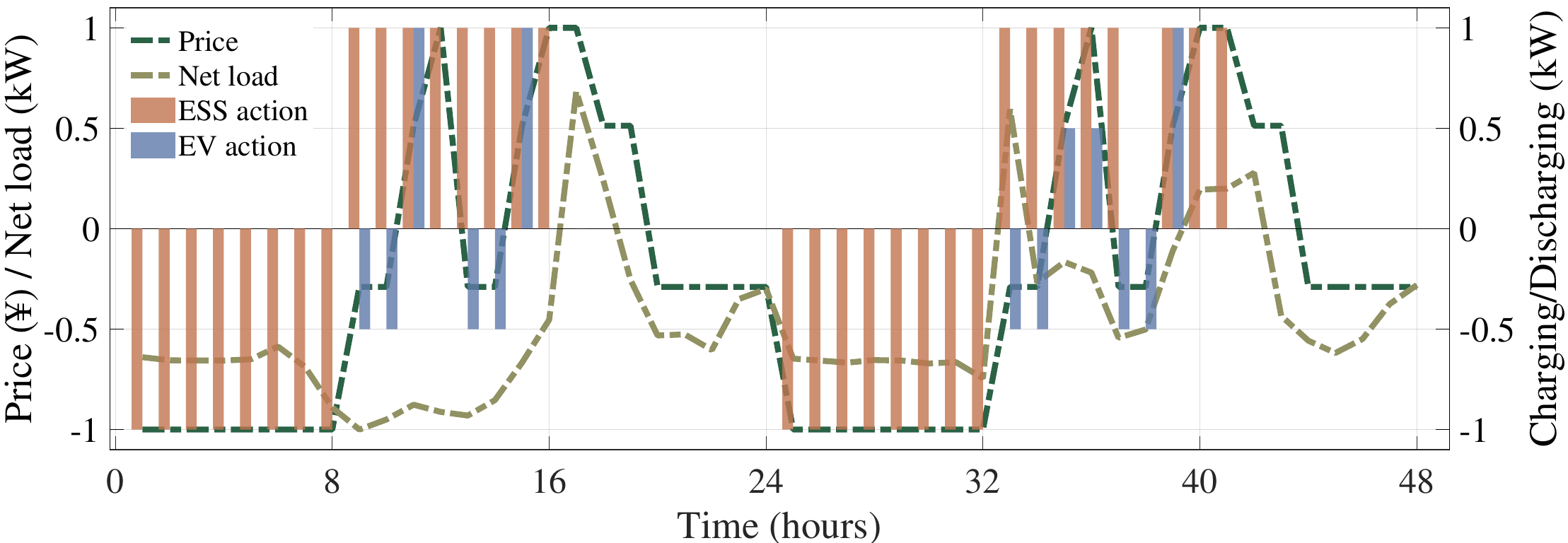}
        \label{base_action}
    }\\
    \subfloat[]{
        \includegraphics[width=14cm]{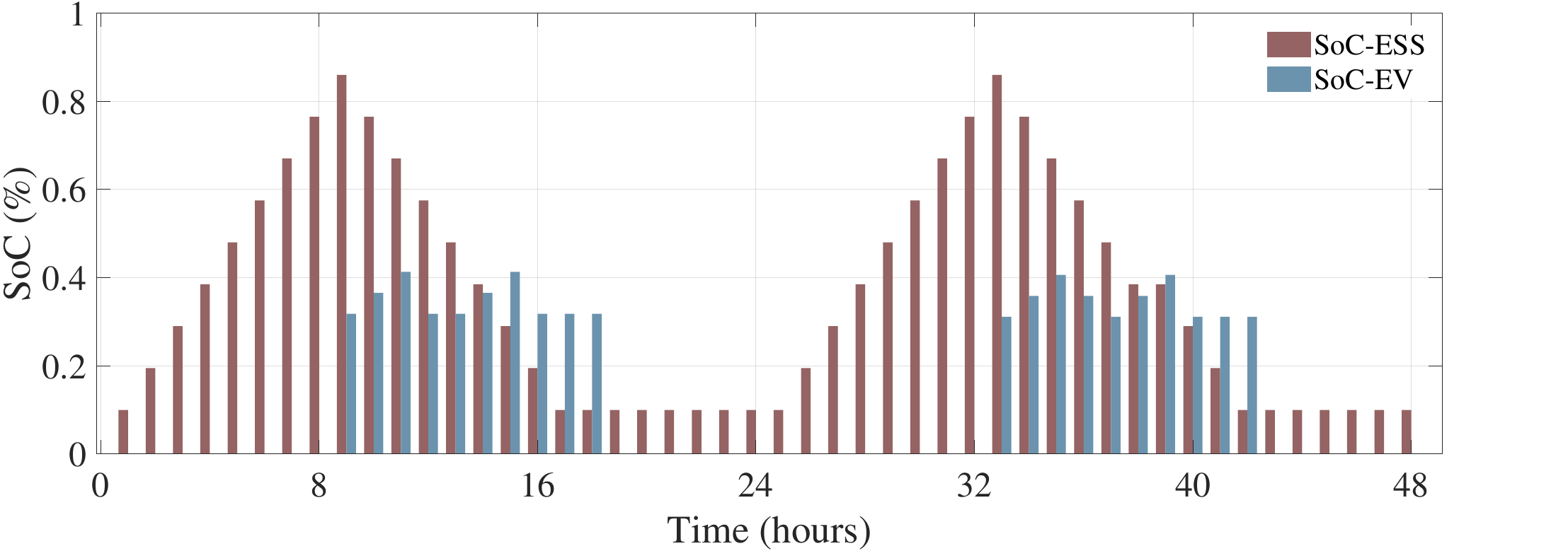}
        \label{base_soc}
    }\\
    \caption{The charging/discharging control results of the proposed method without prediction in summer. (a) The blue bar and orange bar represent the charging(-)/discharging(+) control actions of EV and ESS, the green line and yellow line denote the electricity price and net load. (b) The blue bar and red bar represent the SoC of EV and ESS.}
    \label{fig:base_action_soc}
\end{figure}

We investigated the operation of BEMS in the absence of predictive information, as illustrated in Figs. \ref{fig:pre_comparison} and \ref{fig:base_action_soc}. As shown in Fig. \ref{fig:pre_comparison}, the absence of the prediction module leads to a significant increase in system operating costs, rising by 31.08\% in summer and 31.75\% in winter. The control results from two episodes without net load prediction information, as depicted in Fig. \ref{fig:base_action_soc}, partially explain the decline in scheduling performance. Due to space constraints, only the scheduling results of summer are displayed. From Fig. \ref{fig:base_action_soc}, it is evident that even without net load variations, the agent can still issue appropriate actions, such as charging during valley tariffs and discharging during peak tariffs based on price signals. However, the BEMS responds insensitively to changes in net load. When both the net load and electricity price are high, it is more optimal for the CBS to supply power to the building to avoid purchasing energy from the grid at higher prices. Yet, the agent tends to start discharging at a high power level as soon as prices begin to rise at 09:00. During the second peak period of energy prices and net load, roughly between 16:00 and 19:00, the energy in the CBS is depleted, preventing it from conducting arbitrage or supplying the building load during this profitable period. In summary, the agent struggles to make decisions that align with long-term returns due to the lack of critical predictive information. It fails to accurately learn the price difference between $c^{b}$ and $c^{s}$, as well as the underlying cost optimization potential, leading to increased operational costs for the system. 

\subsubsection{Analysis of energy allocation}
%给出2天内的CBS放电分配的柱状图 注意修正负荷（太多了超页数了 没必要）%分析能量分配的有效性  因为电价不同  看情况图/表  （给出0.9 0.8 0.7 0.6 多种电价情况）

\begin{figure*}[hbt]
    \centering
    \subfloat[]{\includegraphics[width=0.45\textwidth]{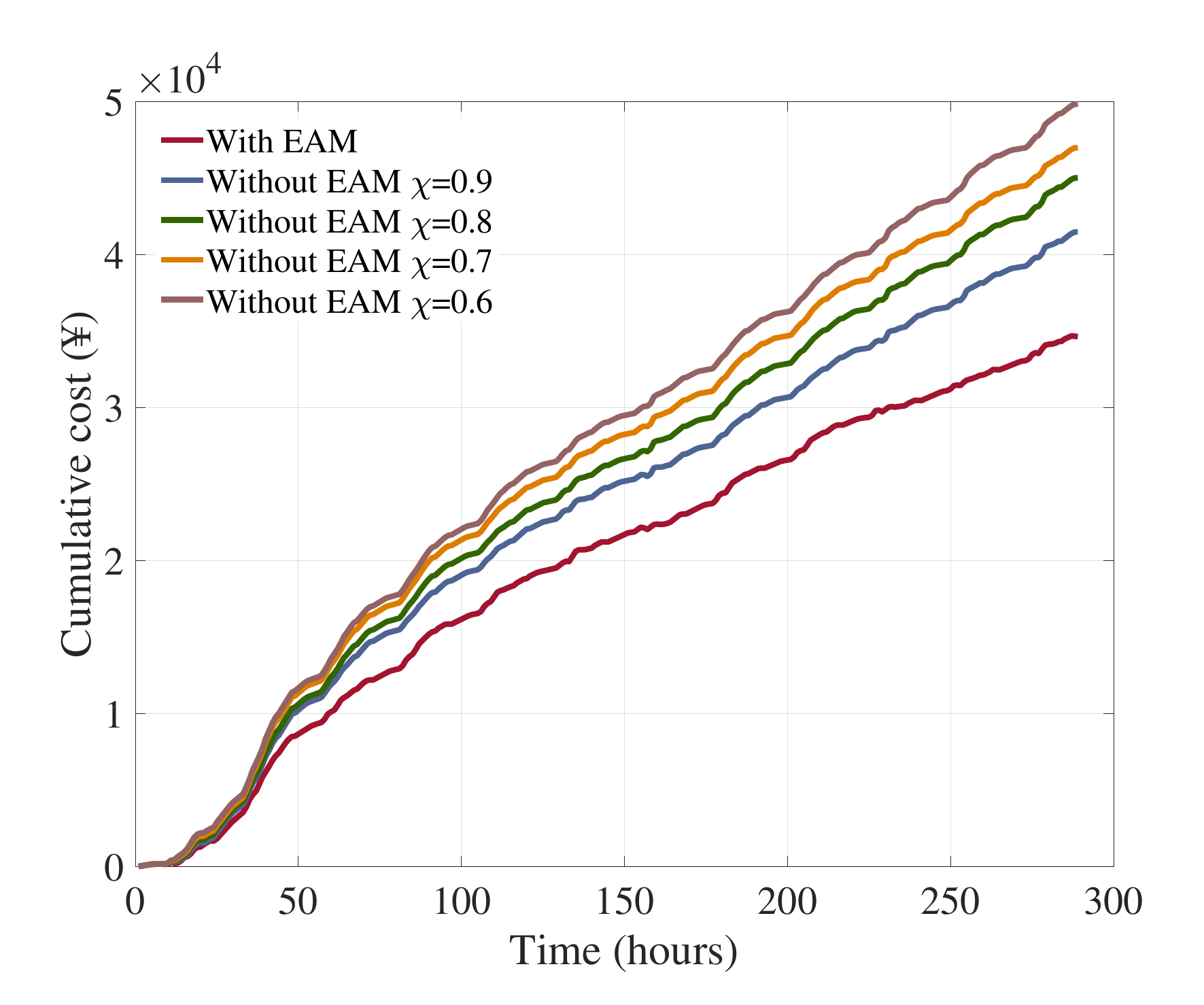}%
        \label{EAM_summer}}
    \hfil
    \subfloat[]{\includegraphics[width=0.45\textwidth]{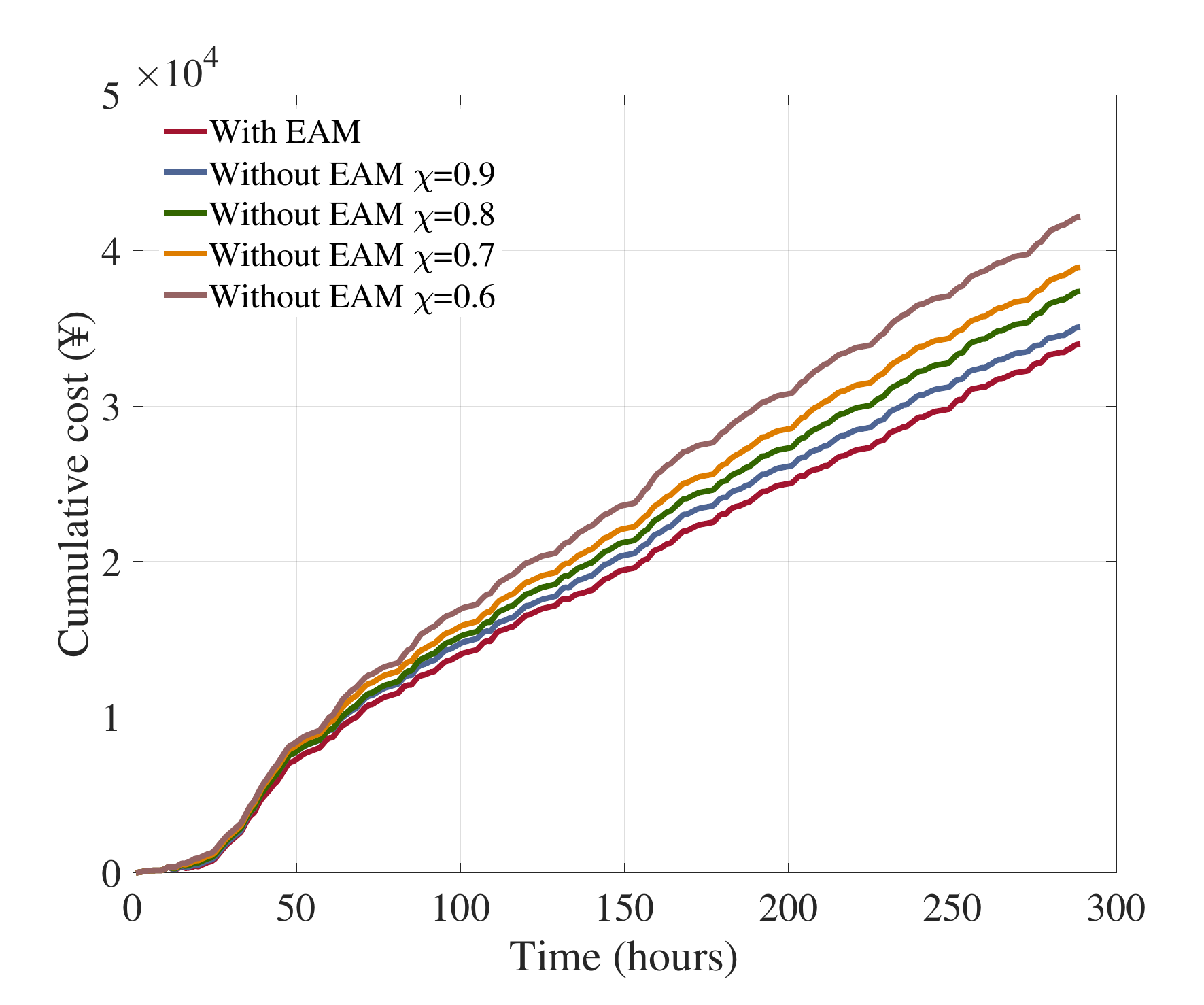}%
        \label{EAM_winter}}
    \caption{Comparison of operating costs without EAM in different seasons. (a) Summer; (b) Winter.}
    \label{fig:EAM}
\end{figure*}

We analyze the impact of the proposed EAM on system operations, as shown in Fig. \ref{fig:EAM}. It is evident that the system operates more profitably with EAM compared to without it. As $\beta_\text{pri}$ gradually decreases, the system costs exhibit an upward trend. Specifically, when $\beta_\text{pri}=0.9$, the operational costs without EAM increase by 19.70\% and 3.22\% in summer and winter, respectively, compared to those with EAM. As $\beta_\text{pri}$ decreases from 0.8 to 0.6, the operational costs without EAM increase by 29.94\%, 35.63\%, and 43.89\% in summer, and 9.99\%, 14.57\%, and 24.11\% in winter, respectively.

These experimental results are foreseeable. Since the cost of purchasing energy from the PG, $c^{b}$, is higher than the price of selling energy back to the PG, $c^{s}$, prioritizing the supply of energy to the buildings before selling power to the PG is the optimal decision for the CBS. Specifically, the system employing EAM can avoid purchasing energy at a higher price $c^{b}$ from the grid to meet building load deficits.  If there is surplus energy, arbitrage can occur at price $c^{s}$, generating additional revenue. For example, in summer, the difference between peak and valley tariffs can reach 1.34 yuan. Even with an energy purchasing discount at $\beta_\text{pri}=0.9$, the system still has significant arbitrage potential. Regarding the differences in operating costs across seasons, summer features abundant sunlight and high PV generation compared to winter. Without EAM, a significant amount of PV power is sold back to PG at price $c^{s}$ instead of supplying GB. The system then has to purchase energy from PG at the higher price $c^{b}$, which considerably increases operating costs. $\beta_\text{pri}$ is directly proportional to the system's arbitrage revenue, reflecting the adjustment of the market price mechanism to the supply and demand dynamics of electricity. Importantly, EAM promotes the local consumption of energy in GB microgrid systems, reducing power transmission losses, and facilitating real-time matching of supply and demand, which contributes to the stable operation of smart grids.

\subsection{Operational analysis of CBS}
\subsubsection{Arbitrage analysis of CBS}
%分析刨除负荷后的cbs盈利能力 五个方法对比 表格 证明算法好且CBS的对EMS至关重要。
\begin{table}[htbp]
    \centering
    \begin{threeparttable}
        \caption{ Comparison of additional costs without CBS under different operating durations}
        \label{CBScost}
        \begin{tabular}{lccc}
        \toprule
        \multirow{2}{*}{Season} & \multicolumn{3}{c}{Operating cost (¥)} \\
        \cmidrule(lr){2-4} 
        & 288 (hours) & 576 (hours) & 1152 (hours) \\
        \midrule
        Summer & 8885.59 & 12869.34& 32757.02 \\
        Winter & 5518.03 & 11282.26 & 31506.31 \\
        Mean & 7201.81 & 12075.80 & 32131.67 \\
        \bottomrule
        \end{tabular}
    \end{threeparttable}
\end{table}

\begin{figure*}[hbt]
    \centering
    \subfloat[]{\includegraphics[width=0.45\textwidth]{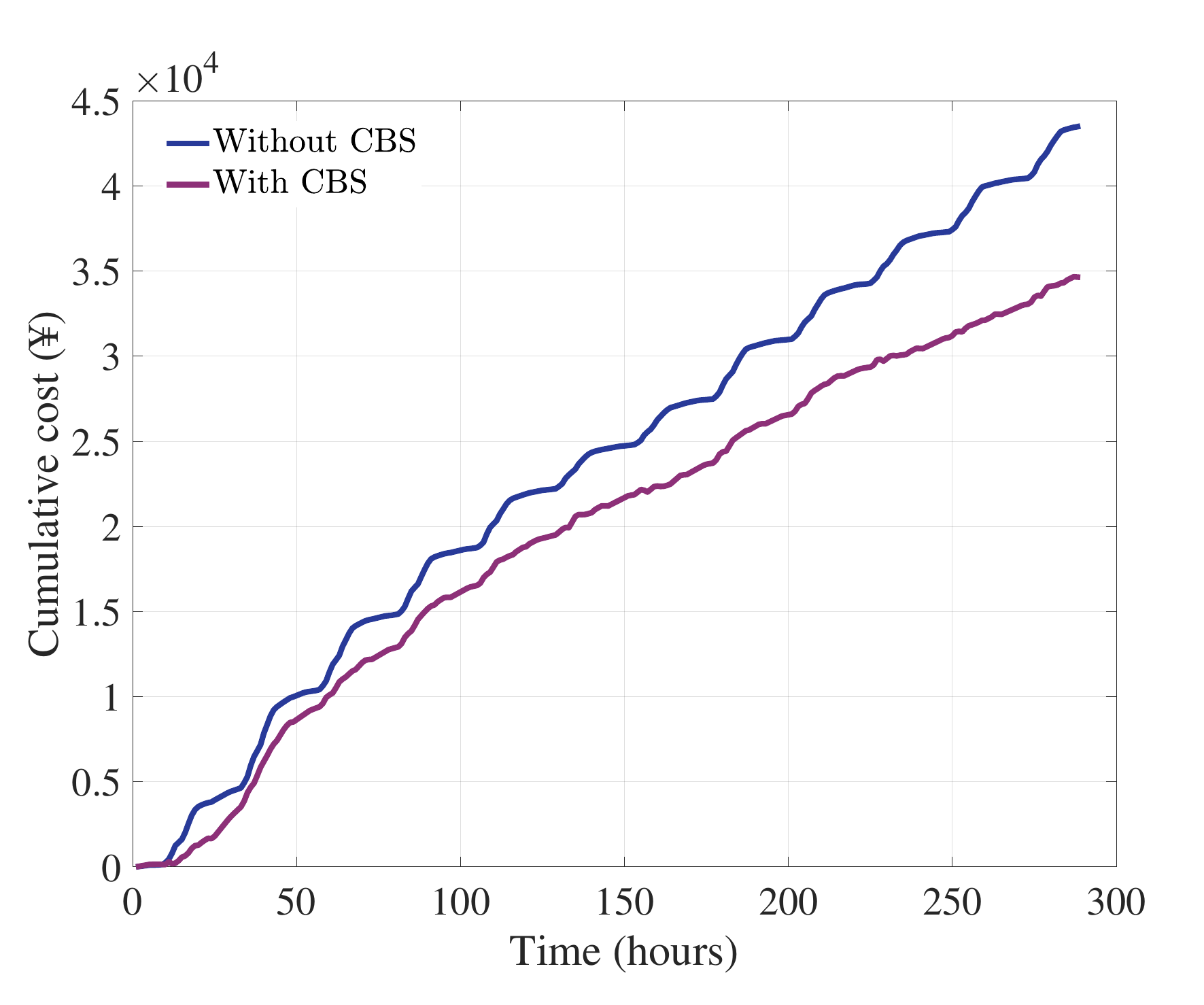}%
        \label{CBS_summer}}
    \hfil
    \subfloat[]{\includegraphics[width=0.45\textwidth]{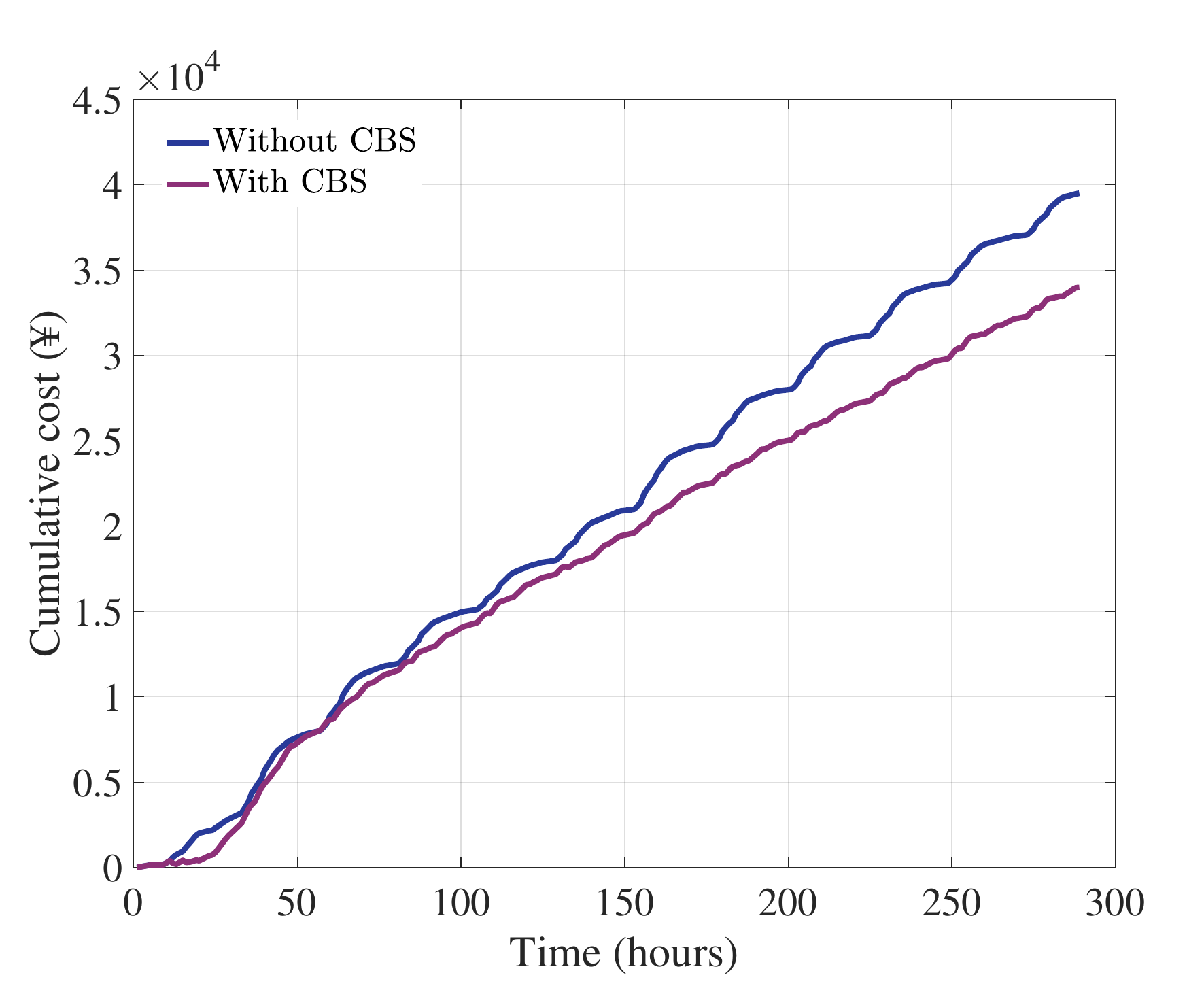}%
        \label{CBS_winter}}
    \caption{Comparison of operating costs without CBS in different seasons. (a) Summer; (b) Winter.}
    \label{fig:CBS}
\end{figure*}

We analyze the importance of the CBS for system operations. Fig. \ref{fig:CBS} presents the operational costs of the proposed method when scheduling the system without CBS component. The impact of CBS on the energy management process is primarily reflected in the following aspects. Firstly, the revenue generated by the direct sale of power to the PG by the CBS. Secondly, the cost savings for the BEMS by the CBS supplying energy to the buildings, which can be equated to revenue. The operation of the CBS incurs costs, including the cycle aging costs of the ESS and EV, as well as the calendar aging costs of the ESS. Removing the CBS, the net load demand of the system is entirely met by the energy purchased from the PG. Fig. \ref{fig:CBS}, it can be seen that compared to with CBS, the operational costs significantly increase without CBS, rising by 25.65\% and 16.24\% in summer and winter, respectively. This additional operational cost shows an increasing trend with extended system operation time. Table \ref{CBScost} provides the additional operational costs incurred by the system running continuously without the CBS under the proposed scheduling method. For the average additional cost in the last row of Table \ref{CBScost}, the system running without CBS incurs increased additional costs of 7201.81 yuan, 12075.80 yuan, and 32131.67 yuan after running for 288, 576, and 1152 hours, respectively. 

These results underscore the economic significance of the CBS for BEMS operations. It not only allows for arbitrage based on electricity price signals but also saves costs by prioritizing meeting building loads. Moreover, the CBS plays numerous critical roles in BEMS that are challenging to quantify in purely economic terms. For instance, it enhances the flexibility and resilience of building microgrids, acts as a backup power source to improve building power supply reliability, and ensures the balance and stable operation of supply and demand for the PG.

% From the building GB and reliability and stable energy supply, the agent can make a greet option of charge and discharge.
\subsubsection{Analysis of the operation by individual energy storage device}
%只有单电池系统运行的盈利能力 和 双电池比较  图  证明EV加入的必要性。
\begin{table}[htbp]
    \centering
    \begin{threeparttable}
        \caption{Comparison of additional costs without EV under different operating durations}
        \label{tab:EV_operating_cost}
        \begin{tabular}{lccc}
        \toprule
        \multirow{2}{*}{Season} & \multicolumn{3}{c}{Operating cost (¥)} \\
        \cmidrule(lr){2-4}
        & 288 (hours) & 576 (hours) & 1152 (hours) \\
        \midrule
        Summer & 1020.58 & 1992.82 & 3985.64 \\
        Winter & 1905.85 & 2591.70 & 5182.76\\
        Mean & 1463.21 & 2292.26 & 4584.20 \\
        \bottomrule
        \end{tabular}
    \end{threeparttable}
\end{table}

\begin{figure*}[hbt]
    \centering
    \subfloat[]{\includegraphics[width=0.45\textwidth]{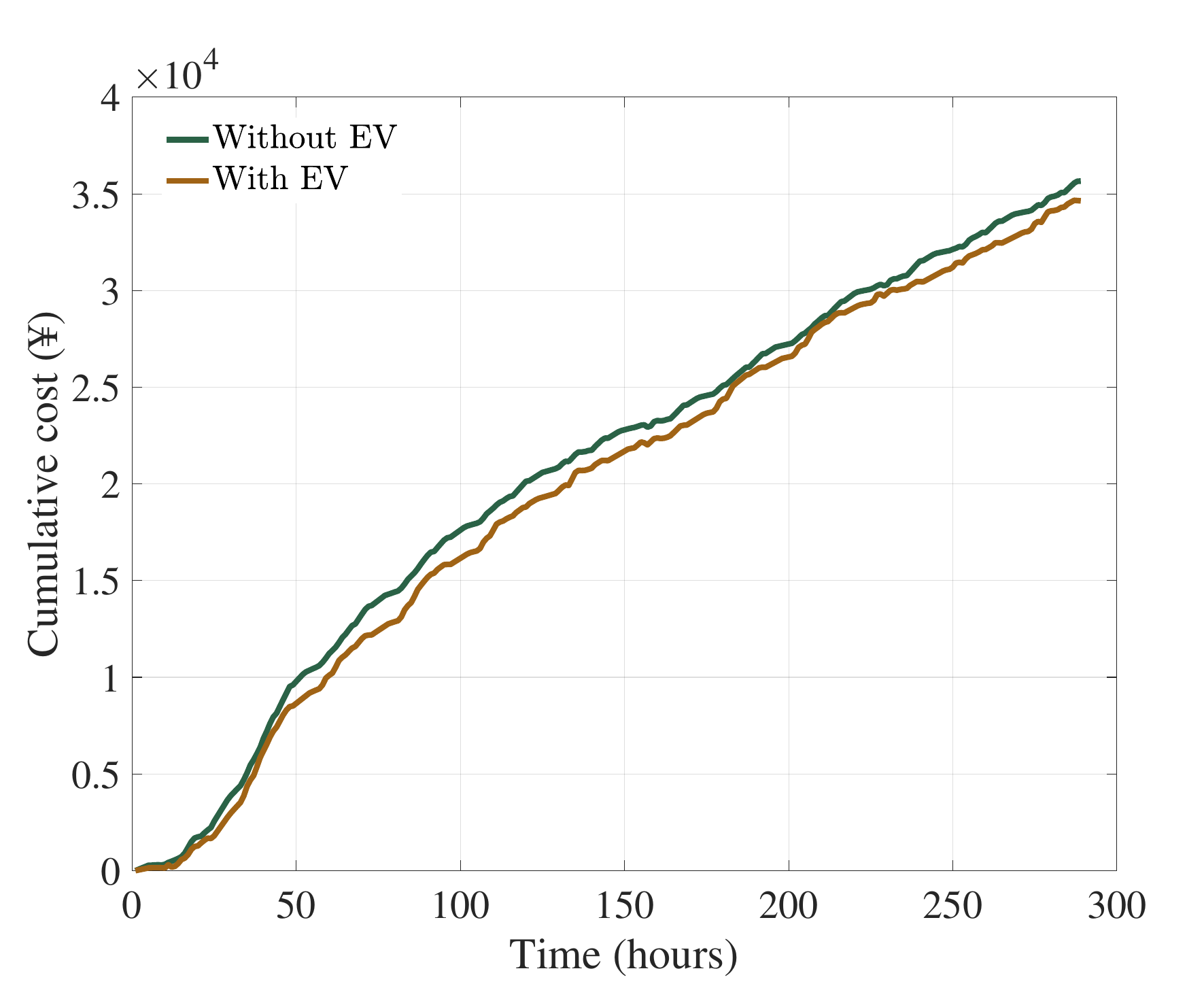}%
        \label{EV_summer}}
    \hfil
    \subfloat[]{\includegraphics[width=0.45\textwidth]{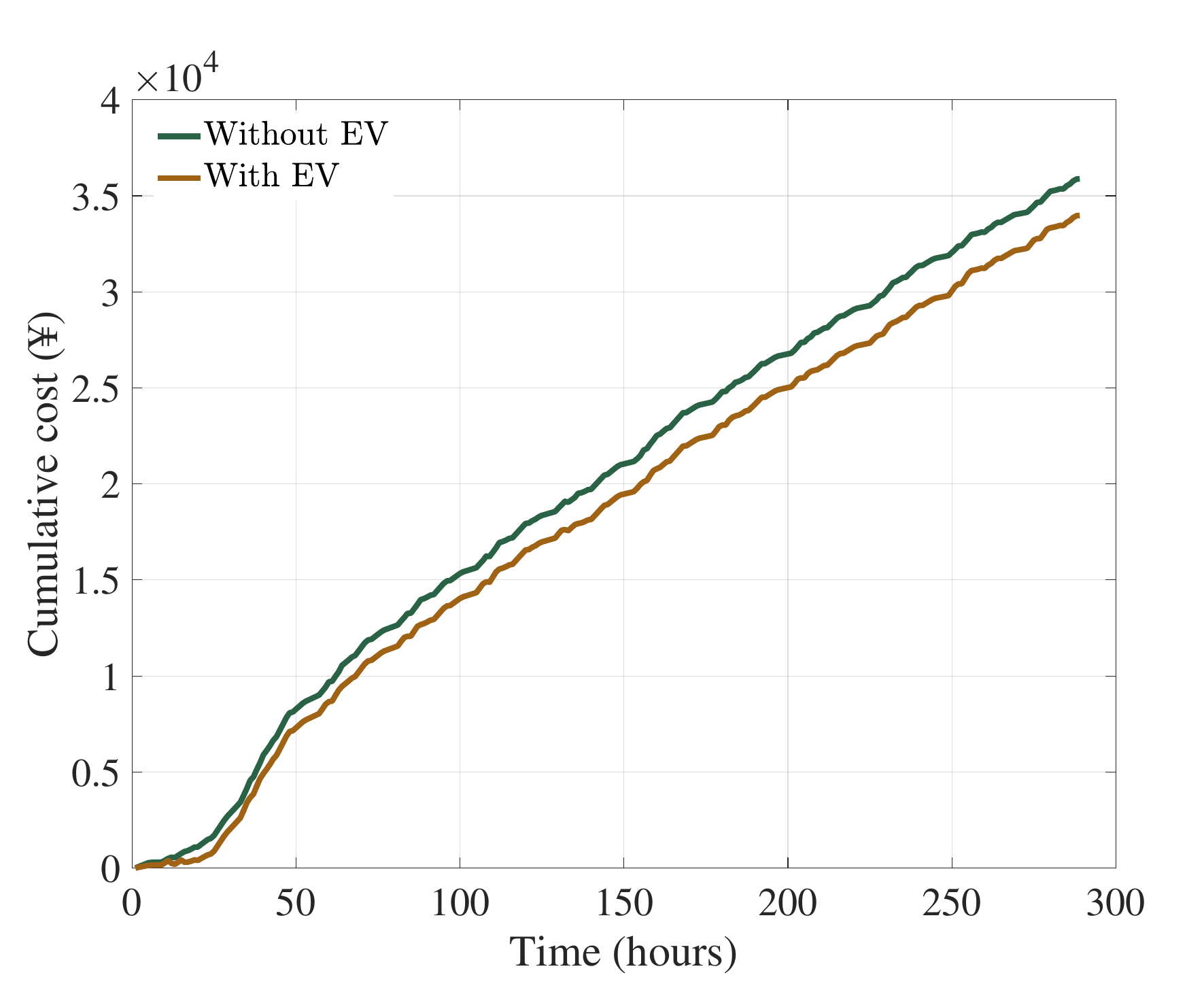}%
        \label{EV_winter}}
    \caption{Comparison of operating costs without EV in different seasons. (a) Summer; (b) Winter.}
    \label{fig:EV}
\end{figure*}

Based on the analysis of the CBS on system operations, we further examine the impact of EV on system operating costs. EV engage in arbitrage based on peak and valley tariffs while incurring battery cycle degradation costs. When EV component is removed, the system formulates charging and discharging control strategies based solely on an ESS single-battery system. In Fig. \ref{fig:EV}, the yellow line representing the system with EV shows lower operating costs compared to the green line for the system without EV. In summer and winter, the costs without EV are 2.95\% and 5.61\% higher than those with EV, respectively. This highlights the necessity of EV participation in BEMS scheduling, indicating that EV involvement can effectively reduce the system's operating costs. 

Notably, the additional revenue gained from EV participation is not as substantial as anticipated. Especially in summer, the cost evolution curves for the two scenarios are very close. This may be attributed to the design of our system model. EV users and building managers belong to two distinct interest entities. From the building manager's perspective, energy scheduling must prioritize the willingness of EV users to participate. The costs associated with battery degradation and driving range are the most significant factors for EV users when deciding to engage in scheduling. To ensure the reasonableness and enthusiasm of EV participation in scheduling, we consider not only the state of health (SoH) of EVs but also account for cycle aging costs and range anxiety in the joint energy scheduling model. Consequently, under the strict constraints of these factors, the available controllable capacity from EV is limited, impacting the scheduling benefits.

To further analyze the cost-saving potential of EV participation in scheduling, we calculated the additional costs of system operation without EV, as shown in Table \ref{tab:EV_operating_cost}. It can be observed that with continued scheduling, the additional costs without EV gradually increase. In winter, after running for 1152 hours without EV, the additional costs reach 5182.76 yuan. The results in Table \ref{tab:EV_operating_cost} highlight the potential for sustained profitability from EV participation in scheduling. In practical EV scheduling scenarios, some users may be less sensitive to range anxiety, potentially enhancing revenue from EV involvement. Similar to the CBS, EV participation plays a crucial role in enhancing the flexibility of microgrids and the stability of the power grid. Moreover, effectively utilizing EV flexible storage resources can reduce the budget for ESS capacity configuration.

\subsection{Analysis of battery degradation}
%分析ESS和EV电池的调度损耗成本和SoH变化  证明智能体学到了多调度ess，并且两种电池的损耗SoH都小，不影响用户参与调度的积极性。

\subsubsection{Ablation study of the degradation model}
\begin{figure*}[hbt]
    \centering
    \subfloat[]{\includegraphics[width=0.45\textwidth]{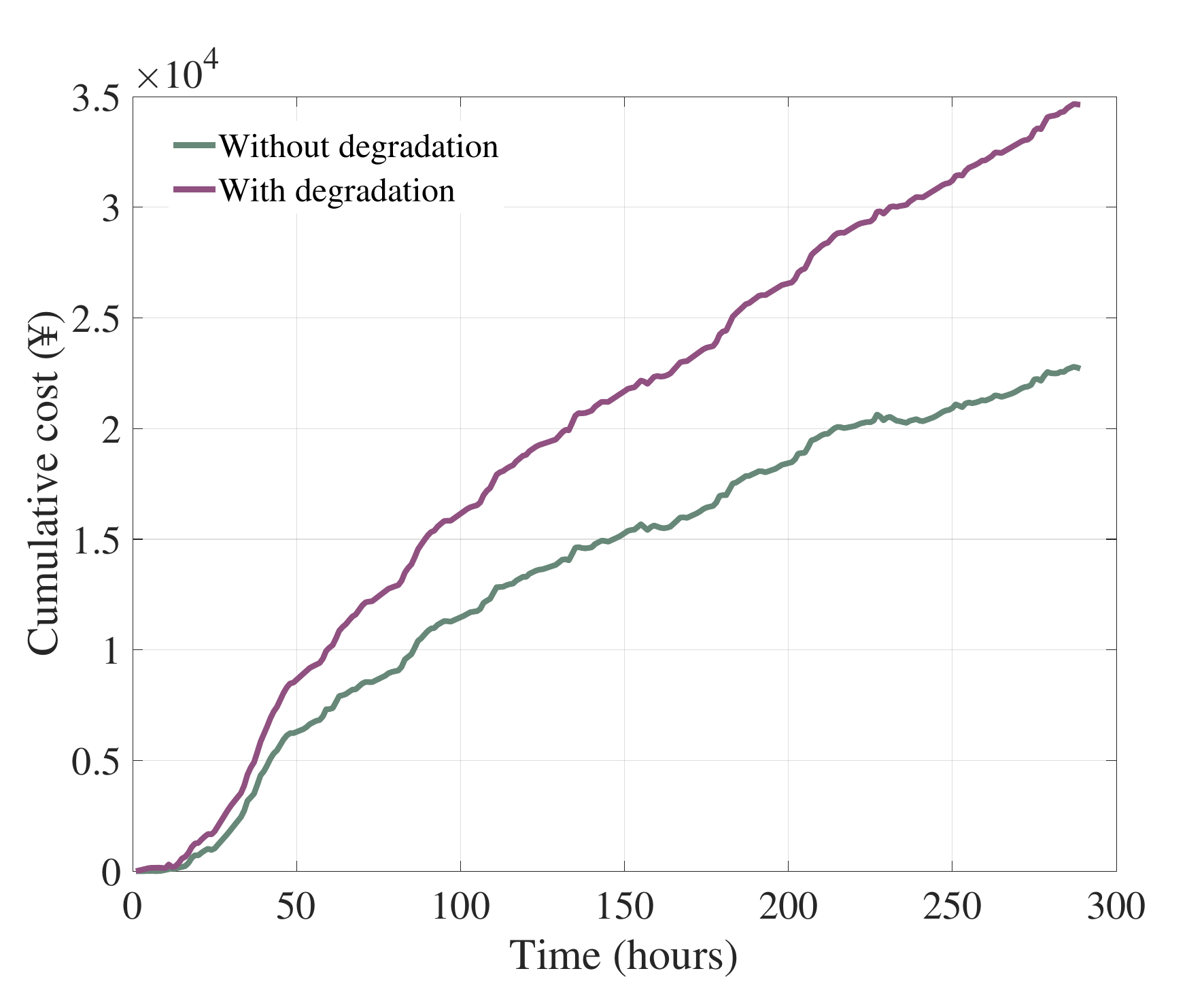}%
        \label{deg_summer}}
    \hfil
    \subfloat[]{\includegraphics[width=0.45\textwidth]{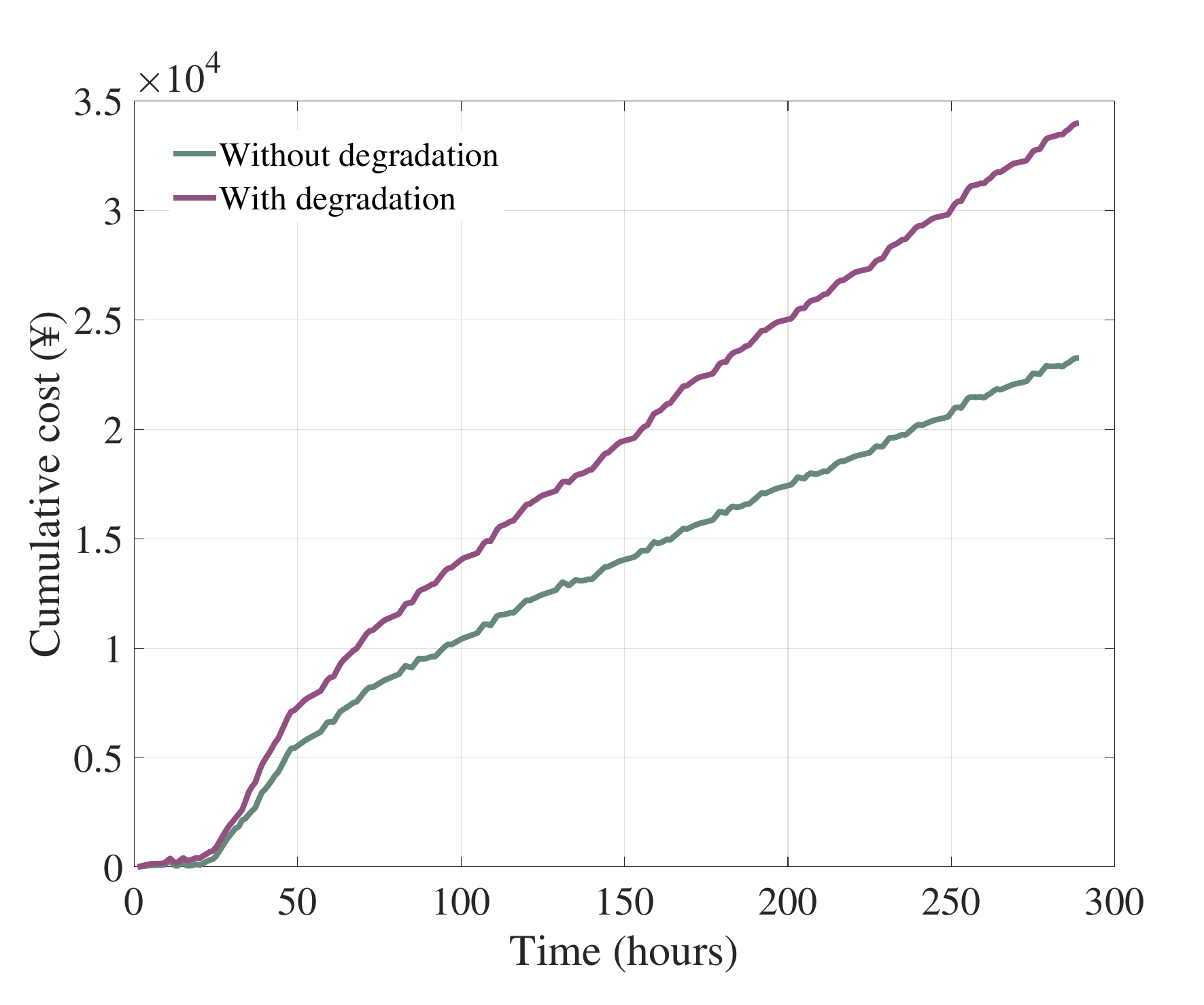}%
        \label{deg_winter}}
    \caption{Comparison of operating costs without considering battery degradation in different seasons. (a) Summer; (b) Winter.}
    \label{fig:deg_trend}
\end{figure*}
The influence of the accurate multi-battery degradation model on the estimation of system arbitrage costs is discussed. Fig. \ref{fig:deg_trend} illustrates the operational costs of ESS and EVs in different seasons without accounting for battery degradation costs. As shown, the system's operating costs significantly decrease when degradation costs are ignored, with reductions of 34.44\% in summer and 31.58\% in winter. However, this overestimation of CBS arbitrage potential is unrealistic. In addition, the development of scheduling plans that do not consider degradation costs can lead to numerous issues. On one hand, energy scheduling without considering battery degradation costs significantly impact battery lifespan. On the other hand, EVs and ESS are associated with different stakeholders. For EVs, which primarily serve as transportation tools, overly frequent charging and discharging control can compromise their utility. As a result, the willingness of EV users to participate in scheduling may decrease significantly, raising concerns about the rationality of CBS operations. Therefore, accurately modeling the degradation characteristics of different ESDs within the DRL-based scheduling framework is essential. This approach provides decision support for a reasonable and efficient energy management plan.

\subsubsection{Evaluation of degradation costs for batteries participating in scheduling}

\begin{figure*}[hbt]
    \centering
    \subfloat[Summer]{\includegraphics[width=0.45\textwidth]{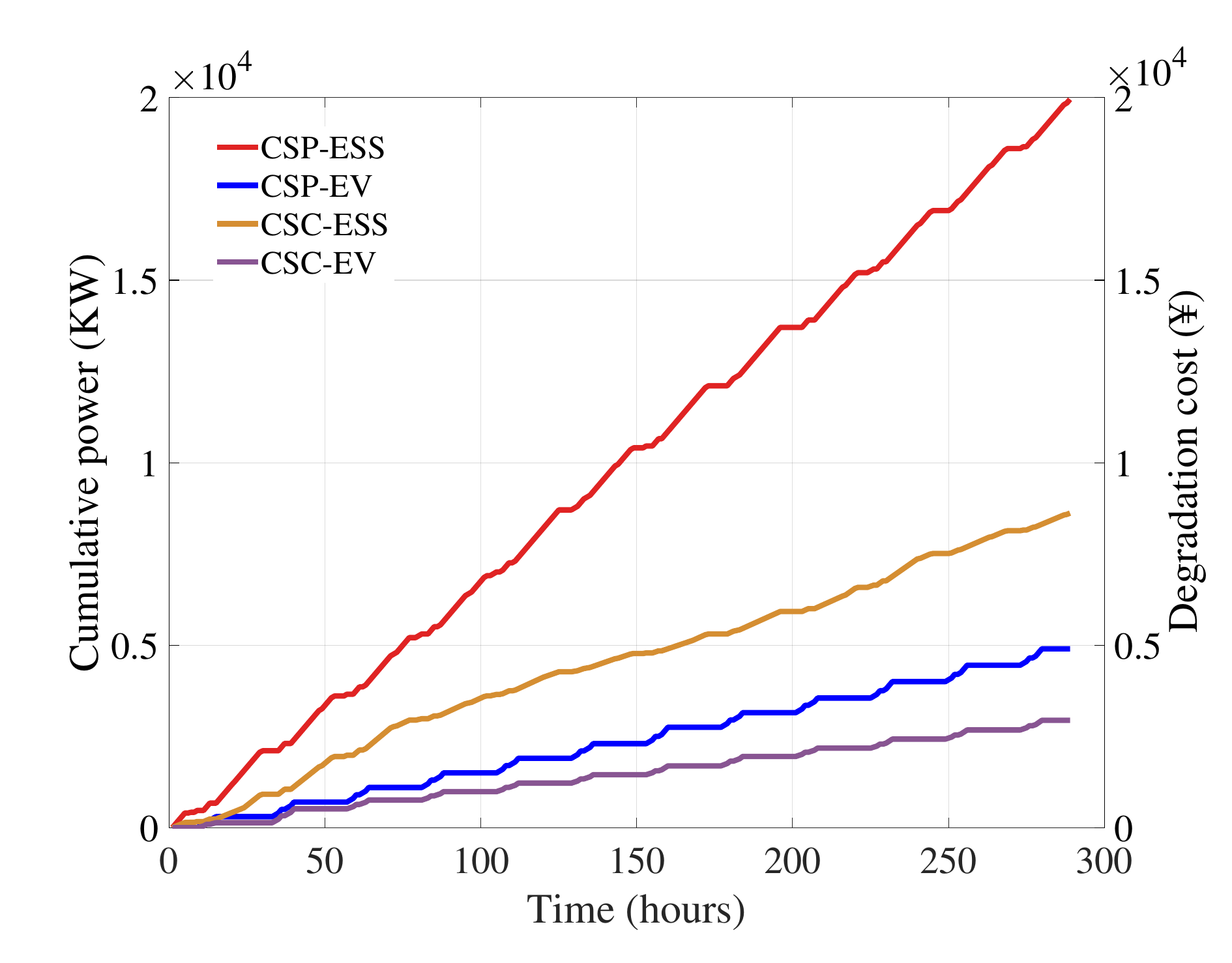}%
        \label{deg_cost_summer}}
    \hfil
    \subfloat[Winter]{\includegraphics[width=0.45\textwidth]{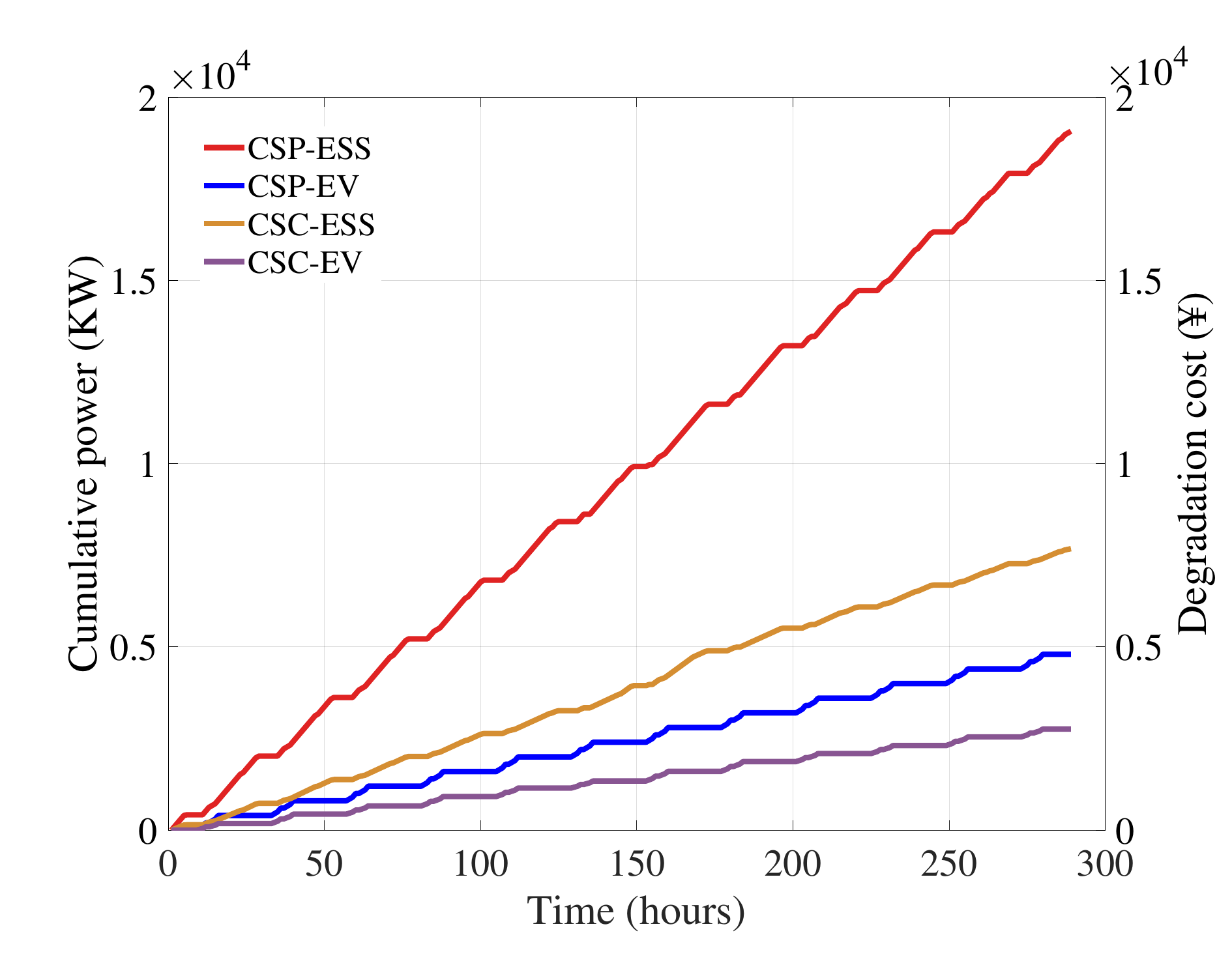}%
        \label{deg_cost_winter}}
    \caption{The cumulative scheduling power and cost of batteries in different seasons. (a) Summer; (b) Winter.}
    \label{fig:deg_cost}
\end{figure*}

We investigated the degradation of ESS and EV batteries during the scheduling process, and further analyzed the effectiveness of joint scheduling strategies. Fig. \ref{fig:deg_cost} presents the cumulative scheduling power (CSP) and cumulative scheduling cost (CSC) for ESS and EVs. It is clear that the CSP of ESS is significantly higher than that of EVs, which can be attributed to two primary reasons. First, EVs participate in BEMS scheduling only during parking periods. When EVs are not in the parking lot during non-working hours, they cannot be scheduled, leading to a reduced total CSP. Second, the cycle aging cost of EV batteries is higher than that of ESS batteries. To protect the lifespan of EV batteries and maintain user enthusiasm for scheduling, we have incorporated coefficient $w^\text{EV}$ in the reward function. This incentivizes the agent to schedule EVs only when sufficient profit prospects exist, thereby reducing the CSP for EVs.

\begin{figure*}[hbt]
    \centering
    \subfloat[]{\includegraphics[width=0.45\textwidth]{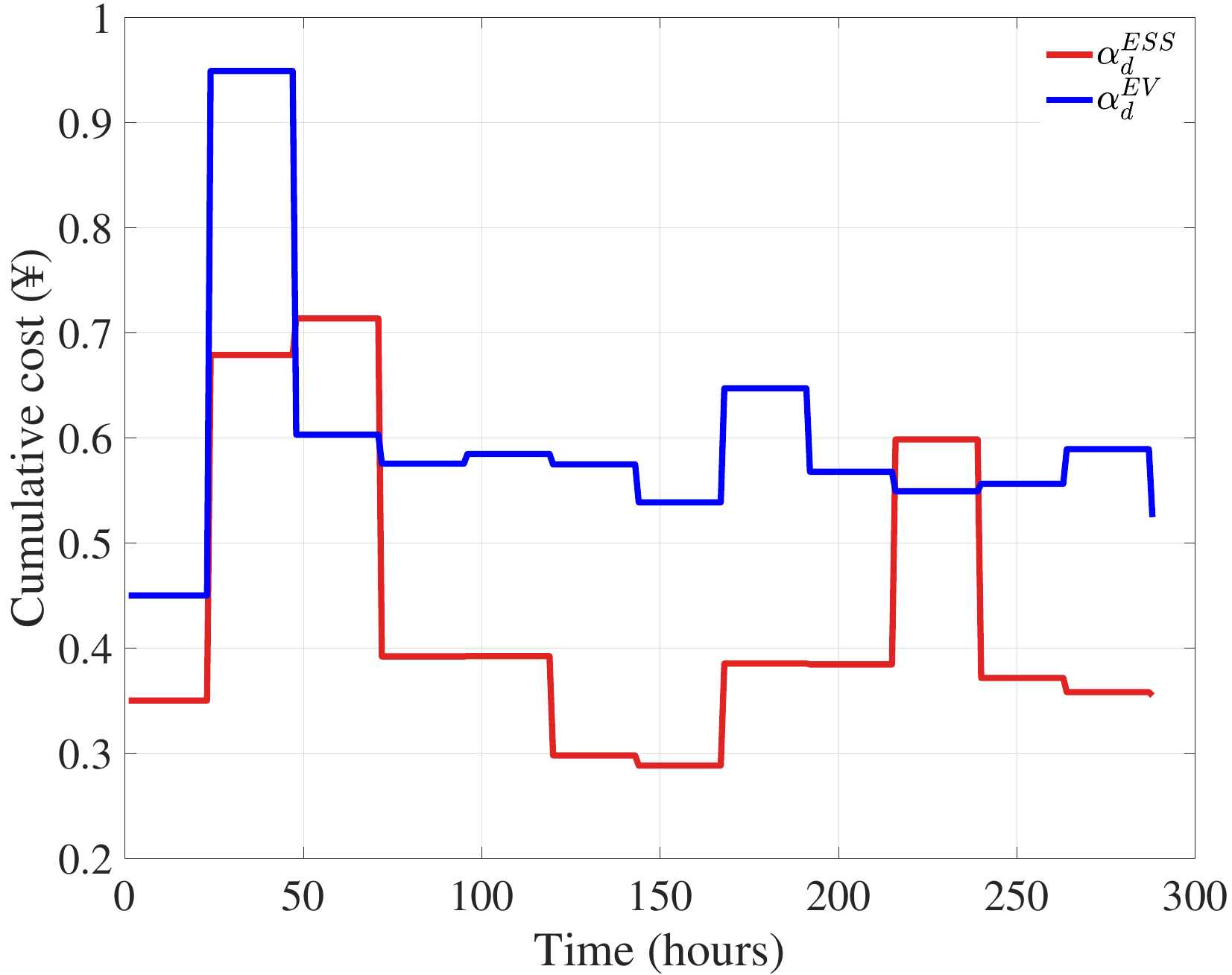}%
        \label{deg_coe_summer}}
    \hfil
    \subfloat[]{\includegraphics[width=0.45\textwidth]{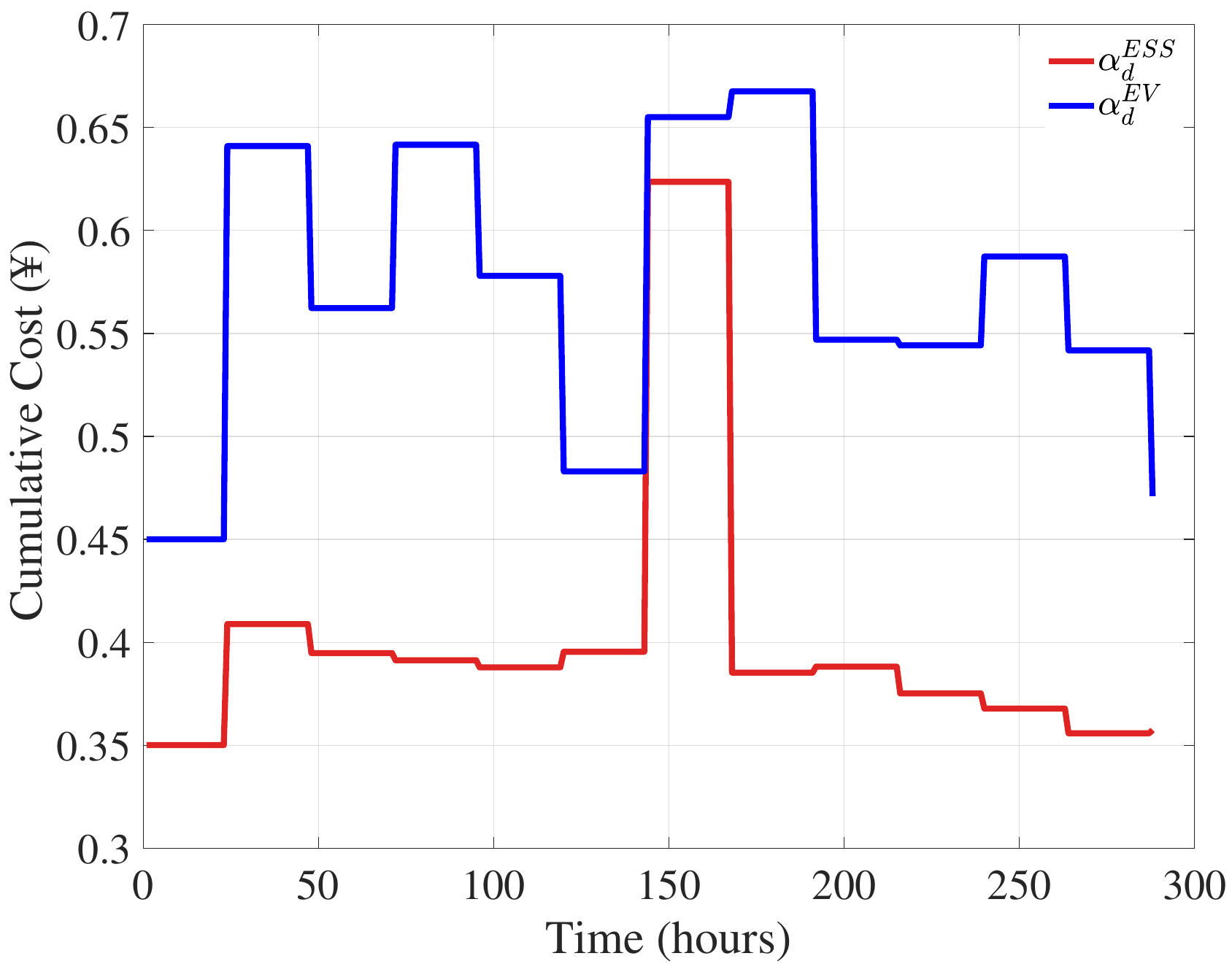}%
        \label{deg_coe_winter}}
    \caption{The evolution of battery degradation coefficient in different seasons. (a) Summer; (b) Winter.}
    \label{fig:deg_coe}
\end{figure*}

The CSC illustrates the differences degradation costs for ESS and EVs under the optimized joint scheduling strategy. Fig. \ref{fig:deg_cost} shows that, regardless of the season, the CSC of ESS is higher than that of EV, with values being 66.65\% and 73.93\% higher in summer and winter, respectively. Given the significant difference in CSP between ESS and EV, this difference in CSC is acceptable. Fig. \ref{fig:deg_coe} depicts the evolution of $\alpha ^\text{ESS}_{d}$ and $\alpha ^\text{EV}_{d}$. Although the CSP of EV is much lower than that of ESS, due to differences in cycle costs, $\alpha ^\text{EV}_{d}$ is typically greater than $\alpha ^\text{ESS}_{d}$. Notably, in summer, $\alpha^\text{EV}_{d}$ approaches 1 in the early stages of scheduling. As scheduling progresses, $\alpha ^\text{EV}_{d}$ stabilizes around 0.5 across different seasons, mitigating further increases in EV degradation costs. 

The above results reflect the effectiveness of the proposed scheduling method. ESS batteries, primarily use LFP material, offer significant safety, long cycle life, and cost-effectiveness, prioritizing the use of ESS is a reasonable decision. Conversely, EV batteries, mainly utilizing NMC material, emphasize higher power and energy density, enabling efficient charging and discharging to meet the long-range and high-power demands of EVs. However, their cycle life is shorter than that of ESS batteries. This necessitates that the agent comprehensively considers both short-term arbitrage profits and long-term operational costs based on the different characteristics of the batteries. It is an optimal decision for the agent to schedule EVs only when the arbitrage profit is high. In other cases, the degradation costs of EV batteries should be prioritized.

\subsubsection{Evaluation of the SoH of batteries}
\begin{figure*}[hbt]
    \centering
    \subfloat[]{\includegraphics[width=0.45\textwidth]{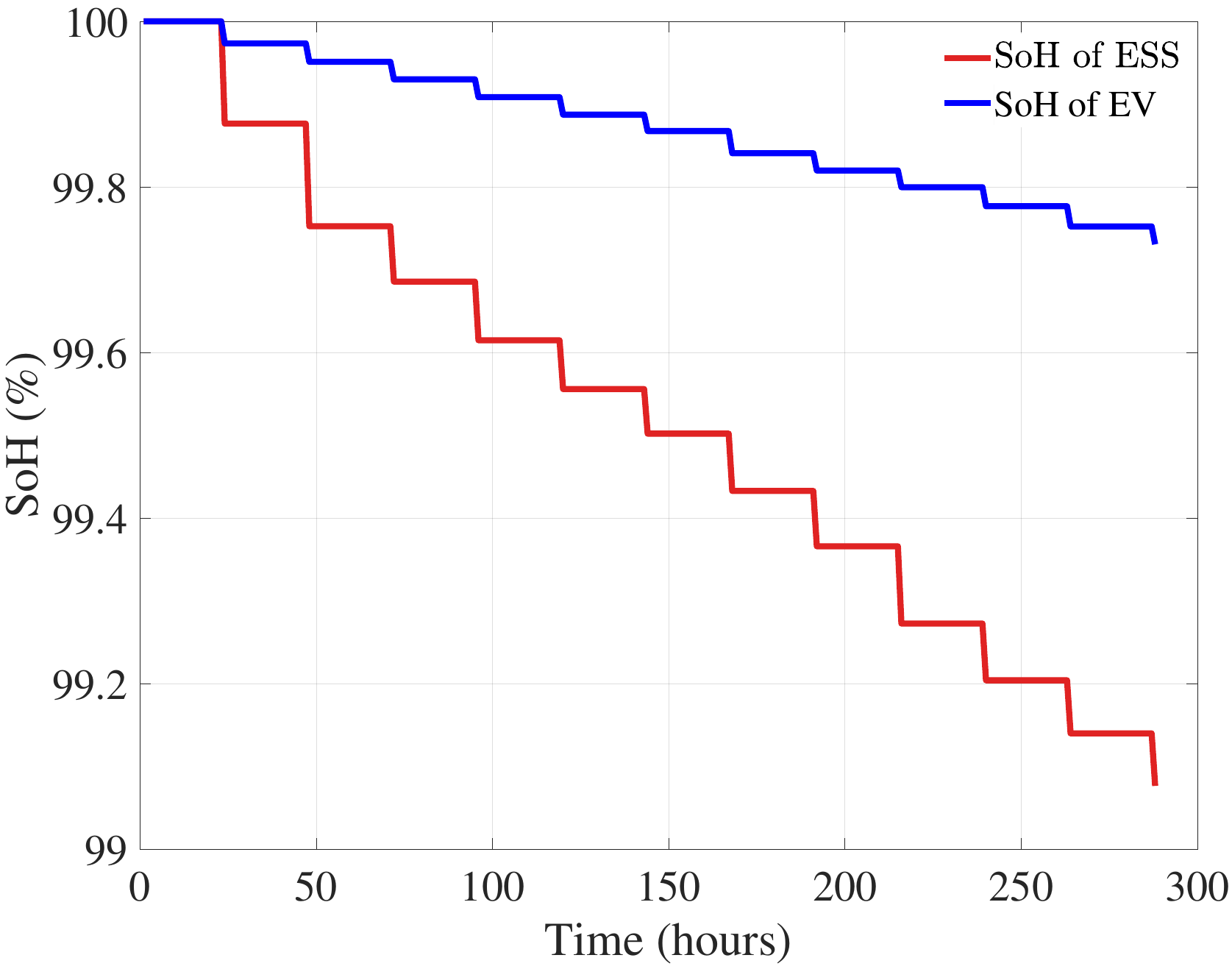}%
        \label{SoH_summer}}
    \hfil
    \subfloat[]{\includegraphics[width=0.45\textwidth]{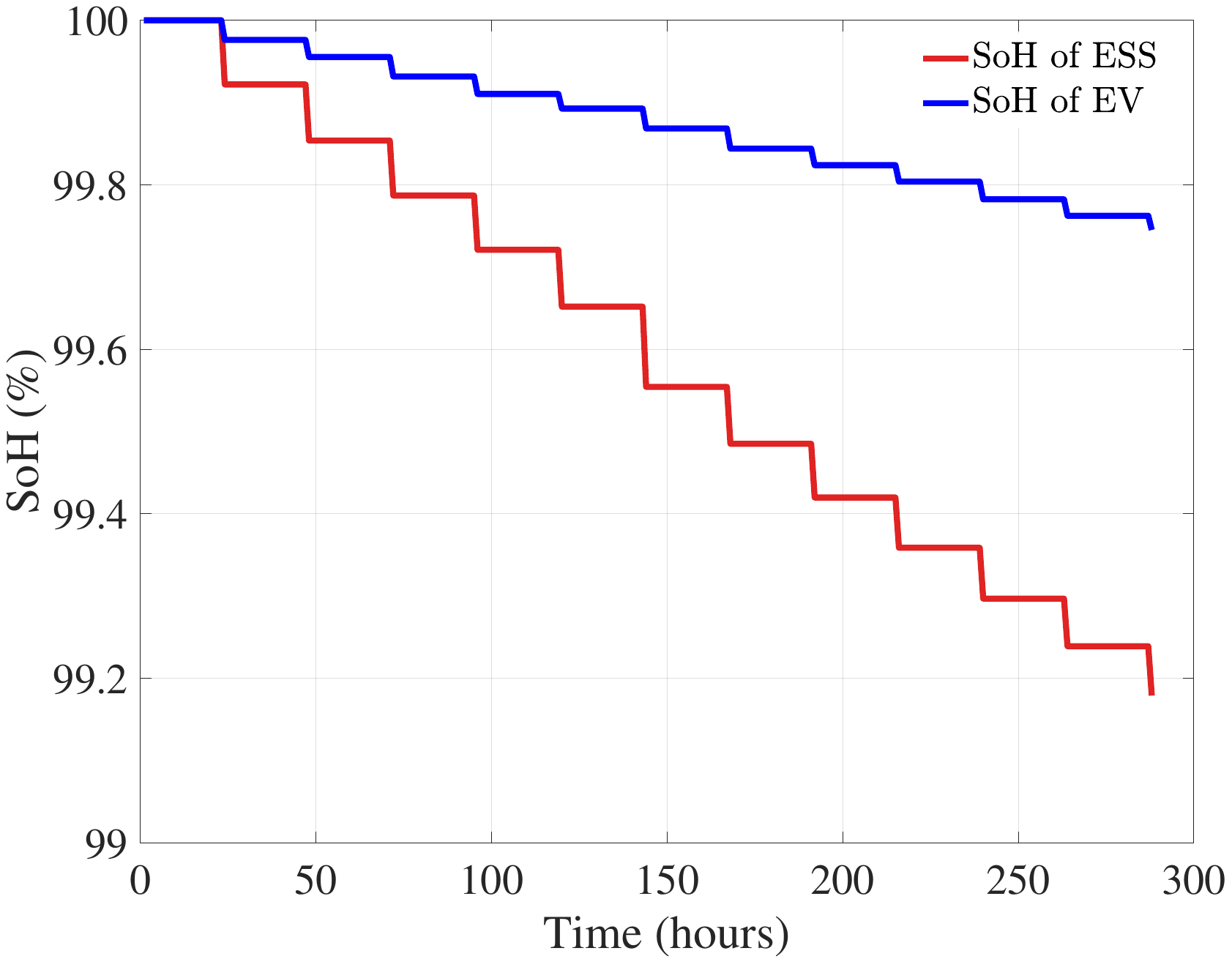}%
        \label{SoH_winter}}
    \caption{The evolution curve of battery SoH in different seasons. (a) Summer; (b) Winter.}
    \label{fig:SoH}
\end{figure*}
We examine the evolution of battery SoH for ESS and EVs participating in BEMS scheduling. Fig. \ref{fig:SoH} presents the evolution trends of the SoH for ESS and EV within the DRL joint real-time scheduling framework. In this study, cycle aging costs and calendar aging costs are calculated at each time step and at the end of each episode, respectively. The cycle aging costs provide real-time feedback for immediate rewards to the system, while the calendar aging costs ensure accuracy. Consequently, the SoH is updated at the end of each episode, resulting in a step-like evolution trend.

The evolution process of SoH indicates that, during scheduling, the SoH of both ESS and EV is maintained at relatively healthy levels. In summer, after scheduling on the test dataset, the SoH values for ESS and EV are 99.07\% and 99.73\%, respectively. In winter, these values are 99.18\% and 99.75\%, respectively. To encourage active participation from EV users in the BEMS scheduling process, the agent prioritizes the SoH of EV batteries, ensuring that capacity loss remains below 0.3\% in both summer and winter. These results demonstrate that the proposed scheduling method effectively accounts for the battery degradation costs of the CBS, enabling participation in system scheduling while maintaining battery health and reducing operational costs. Furthermore, EVs participation in scheduling does not significantly harm battery life.

\section{Conclusion}
In this study, we aim to minimize BEMS operational costs by proposing a model-free real-time energy scheduling framework based on a multi-battery storage system. First, we establish a CBS incorporating both ESS and EVs to enhance system flexibility and energy utilization efficiency. We then conduct an in-depth analysis of the significant differences in battery chemistry between ESS and EVs, discussing aging models and corresponding degradation costs relevant to each type. Next, we optimize the collaborative operation of the multi-battery system, aligning EV scheduling with user travel demands while accounting for operational differences in scheduling windows and degradation costs. Subsequently, a prediction network is introduced in the DRL framework to optimize local energy consumption and operational costs. Lastly, we develop an improved DQN scheduling algorithm to tackle decision-making inefficiencies in complex, dynamic, and coupled system environments. We enhance scheduling decisions by integrating double networks and a dueling mechanism, while PER is employed to boost learning efficiency. The effectiveness of the proposed algorithm is validated in real dynamic settings, leading to the following conclusions.

Given the significant differences in usage and chemical properties between ESS and EVs, differentiated battery degradation modeling is crucial for accurately estimating the system's arbitrage potential. By considering variations in degradation costs and scheduling windows among different batteries, we facilitate more rational scheduling plans. Our approach coordinates the operations of ESS and EVs, carefully scheduling EVs during operational periods to enhance system profitability while ensuring battery health. Notably, the benefits derived from EV scheduling are not as high as those from the CBS, which is expected. As energy components primarily used for transportation, EVs must navigate constraints related to range, SoC limits, scheduling windows, and battery SoH,  all of which directly impact the arbitrage potential of EVs. The introduction of the prediction module and EAM has fostered a more beneficial energy interaction between the CBS and the system. The agent prioritizes building demands based on prediction information, avoiding high-priced electricity purchases from the PG, thus optimizing system costs. Furthermore, this design significantly benefits local energy consumption. Finally, it is necessary to enhance the DRL algorithm to address the specific issues of multi-battery energy systems. The double network and dueling mechanism improve Q-value estimation and action selection in multi-battery environments. The PER promotes efficient sample learning, thereby enhancing algorithm performance.

In future work, we will focus on addressing the uncertainties associated with EVs within the multi-battery DRL energy scheduling framework. By modeling EV user driving habits, we aim to mitigate the impact of the randomness of flexible energy storage devices on the system.

% \section*{CRediT authorship contribution statement}
% {\bf Zhezhuang Xu:} Conceptualization, Investigation, Project administration, Methodology, Writing – Original draft preparation. {\bf Jinlong Wang:} Methodology, Validation, Writing – Original draft preparation, Writing – review \& editing. {\bf Meng Yuan:} Conceptualization, Investigation, Methodology. {\bf Yazhou Yuan:} Conceptualization. {\bf Boyu Chen:} Resources, Data curation.
% {\bf Qingdong Zhang:} Resources, Data curation. {\bf Cailian Chen:} Formal Analysis, Methodology. {\bf Xinping Guan:} Formal Analysis, Methodology.

\section*{Declaration of competing interest}
The authors declare that they have no known competing financial interests or personal relationships that could have appeared to
influence the work reported in this paper.

\section*{Data availability}
Data will be made available on request.

\section*{Acknowledgement}
This work is supported by the National Natural Science Foundation of China under Grant 61973085, the European Union (EU)-funded Marie Skłodowska-Curie Actions (MSCA) Postdoctoral Fellowship under Grant Number 101110832, and the Overseas Qishan Scholars Program at Fuzhou University under Grant 511262.


%\bibliographystyle{cas-model2-names}
%\bibliographystyle{unsrt}
\bibliographystyle{elsarticle-num} 
\bibliography{ref}
% \bibliography{myref}

\end{document}